\documentclass{aa}  
\usepackage{graphicx}
\usepackage{txfonts}
\usepackage[colorlinks=true,citecolor=blue,linkcolor=blue,urlcolor=cyan]{hyperref}
\usepackage{pdflscape}
\newcommand{\nodata}{}
\renewcommand\labelitemii{$\m@th\bullet$}
\begin{document} 

\title{Chamaeleon DANCe}
\subtitle{Revisiting the stellar populations of Chamaeleon~I and Chamaeleon~II \\with Gaia-DR2 data\thanks{Tables \ref{tab_members_Cha1}, \ref{tab_members_Cha2}, \ref{tab_prob_Cha1}, \ref{tab_prob_Cha2}, \ref{tab_isochrone_Cha1}, \ref{tab_isochrone_Cha2}, and \ref{tab_HRD} are only available in electronic form at the CDS via anonymous ftp to cdsarc.u-strasbg.fr (130.79.128.5) or via http://cdsweb.u-strasbg.fr/cgi-bin/qcat?J/A+A/}}

\author{
P.A.B.~Galli \inst{1}
\and
H.~Bouy \inst{1}
\and
J.~Olivares\inst{1}
\and 
N.~Miret-Roig\inst{1}
\and
L.M.~Sarro\inst{2}
\and
D.~Barrado\inst{3}
\and
A.~Berihuete\inst{4}
\and
E.~Bertin\inst{5}
\and
J.-C.~Cuillandre\inst{6}
}

\institute{
Laboratoire d’Astrophysique de Bordeaux, Univ. Bordeaux, CNRS, B18N, allée Geoffroy Saint-Hillaire, F-33615 Pessac, France\\
\email{phillip.galli@u-bordeaux.fr}
\and
Depto. de Inteligencia Artificial, UNED, Juan del Rosal, 16, 28040 Madrid, Spain
\and
Centro de Astrobiolog\'ia, Depto. de Astrof\'isica, INTA-CSIC, ESAC Campus, Camino Bajo del Castillo s/n, 28692 Villanueva de la Ca\~nada, Madrid, Spain
\and
Dpto. Estadística e Investigación Operativa, Universidad de Cádiz, Campus Río San Pedro s/n, 11510 Puerto Real, Cádiz, Spain
\and
Sorbonne Universit\'e, CNRS, UMR 7095, Institut d'Astrophysique de Paris, 98 bis bd Arago, 75014 Paris, France.
\and
AIM Paris Saclay, CNRS/INSU, CEA/Irfu, Université Paris Diderot, Orme des Merisiers, France
}

\date{Received XXX; accepted XXX}

 \abstract
{Chamaeleon is the southernmost low-mass star-forming complex within 200~pc from the Sun. Its stellar population has been extensively studied in the past, but the current census of the stellar content is not complete yet and deserves further investigation.}
{We take advantage of the second data release of the \textit{Gaia} space mission to expand the census of stars in Chamaeleon and to revisit the properties of the stellar populations associated to the Chamaeleon~I (Cha~I) and Chamaeleon~II (Cha~II) dark clouds.}
{We perform a membership analysis of the sources in the \textit{Gaia} catalogue over a field of 100~deg$^{2}$ encompassing the Chamaeleon clouds, and use this new census of cluster members to investigate the 6D structure of the complex. }
{We identify 188 and 41 high-probability members of the stellar populations in Cha~I and Cha~II, respectively, including 19 and 7 new members. Our sample covers the magnitude range from $G=6$ to $G=20$~mag in Cha~I, and from $G=12$ to $G=18$~mag in Cha~II. We confirm that the northern and southern subgroups of Cha~I are located at different distances ($191.4^{+0.8}_{-0.8}$~pc and $186.7^{+1.0}_{-1.0}$~pc), but they exhibit the same space motion within the reported uncertainties. Cha~II is located at a distance of $197.5^{+1.0}_{-0.9}$~pc and exhibits a space motion that is consistent with Cha~I within the admittedly large uncertainties on the spatial velocities of the stars that come from radial velocity data. The median age of the stars derived from the Hertzsprung-Russell diagram (HRD) and stellar models is about 1-2~Myr, suggesting that they are somewhat younger than previously thought. We do not detect significant age differences between the Chamaeleon subgroups, but we show that Cha~II exhibits a higher fraction of disc-bearing stars compared to Cha~I.}
{This study provides the most complete sample of cluster members associated to the Chamaeleon clouds that can be produced with \textit{Gaia} data alone. We use this new census of stars to revisit the 6D structure of this region with unprecedented precision.   }

\maketitle

\section{Introduction}\label{section1}

The molecular cloud complex in the southern constellation of  Chamaeleon hosts one of the richest populations of T~Tauri stars in the Solar neighbourhood. It consists of three molecular clouds with angular sizes of a few degrees: Chamaeleon~I (Cha~I), Chamaeleon~II (Cha~II), and Chamaeleon~III (Cha~III). Star formation activity is restricted to Cha~I and Cha~II as no young stars have been detected in Cha~III \citep[see][]{Luhman2008_review}. The cloud morphology has been investigated in previous studies from star counts and extinction maps \citep[see e.g.][]{Gregorio-Hetem1988,Cambresy1997,Mizuno2001,Dobashi2005} revealing moderate extinction levels in Chamaeleon as compared to other nearby star-forming regions such as for example Ophiuchus \citep[][]{Cambresy1999}. The modest extinction, proximity, compact structure and isolated location with respect to other young stellar groups led to Chamaeleon being one of the most targeted regions over the past decades for the study of low-mass star formation. 
 
The first young stellar objects associated to the Chamaeleon clouds were mostly classical T~Tauri stars (CTTSs) identified from objective prism surveys based on their strong H$\alpha$ emission \citep{Henize1973,Schwartz1977,Hartigan1993}. Deeper H$\alpha$ surveys conducted in the years that followed added a number of very low-mass members and brown dwarfs to the region \citep{Comeron1999,Comeron2000,Neuhauser1999}. \textit{ROSAT X-ray} pointed observations led to the discovery of a significant number of weak-emission-line T~Tauri stars (WTTSs) in this region \citep{Feigelson1993,Alcala1995,Alcala2000} and the first brown dwarf detected in \textit{X-}rays \citep{Neuhauser1998}. Additional members were later discovered from data collected with the \textit{XMM-Newton}  \citep{Stelzer2004,Robrade2007} and \textit{Chandra X-ray} observatories \citep{Feigelson2004}. Several optical and infrared surveys followed by ground-based spectroscopic observations expanded the census of Chamaeleon stars and confirmed the most likely members associated to the molecular clouds \citep[see e.g.][]{Hughes1992,Prusti1992,Cambresy1998,Persi2000,Vuong2001,Carpenter2002,Barrado2004,Comeron2004,LopezMarti2004,LopezMarti2005,Luhman2004a,Luhman2004b,Allers2007,Luhman2007}. The most recent census of the stellar population of Cha~I is given by \citet{Esplin2017} and contains 250~stars. Analogously, the list of 63~stars studied by \citet{Alcala2008} and \citet{Spezzi2008} represents the most complete sample of stars in Cha~II to date. 

The distance to Chamaeleon has undergone extensive revision in recent decades. Early studies reported distance estimates to Cha~I in the range from 115 to 215~pc \citep{Grasdalen1975,Rydgren1980,Hyland1982,Schwartz1992}, and suggested that the distance to Cha~II could be as large as 400~pc \citep{Fitzgerald1976,Graham1988}. \citet{Franco1991} investigated the interstellar extinction of field stars projected towards the Chameleon clouds and estimated distances of 140~pc and 158$\pm$40~pc to Cha~I and Cha~II, respectively. A subsequent study conducted by \citet{Hughes1992} based on the same technique derived the distance of 200$\pm$20~pc to Cha~II. In the following years, \citet{Whittet1997} derived the more robust distance estimates of 160$\pm$15~pc and 178$\pm$18~pc for Cha~I and Cha~II, respectively, based on multiple distance indicators. \citet{Bertout1999}  computed the distance of 168$^{+14}_{-12}$~pc to Cha~I from the trigonometric parallaxes delivered by the \textit{Hipparcos} satellite \citep{Hipparcos} for a few stars in this region. A major contribution to the effort in constraining the distance to the Chamaeleon clouds was made by \citet{Voirin2018}, where the authors combined the extinction distribution of field stars projected towards the clouds with the parallaxes delivered by the first data release of the \textit{Gaia} space mission \citep[Gaia-DR1,][]{GaiaDR1} to estimate the distances of $179^{+11+11}_{-10-10}$~pc, $181^{+6+11}_{-5-10}$~pc, $199^{+8+12}_{-7-11}$~pc to Cha~I, Cha~II and Cha~III, respectively. This study put Cha~I about 20~pc further away from previous estimates and returned the first distance determination to the Cha~III molecular cloud. However, the systematic uncertainties of 0.3~mas in the Gaia-DR1 parallaxes \citep{Lindegren2016} largely dominated the distance uncertainties obtained in that study and called for a revision of the results. More recently, \citet{Roccatagliata2018} used the parallaxes from the second data release of the \textit{Gaia} space mission \citep[Gaia-DR2,][]{GaiaDR2} to revisit the distance to Cha~I. These latter authors reported distances of $192.7^{+0.4}_{-0.4}$~pc and $186.5^{+0.7}_{-0.7}$~pc to the northern and southern subgroups of stars in this cloud, respectively. The improved precision level in the distance determination is  related to the more precise parallaxes, but also to the fact that the systematic errors of the Gaia-DR2 catalogue \citep[see e.g.][]{Lindegren2018,Luri2018} were not modelled in that solution. 

Kinematic studies have proven to be fundamental in distinguishing between the different subgroups of the Chamaeleon region and to searching for new cluster members. For example, \citet{LopezMarti2013a} found evidence that the stars in Cha~I and Cha~II have different proper motions, and that they also differ from the adjacent $\epsilon$~Cha and $\eta$~Cha associations that populate the same region of the sky. In a subsequent study, \citet{LopezMarti2013b} identified new kinematic members in Cha~I and Cha~II located in the outskirts of the molecular clouds and argued that this dispersed population could be larger, but more accurate data would be required to confirm this hypothesis. Indeed, the scarcity of trigonometric parallaxes and radial velocity (RV) information for most stars in Chamaeleon has been the main limitation to studying the kinematic properties of this region. This situation has dramatically changed with the advent of the Gaia-DR2 catalogue combined with the spectroscopic observations conducted by the \textit{Gaia}-ESO Survey \citep{Gilmore2012} in the Chamaeleon region \citep[see e.g.][]{Sacco2017}. Altogether, this puts us in a timely position to investigate the 3D structure and 3D space motion of the Chamaeleon star-forming complex with unprecedented precision as discussed throughout this study.  

This paper is one in a series conducted in the context of the Dynamical Analysis of Nearby Clusters project \citep[DANCe,][]{Bouy2013}. Here, we investigate the census of stars, and the structure and kinematic properties of the Chamaeleon star-forming region in light of Gaia-DR2 data. The paper is organised as follows. In Section~\ref{section2} we compile the lists of stars in Cha~I and Cha~II published in the literature, and perform a new membership analysis to confirm the historical members associated with these clouds and discover new ones. In Section~\ref{section3} we revisit the distance, spatial velocity, and age of the Chamaeleon subgroups based on our new sample of cluster members selected in this study. Finally, we summarise our results and conclusions in Section~\ref{section4}.

\section{Membership analysis}\label{section2}

Our strategy to assess membership is based on the methodology previously developed by our team \citep{Sarro2014,Olivares2019}. In this section, we describe the main steps of our membership analysis applied to the Chamaeleon star-forming region and we refer the reader to the original papers for more details on the performance and implementation of our classifier.  

\subsection{Field and cluster models}\label{section2.1}

The representation space (i.e., set of observables) that we use in the membership analysis includes the astrometric and photometric features of the stars provided in the Gaia-DR2 catalogue. We do not include the blue photometry $G_{BP}$ in the analysis because of the calibration problems in this band as reported in the literature \citep[see e.g.][]{Apellaniz2018} which can affect our selection of cluster members particularly in the faint end. We therefore restrict the membership analysis to the space of observables defined by $\mu_{\alpha}\cos\delta$, $\mu_{\delta}$, $\varpi$, $G_{RP}$ and $G-G_{RP}$. We then downloaded the Gaia-DR2 catalogue in the region of the sky defined by $295^{\circ}\leq l \leq 305^{\circ}$ and $-20^{\circ}\leq b \leq -10^{\circ}$ in Galactic coordinates which encompasses the three molecular clouds of the complex (Cha~I, Cha~II, and Cha~III). This field includes 4\,433\,409~sources and 3\,904\,492 of them have complete data in our representation space. 

\begin{figure*}[!h]
\begin{center}
\includegraphics[width=0.33\textwidth]{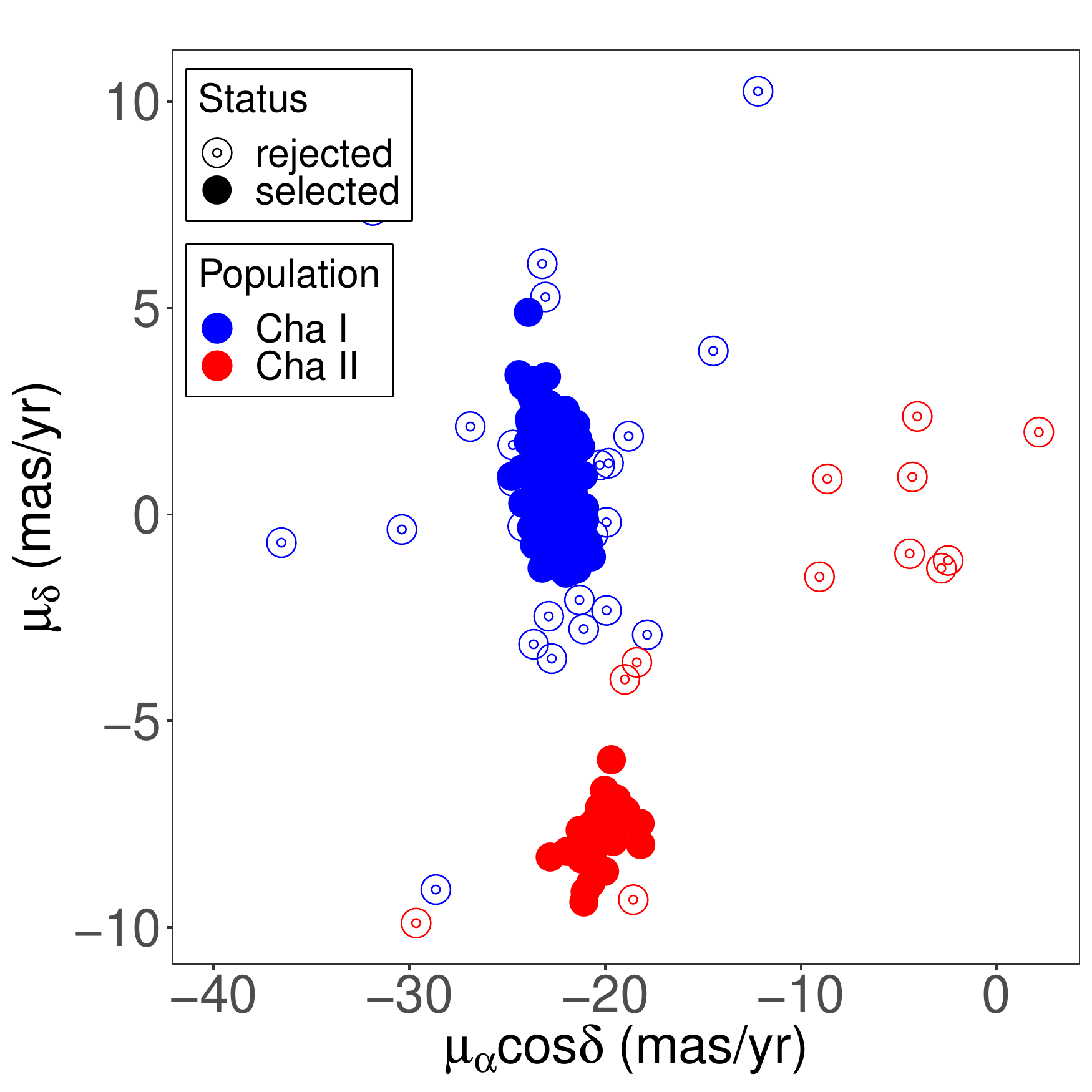}
\includegraphics[width=0.33\textwidth]{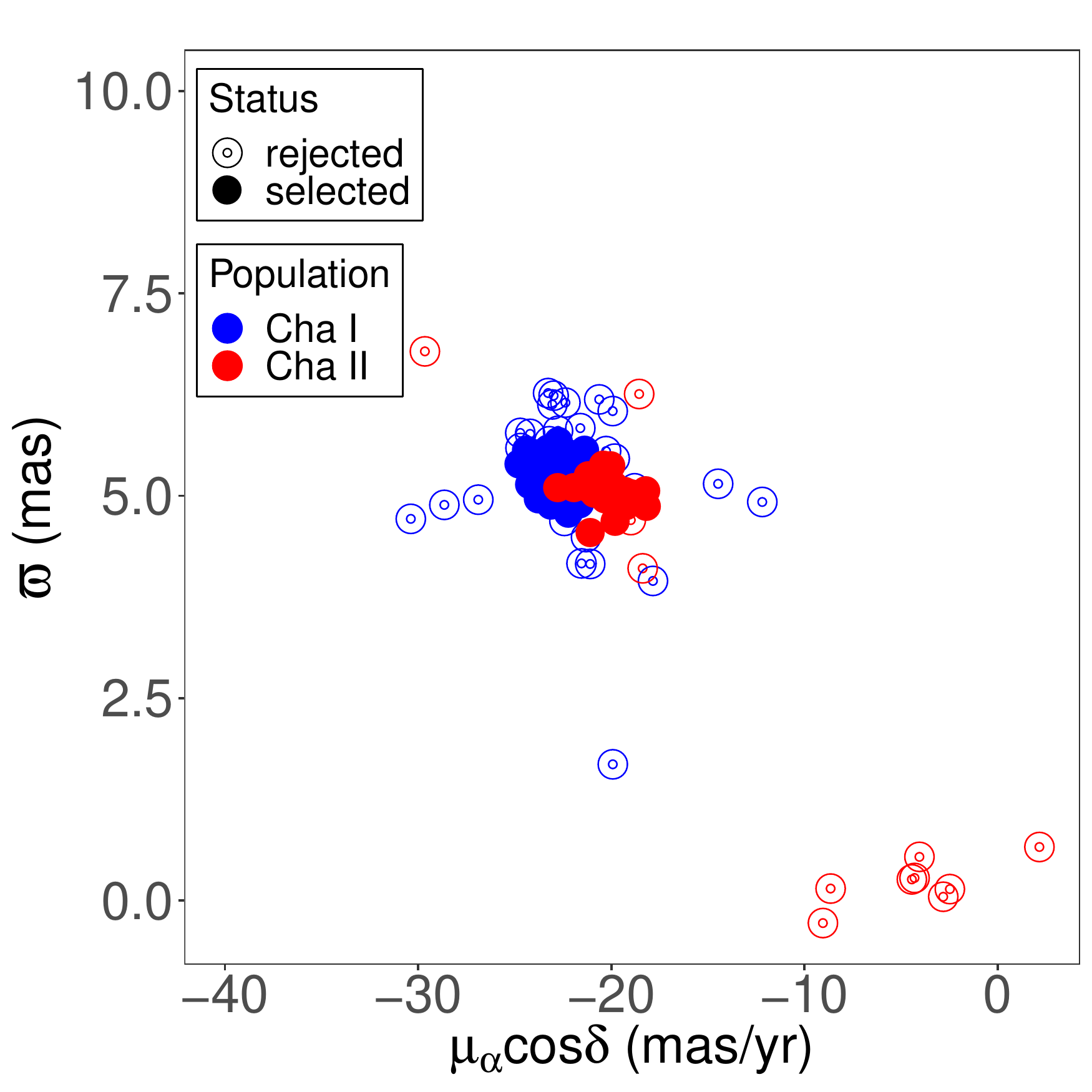}
\includegraphics[width=0.33\textwidth]{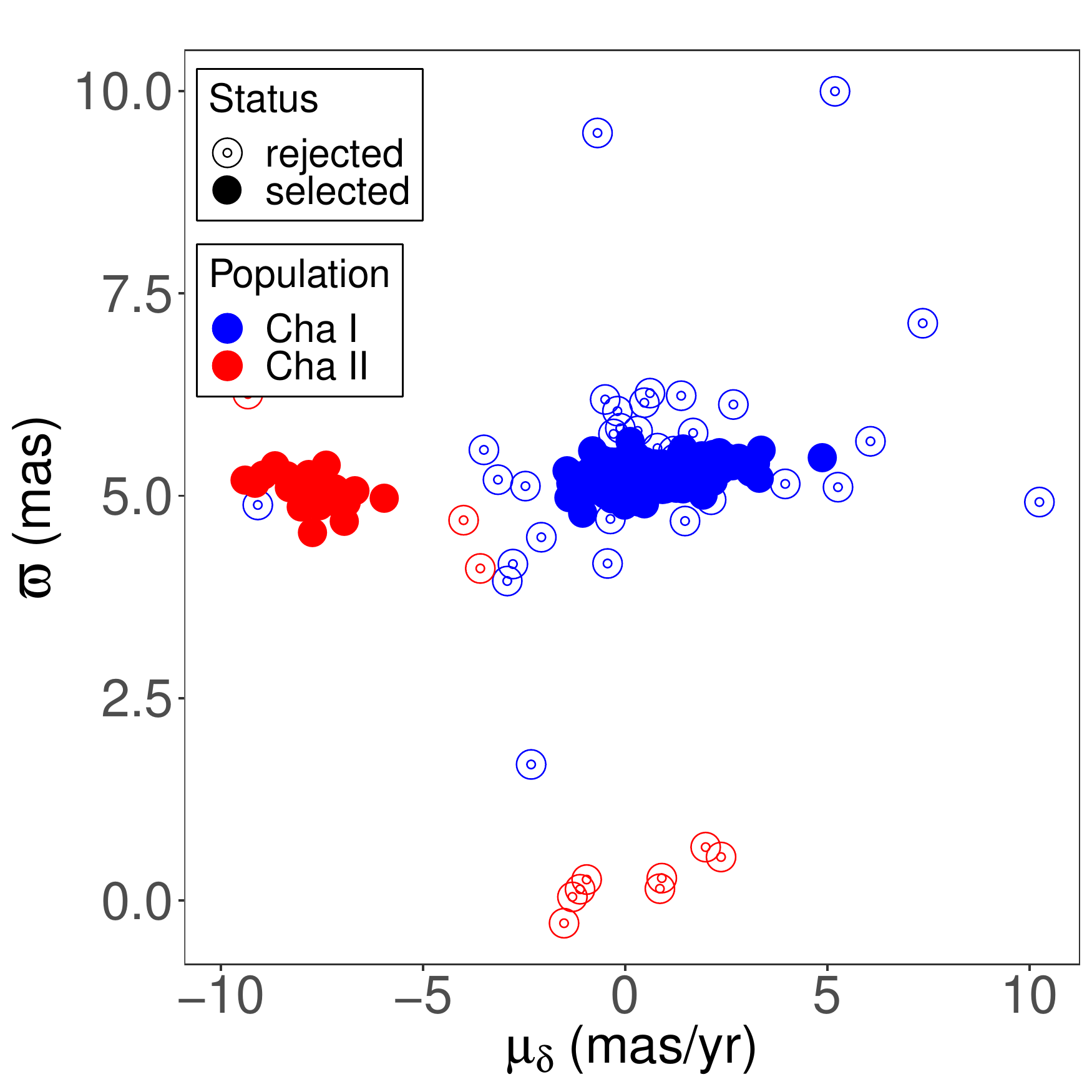}
\caption{Proper motions and parallaxes of the stars in Cha~I and Cha~II in the samples of \citet{Esplin2017} and \citet{Alcala2008}, respectively. Open symbols denote potential outliers in these samples based on Gaia-DR2 data identified from the MCD estimator (see Sect.~\ref{section2}). 
\label{fig_pmra_pmdec_init_list} 
}
\end{center}
\end{figure*}

The field population was modelled using Gaussian Mixture Models (GMMs). We tested the field model with a random sample of 10$^{6}$ sources using GMMs with 20, 40, 60, 80, 100, 120, 140 and 160 components, and we chose the model with 80 components, which returns the smallest Bayesian information criteria (BIC) value. The field model was computed only once at the beginning while the cluster model was built iteratively during the process based on an initial list of members (see below). The cluster model uses the inferred parameters from the initial list of members to define the cluster locus in the space of the astrometric features using GMMs and defines the cluster sequence in the photometric space as a principal curve with a spread at any point along the curve given by a multivariate Gaussian (both the principal curve and its spread are initialised with a fit to the initial list of members). The method then computes membership probabilities for all sources in the field using the fraction of sources in each category (member vs. non-member) obtained in the previous iteration to estimate the marginal class probabilities. The sources are classified into members and non-members based on a probability threshold $p_{in}$ predefined by the user. The list of stars that results from this process is used as input for the next iteration and this procedure is repeated until convergence. The solution is said to converge when the list of cluster members remains fixed after successive iterations. 

The final step of the membership analysis consists in evaluating the performance of our classifier. We generate synthetic data based on the cluster and field properties inferred from the previous steps, and measure the quality of the classifier to define an optimum probability threshold $p_{opt}$ as described in Sect.~4.2.7 of \citet{Olivares2019}. The sources in the catalogue are finally re-classified as members ($p\geq p_{opt}$) and non-members. 

The samples of 250 stars from \citet{Esplin2017} and 63~stars from \citet{Alcala2008} represent the most complete censuses of the stellar population in Cha~I and Cha~II, respectively, known to date. We found Gaia-DR2 astrometry for 194 and 48~stars, respectively, where we note the existence of a few stars with discrepant proper motion and parallax in these samples as shown in Figure~\ref{fig_pmra_pmdec_init_list}. We computed robust distances based on the covariance matrix obtained from the minimum covariance determinant \citep[MCD,][]{Rousseeuw1999} estimator and removed potential outliers from these samples as described in Section~2.1 of \citet{Galli2020}. This step reduces the lists of stars (with Gaia-DR2 data) in Cha~I and Cha~II to 161 and 36~stars, respectively. We use these clean samples of stars as the initial list of members in our membership analysis. The stars in Cha~I and Cha~II exhibit distinct properties as discussed throughout this paper (see also Figure~\ref{fig_pmra_pmdec_init_list}) despite them being part of the same molecular cloud complex. We therefore decided to run two independent membership analyses (one for each cluster) using the same catalogue of Gaia-DR2 sources downloaded for this sky region, representation space, and strategy to select the most likely cluster members, as described above, but using different input lists of stars that are specific for each case. 

In Table~\ref{tab_comp_pin} we compare the solutions obtained for Cha~I and Cha~II using different values of the user predefined probability threshold $p_{in}$. We compute the true positive rate (TPR, i.e. the fraction of synthetic cluster members recovered by our methodology) and contamination rate (CR, i.e. the fraction of synthetic field stars identified by our model as cluster members) of the classifier to better evaluate our results obtained with different $p_{in}$ values. These numbers were obtained from synthetic data sets sampled from the inferred model for each cluster, and so they represent only rough estimates of these indicators computed in the absence of the true distributions. 

\subsection{Projection effects}\label{section2.2}

Our membership analysis conducted over the relatively large field that encompasses the Chamaeleon molecular clouds (as defined in Sect.~\ref{section2.1}) identifies a few more dispersed field sources as cluster members independent of the adopted $p_{in}$ threshold. In particular, we note the existence of one source (namely Gaia DR2 5789232155389250304) that is closer to the Cha~II molecular clouds, but exhibits proper motion and parallax that are consistent with membership in Cha~I. The RV of this source published in the literature yields a spatial velocity ($U=-11.0\pm1.5$~km/s, $V=-21.6\pm2.0$~km/s, $W=-1.8\pm0.6$~km/s) that is not consistent with the space motion of Cha~I stars (see Sect.~\ref{section3.3}). This is due to projection effects that render proper motion and parallax consistent with membership in Cha~I at the spatial location of this source despite the different space motion. We proceed as follows to minimise the existence of potential contaminants in our solution due to projection effects.

\begin{figure*}[!h]
\begin{center}
\includegraphics[width=0.49\textwidth]{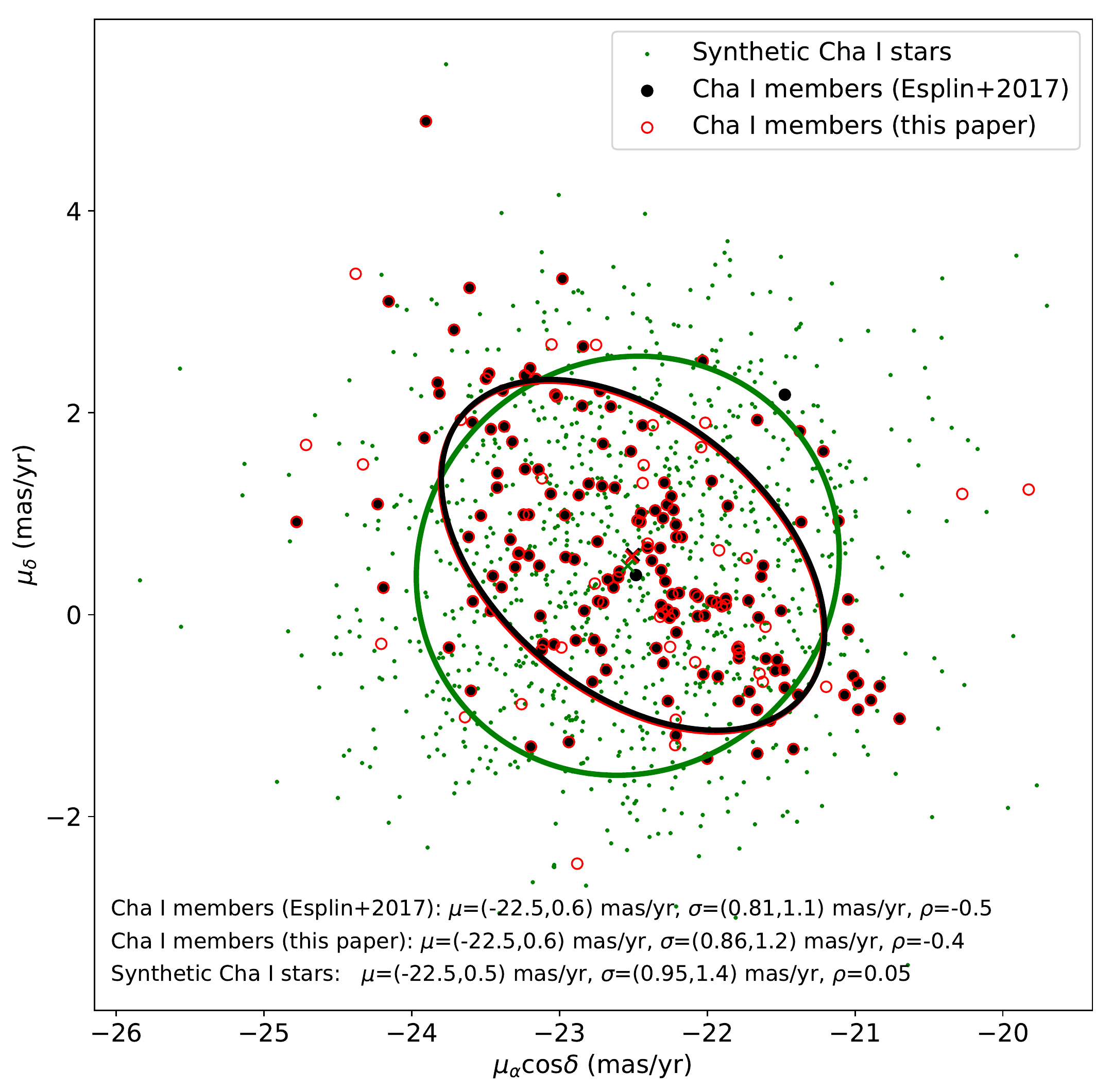}
\includegraphics[width=0.49\textwidth]{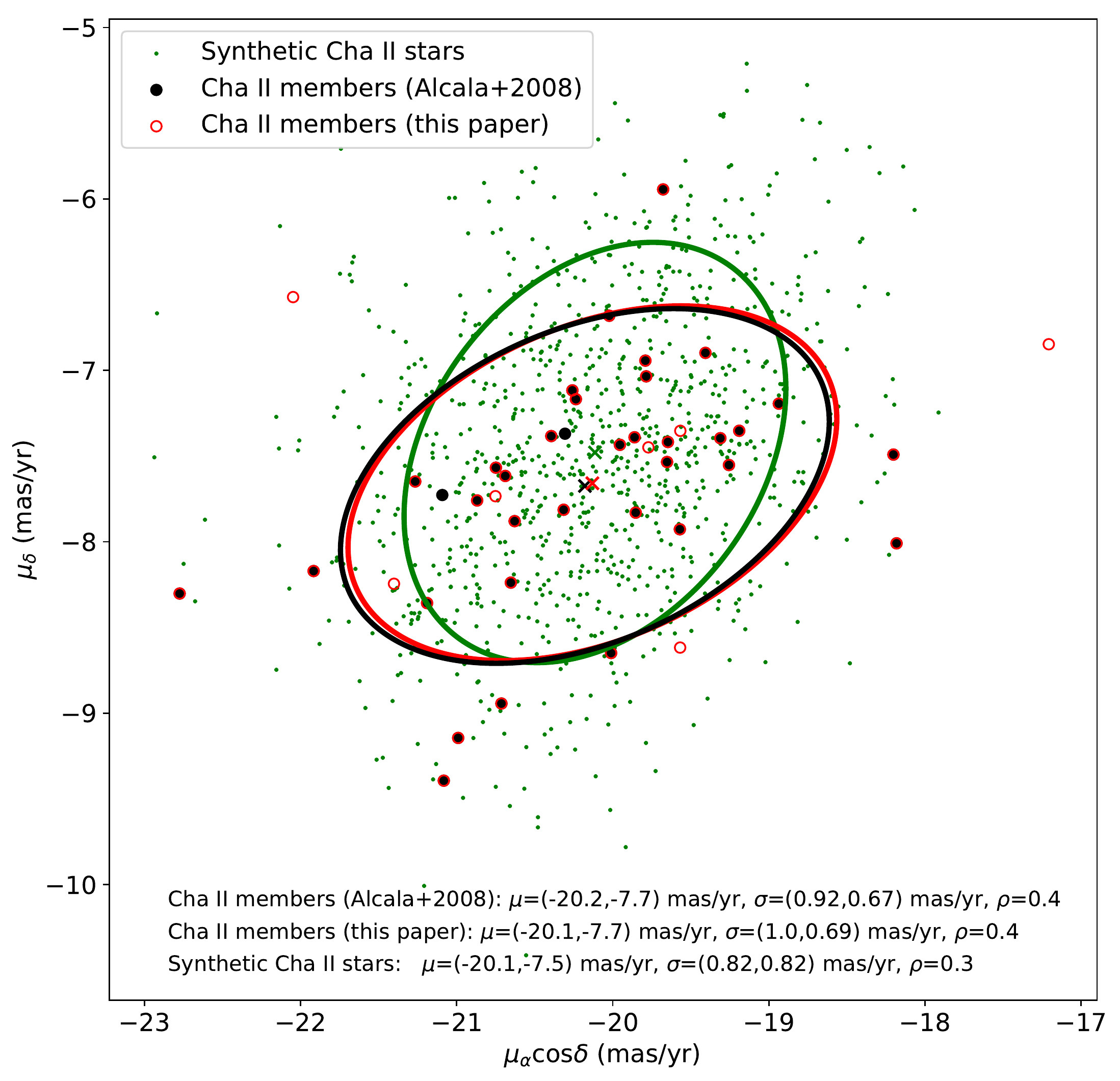}
\includegraphics[width=0.49\textwidth]{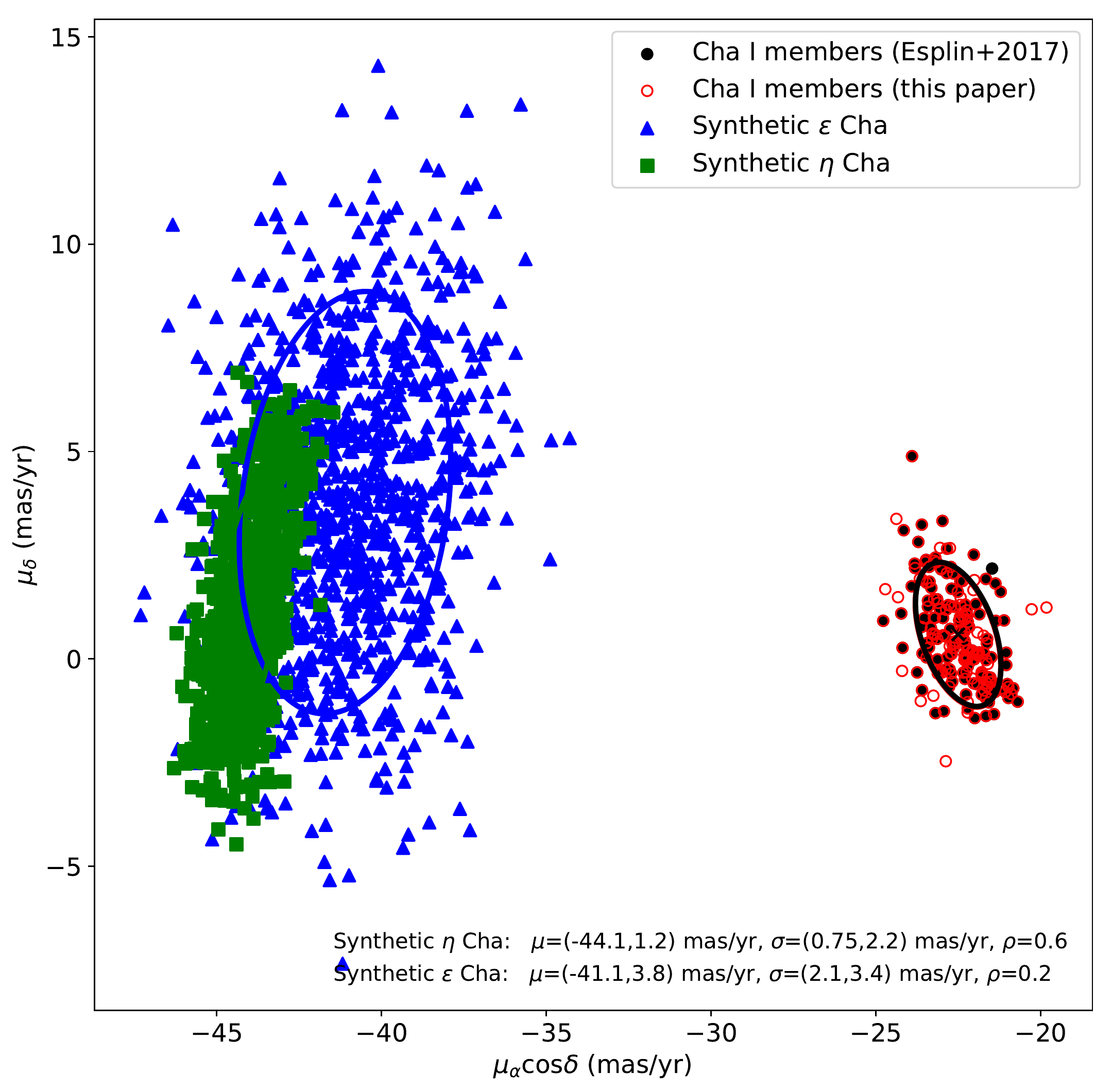}
\includegraphics[width=0.49\textwidth]{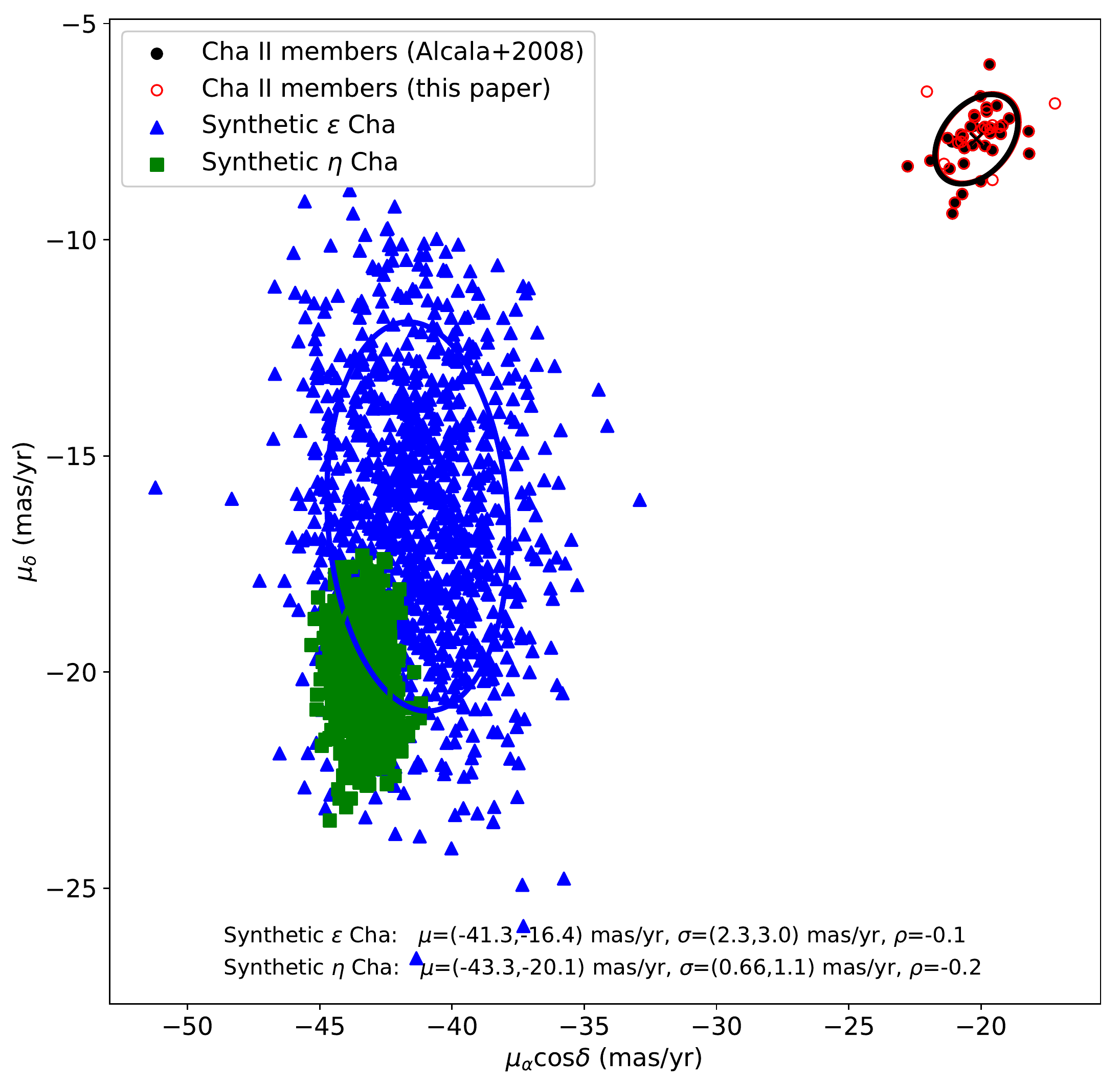}
\caption{Projection effects on the stellar proper motion of Chamaeleon stars. Upper panels show the proper motion distribution of synthetic stars randomly distributed over the fields used in our membership analysis moving with the same space motion of Cha~I and Cha~II stars. Lower panels illustrate the expected proper motion distribution of synthetic $\epsilon$~Cha and $\eta$~Cha located in the same fields. We generated a total of 1\,000 sources in our simulations. We compare the proper motion of the synthetic stars with the observed values for the candidate members previously identified in the literature and members identified in this study (in this case we use the solution with $p_{in}=0.9$, see Table~\ref{tab_comp_pin}). The mean, dispersion and correlation coefficient obtained from each proper motion distribution are indicated in the panels. The solid lines denote the fitted proper motion distributions and the crosses mark the mean proper motion in each case. 
\label{fig_projection_effects} 
}
\end{center}
\end{figure*}

First, we conduct an independent membership analysis following the same methodology described before but in two smaller regions centred around the Cha~I and Cha~II molecular clouds. Our results obtained in these regions are labelled with the term `central region' which we use hereafter to distinguish from the membership analysis conducted over the entire Chamaeleon complex. The size of each field is defined based on the position of the candidate members previously identified in the literature with available Gaia-DR2 data that constitute our input list for the membership analysis (see Sect.~\ref{section2.1}). Thus, the field for the membership analysis in Cha~I covers the sky region defined by  $295.9^{\circ}\leq l \leq 297.8^{\circ}$, $-16.1^{\circ}\leq b \leq -13.1^{\circ}$ in Galactic coordinates and includes 196\, 330 sources. The field centred around Cha~II is defined by $303.2^{\circ}\leq l \leq 304.2^{\circ}$, $-15.1^{\circ}\leq b \leq -13.6^{\circ}$ and includes 56\,100 sources. 

Second, we generated synthetic cluster members located in the fields surveyed by our membership analysis from the velocity and distance distribution of the Cha~I and Cha~II candidate members previously identified in the literature. The 3D positions and spatial velocities of the stars were transformed into the observable proper motions in the corresponding sky regions that we use to discuss the importance of projection effects in our analysis. As illustrated in Figure~\ref{fig_projection_effects} the location and scatter of the proper motion distribution for synthetic stars, candidate members from the literature, and members identified in this study are consistent between themselves. The orientation of the proper motion distribution generated in our simulations is different because we allow the synthetic cluster members to be randomly distributed in the fields covered by our membership analysis. This effect is more apparent in the case of Cha~I. Therefore, our methodology employed for the membership analysis could be missing some cluster members if we assume that the stars are equally probable to be found at any sky position of the field. However, what we observe in practice is that the true members are mostly projected towards the molecular clouds (and in their immediate vicinity) which explains the different orientation observed for the simulated proper motions. The proper motion distribution of the members identified in this study is in good agreement with the results given in the literature. This confirms that our results for the membership analysis in the central regions are not affected by projection effects.  

Third, we performed additional simulations to investigate the existence of contaminants from other young stellar groups due to projection effects. As described in Sect.~\ref{section2.1}, our membership analysis uses not only the astrometric features of the stars, but also their photometry to select cluster members with similar ages. The only two young stellar groups in the Chamaeleon complex that have similar ages to the stellar populations in the molecular clouds are the $\epsilon$~Cha and $\eta$~Cha associations. We generated synthetic stars (as described above) using the distance and space motion of these stellar groups given in the literature \citep[see e.g.][]{Murphy2013} in the fields covered by our membership analysis of Cha~I and Cha~II. Figure~\ref{fig_projection_effects} shows that the proper motions of the synthetic $\epsilon$~Cha and $\eta$~Cha stars are significantly different from the observed proper motion distribution of Cha~I and Cha~II cluster members. We therefore conclude that our sample of members in the central regions is not contaminated by the other young stellar groups of the Chamaeleon complex due to projection effects.

\subsection{Final list of cluster members}\label{section2.3}

The high TPRs and low CRs for all solutions given in Table~\ref{tab_comp_pin} confirm the robustness of our results. In particular, we note that our results for the membership analysis in the central regions have higher TPRs and lower CRs as compared to the ones obtained over the whole Chamaeleon complex. We see little variation in sample size in the two cases and confirm that running the membership analysis over extended regions of the Chamaeleon complex will not allow us to detect more cluster members (but will lead to the inclusion of more contaminants due to projection effects). On the contrary, performing the membership analysis over smaller regions allowed us to recover a few members that have been overlooked in the first analysis. We therefore decided to report the solutions obtained with $p_{in}=0.9$ in the central regions as our final lists of clusters members in Cha~I (188~stars) and Cha~II (41~stars). The cluster members in each population are listed in Tables~\ref{tab_members_Cha1} and \ref{tab_members_Cha2}. We provide in Tables~\ref{tab_prob_Cha1} and \ref{tab_prob_Cha2} the membership probabilities obtained with different $p_{in}$ values for all sources in the field for the analyses conducted in Cha~I and Cha~II, respectively, to allow the reader to choose a different probability threshold that is more suited to their study.    

\begin{table*}[!h]
\renewcommand\thetable{1} 
\centering
\caption{Comparison of membership results in Cha~I and Cha~II using different values for the probability threshold $p_{in}$.
\label{tab_comp_pin}}
\begin{tabular}{ccccccccc}
\hline\hline
&\multicolumn{4}{c}{Cha~I (large field)}&\multicolumn{4}{c}{Cha~II (large field)}\\
\hline\hline
$p_{in}$&$p_{opt}$&Members&TPR&CR&$p_{opt}$&Members&TPR&CR\\
\hline
0.5&0.80&189&$0.964\pm0.010$&$0.040\pm0.017$&0.74&48&$0.978\pm0.018$&$0.039\pm0.004$\\
0.6&0.90&183&$0.947\pm0.005$&$0.024\pm0.011$&0.78&47&$0.982\pm0.013$&$0.030\pm0.005$\\
0.7&0.80&189&$0.962\pm0.014$&$0.052\pm0.024$&0.59&47&$0.977\pm0.006$&$0.030\pm0.016$\\
0.8&0.84&184&$0.972\pm0.006$&$0.037\pm0.016$&0.83&40&$0.970\pm0.023$&$0.030\pm0.005$\\
0.9&0.81&187&$0.972\pm0.020$&$0.034\pm0.011$&0.81&38&$0.983\pm0.010$&$0.021\pm0.011$\\
\hline\hline
&\multicolumn{4}{c}{Cha~I (central region)}&\multicolumn{4}{c}{Cha~II (central region)}\\
\hline\hline
$p_{in}$&$p_{opt}$&Members&TPR&CR&$p_{opt}$&Members&TPR&CR\\
\hline
0.5&0.80&192&$0.995\pm0.005$&$0.021\pm0.013$&0.74&43&$0.980\pm0.017$&$0.022\pm0.003$\\
0.6&0.84&191&$0.996\pm0.003$&$0.014\pm0.008$&0.77&41&$0.988\pm0.006$&$0.012\pm0.003$\\
0.7&0.69&192&$0.998\pm0.003$&$0.012\pm0.003$&0.76&41&$0.985\pm0.013$&$0.008\pm0.008$\\
0.8&0.86&190&$0.993\pm0.008$&$0.007\pm0.012$&0.78&41&$0.983\pm0.016$&$0.008\pm0.003$\\
0.9&0.90&188&$0.991\pm0.008$&$0.002\pm0.003$&0.79&41&$0.985\pm0.010$&$0.017\pm0.011$\\
\hline\hline
\end{tabular}
\tablefoot{We provide the optimum probability threshold, the number of cluster members, the true positive rate (TPR) and contamination rate (CR) obtained for each solution and cluster. We present the results of our membership analysis performed over a large field that encompasses the entire Chamaeleon complex (see Sect.~\ref{section2.1}) and the central regions around the Cha~I and Cha~II molecular clouds (see Sect.~\ref{section2.2}).}
\end{table*}

\begin{figure*}[!h]
\begin{center}
\includegraphics[width=1.0\textwidth]{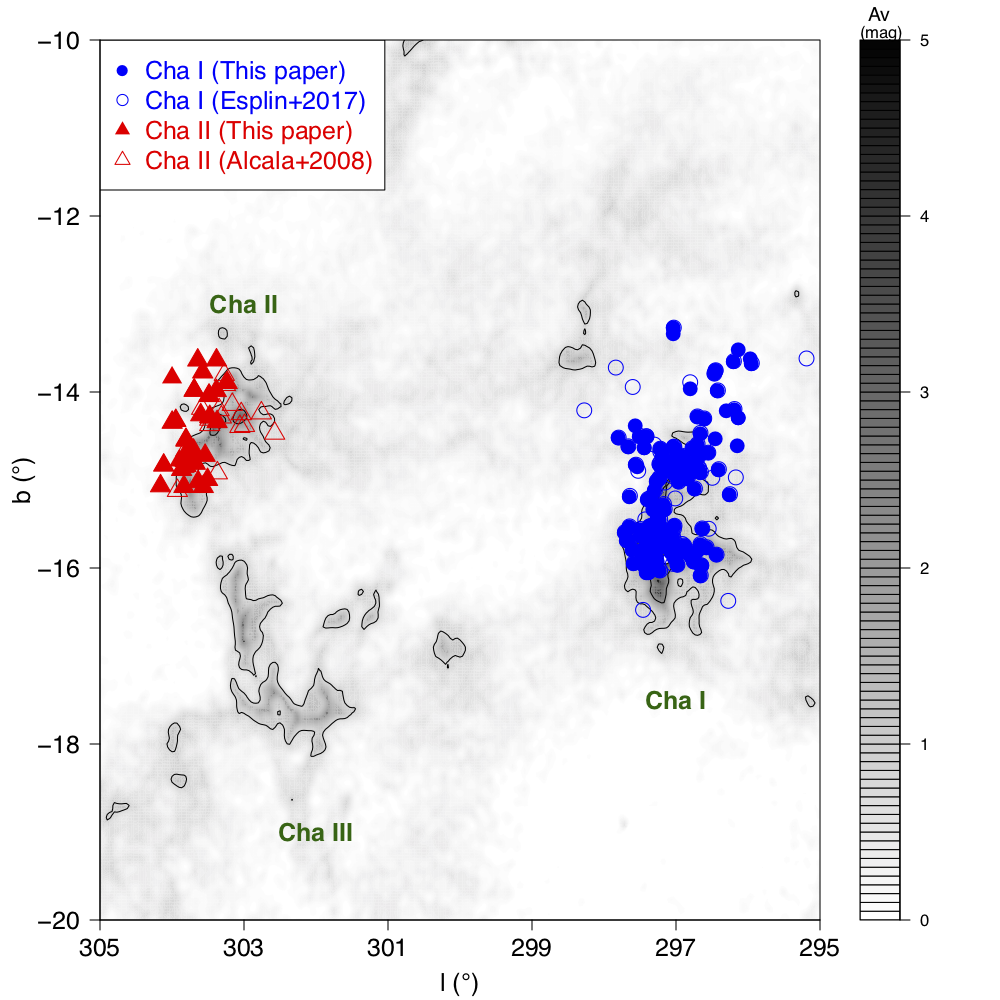}
\caption{Location of the Chamaeleon stars overlaid on the extinction map of \citet{Dobashi2005} in Galactic coordinates. Open and filled symbols indicate the position of the stars from the literature \citep{Alcala2008,Esplin2017} and cluster members identified in this study, respectively. The blue and red colours denote the stars in the Cha~I and Cha~II molecular clouds, respectively. 
\label{fig_location} 
}
\end{center}
\end{figure*}

\begin{figure*}[!h]
\begin{center}
\includegraphics[width=0.33\textwidth]{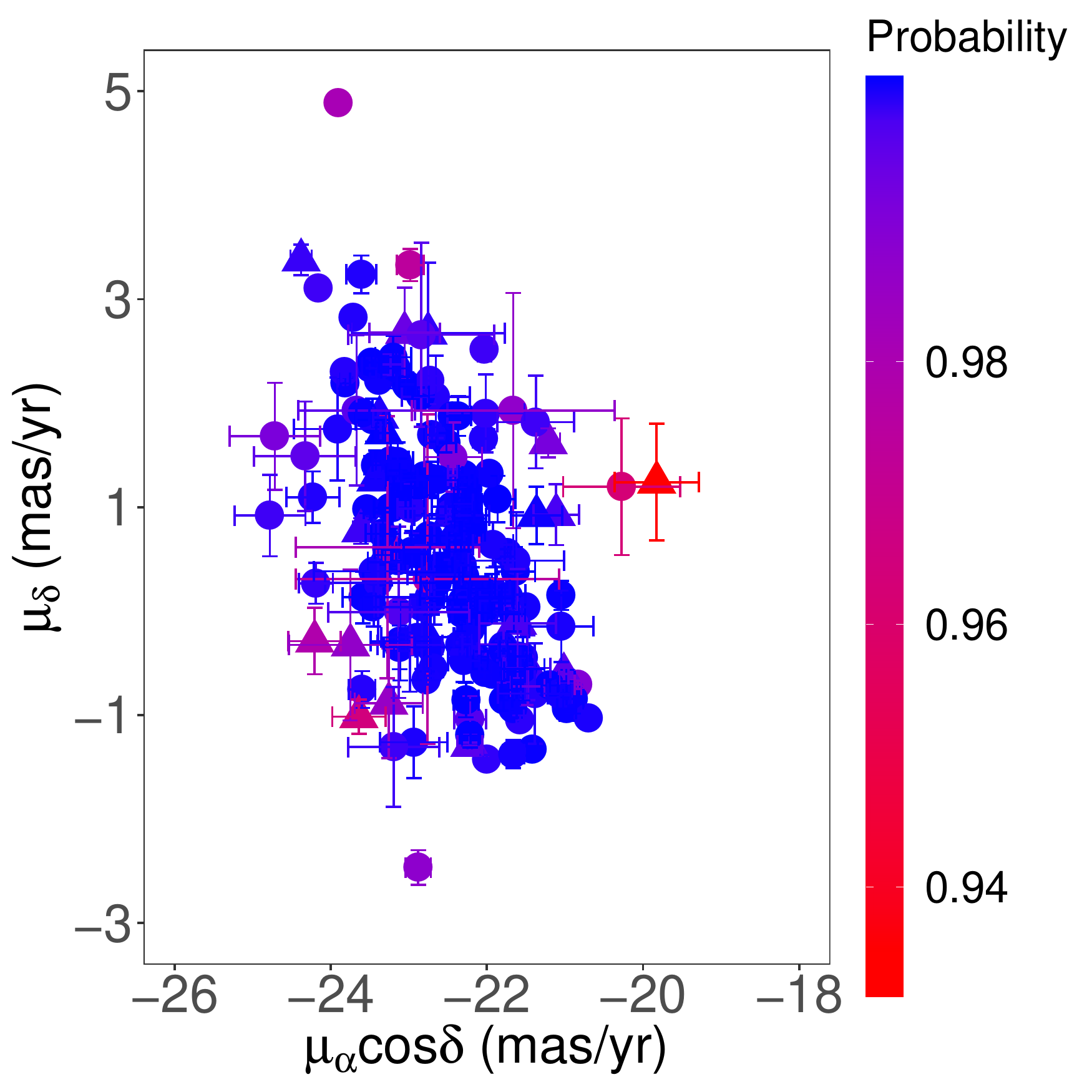}
\includegraphics[width=0.33\textwidth]{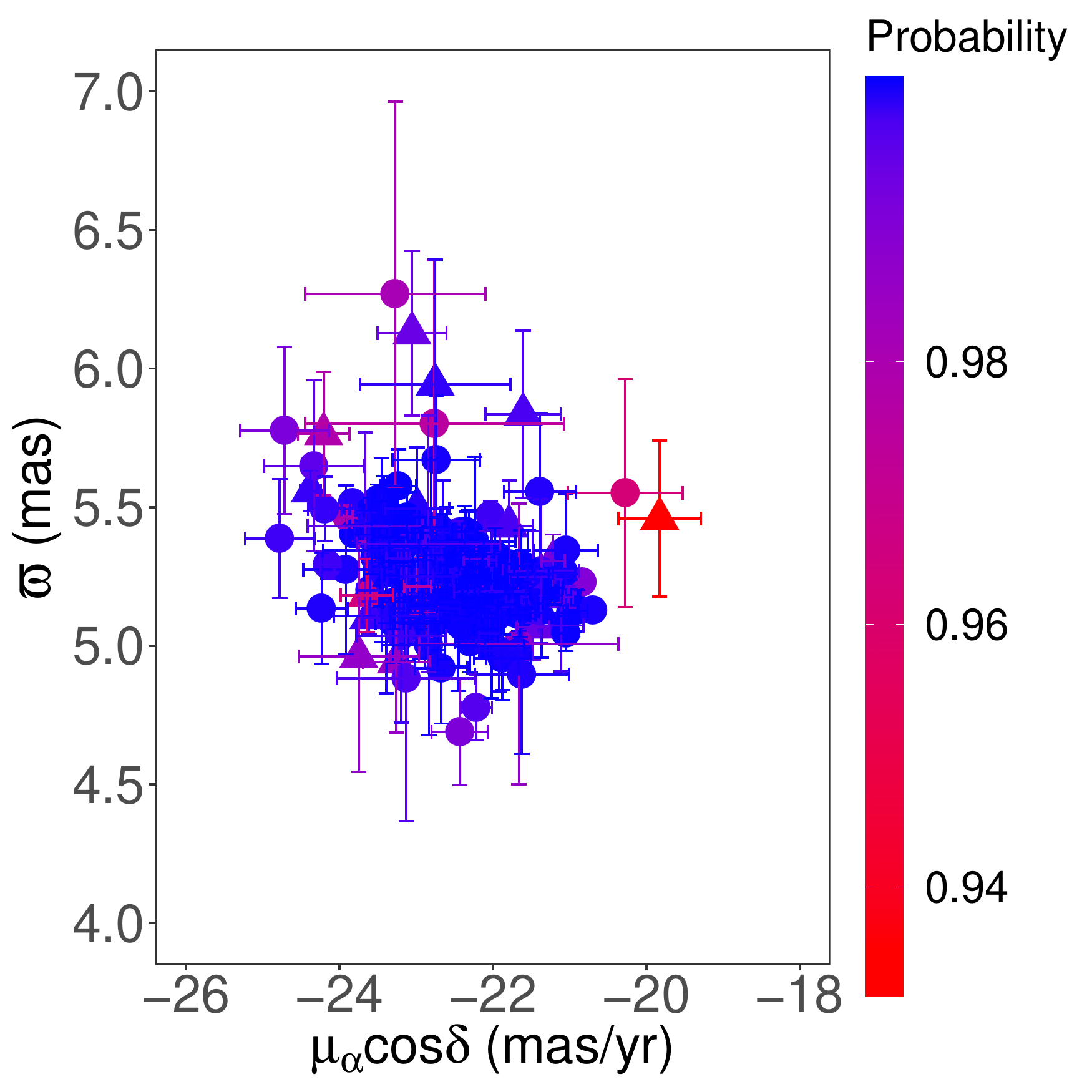}
\includegraphics[width=0.33\textwidth]{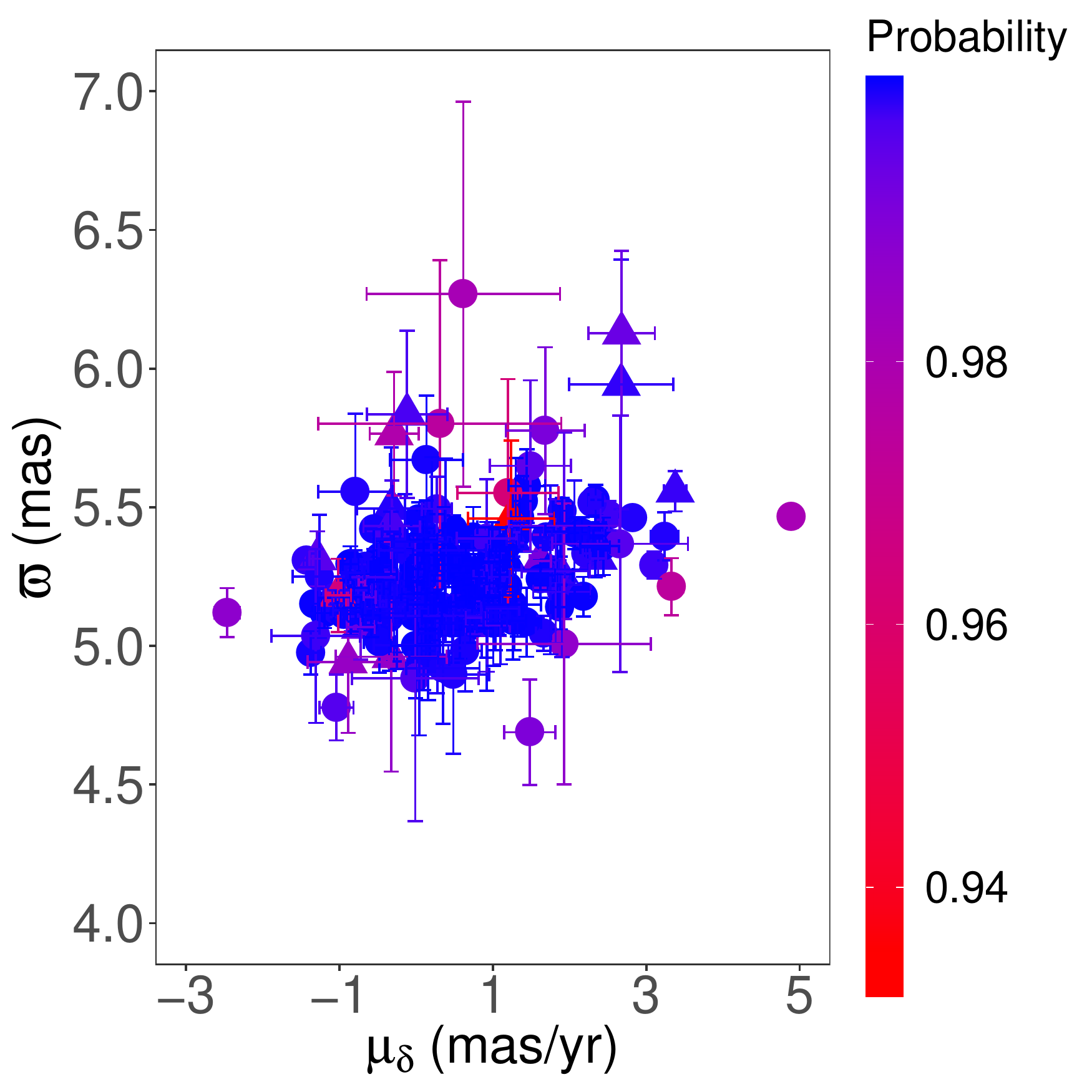}
\caption{Proper motions and parallaxes of the 188 stars in Cha~I identified in our membership analysis. The stars are colour-coded based on their membership probabilities which are scaled from zero to one. Triangles indicate the stars with RUWE $\geq$ 1.4 (see Sect.~\ref{section3.1}). 
\label{fig_pm_plx_cha1} 
}
\end{center}
\end{figure*}

\begin{figure*}[!h]
\begin{center}
\includegraphics[width=0.33\textwidth]{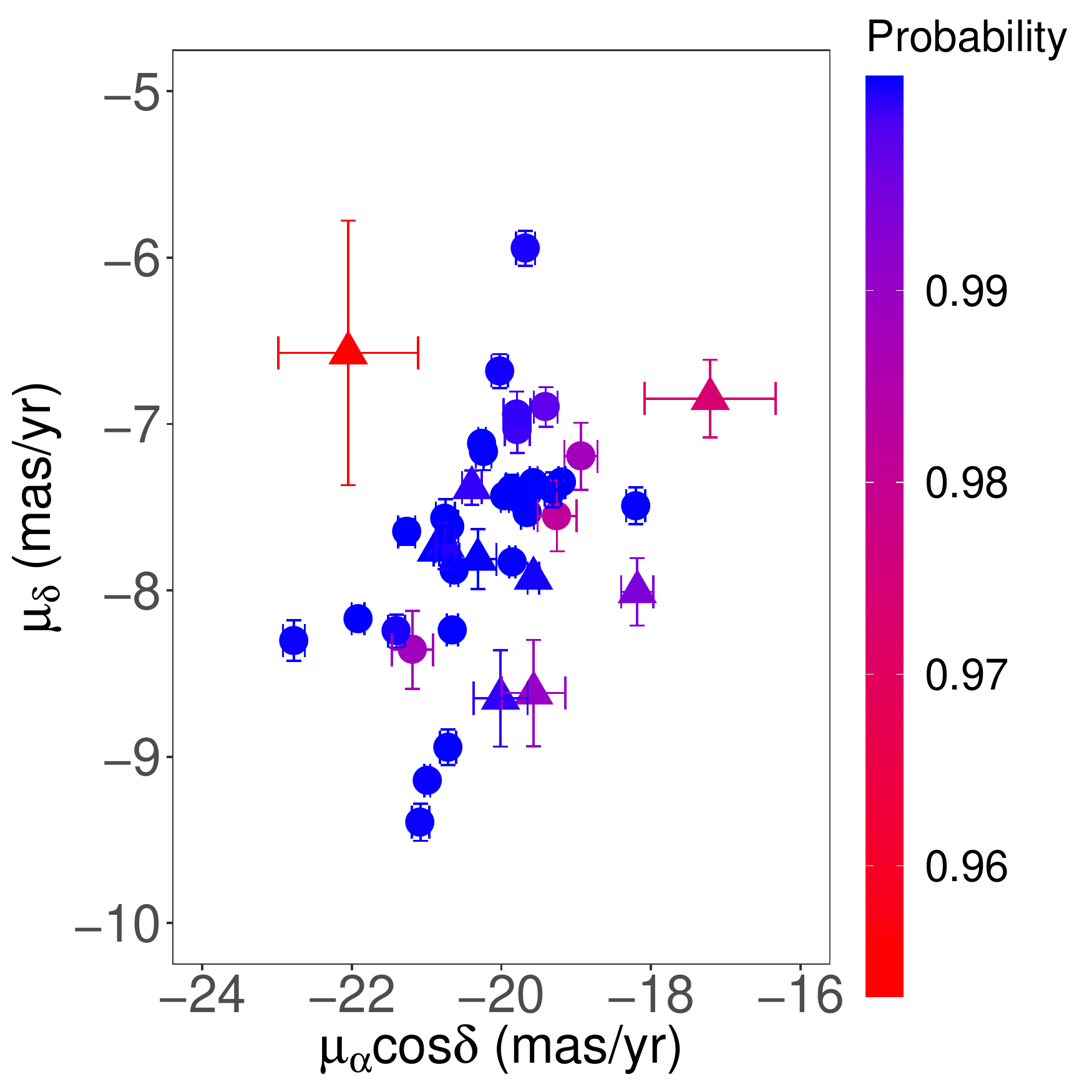}
\includegraphics[width=0.33\textwidth]{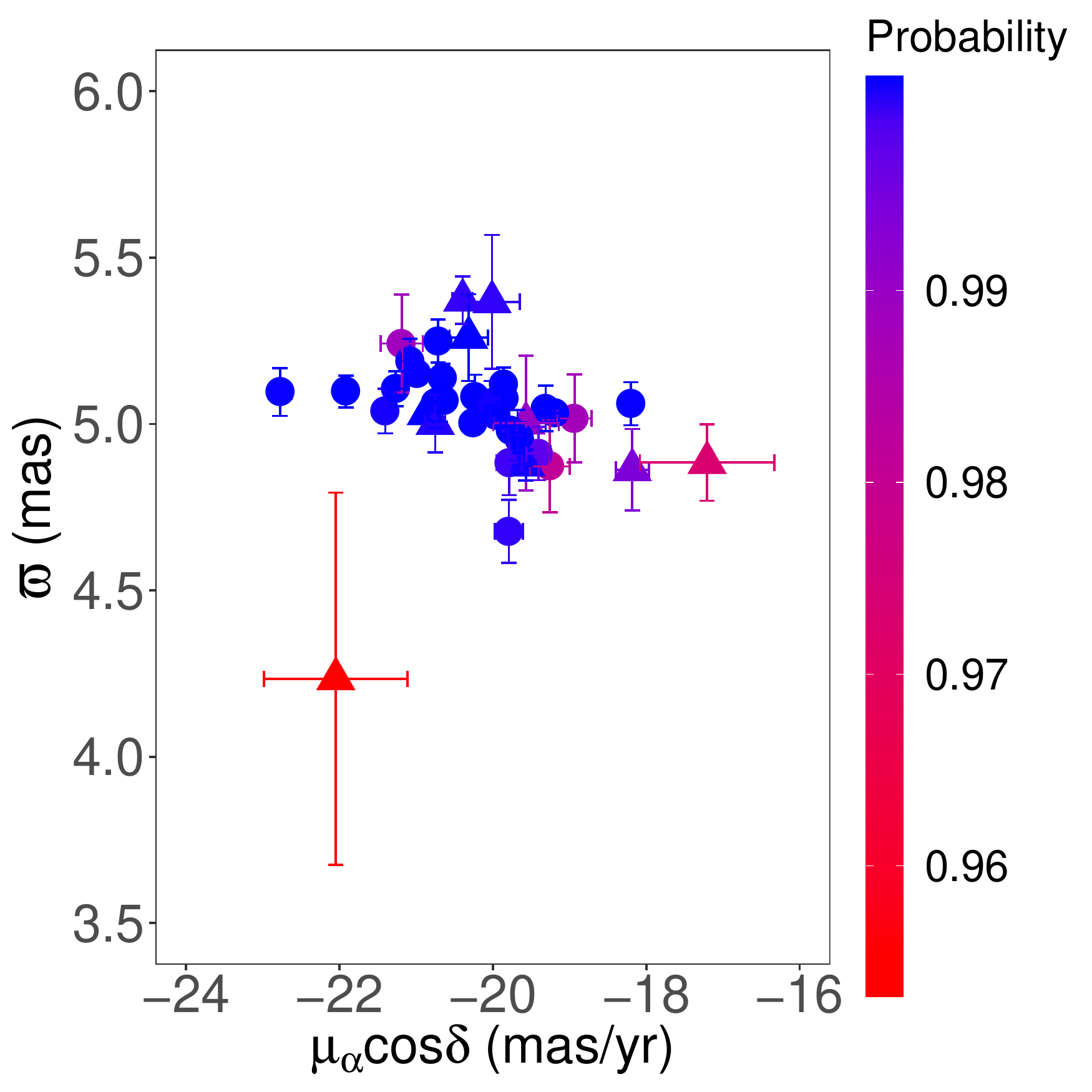}
\includegraphics[width=0.33\textwidth]{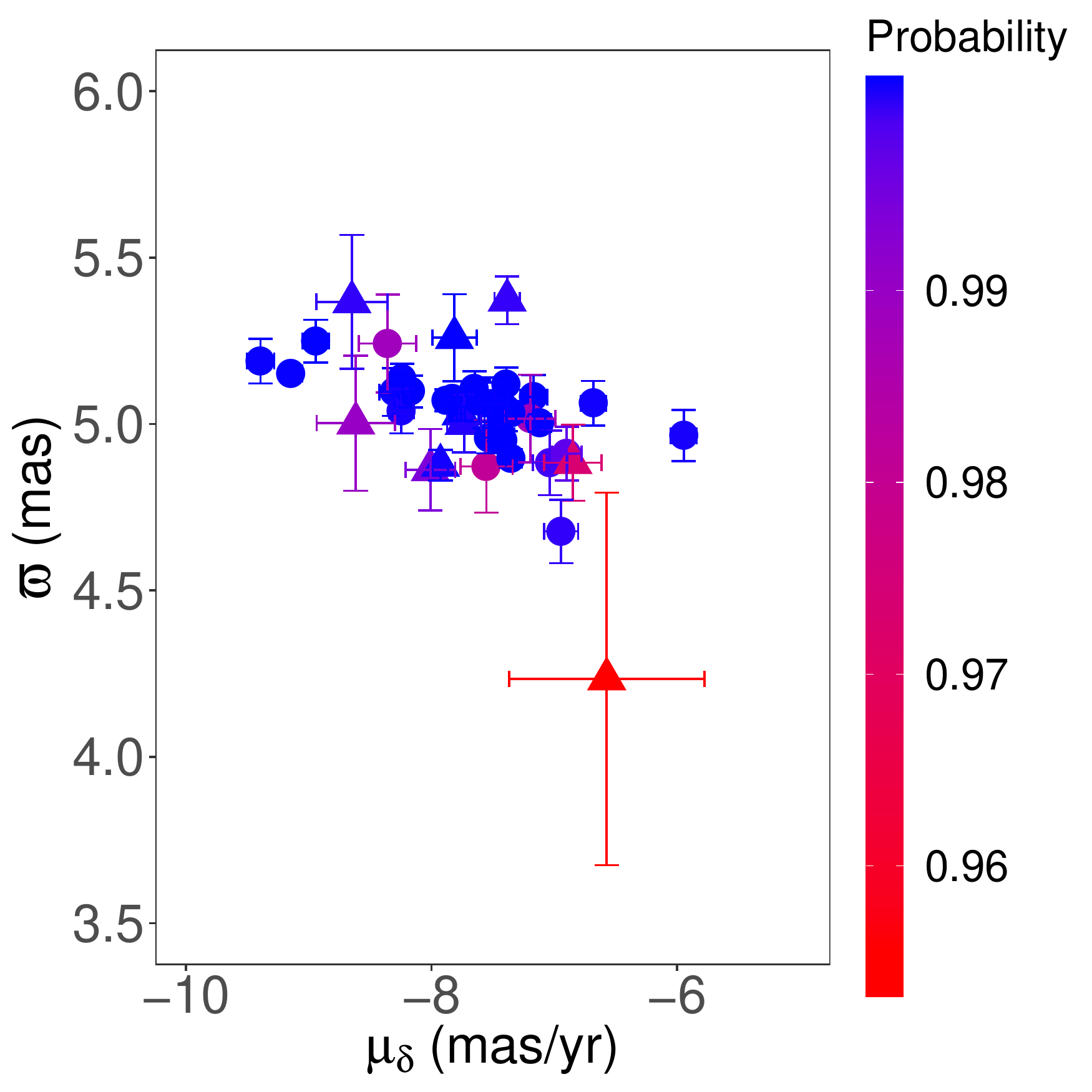}
\caption{Proper motions and parallaxes of the 41 stars in Cha~II identified in our membership analysis. The stars are colour-coded based on their membership probabilities which are scaled from zero to one. Triangles indicate the stars with RUWE $\geq$ 1.4 (see Sect.~\ref{section3.1}).
\label{fig_pm_plx_cha2} 
}
\end{center}
\end{figure*}

\begin{figure*}[!h]
\begin{center}
\includegraphics[width=0.45\textwidth]{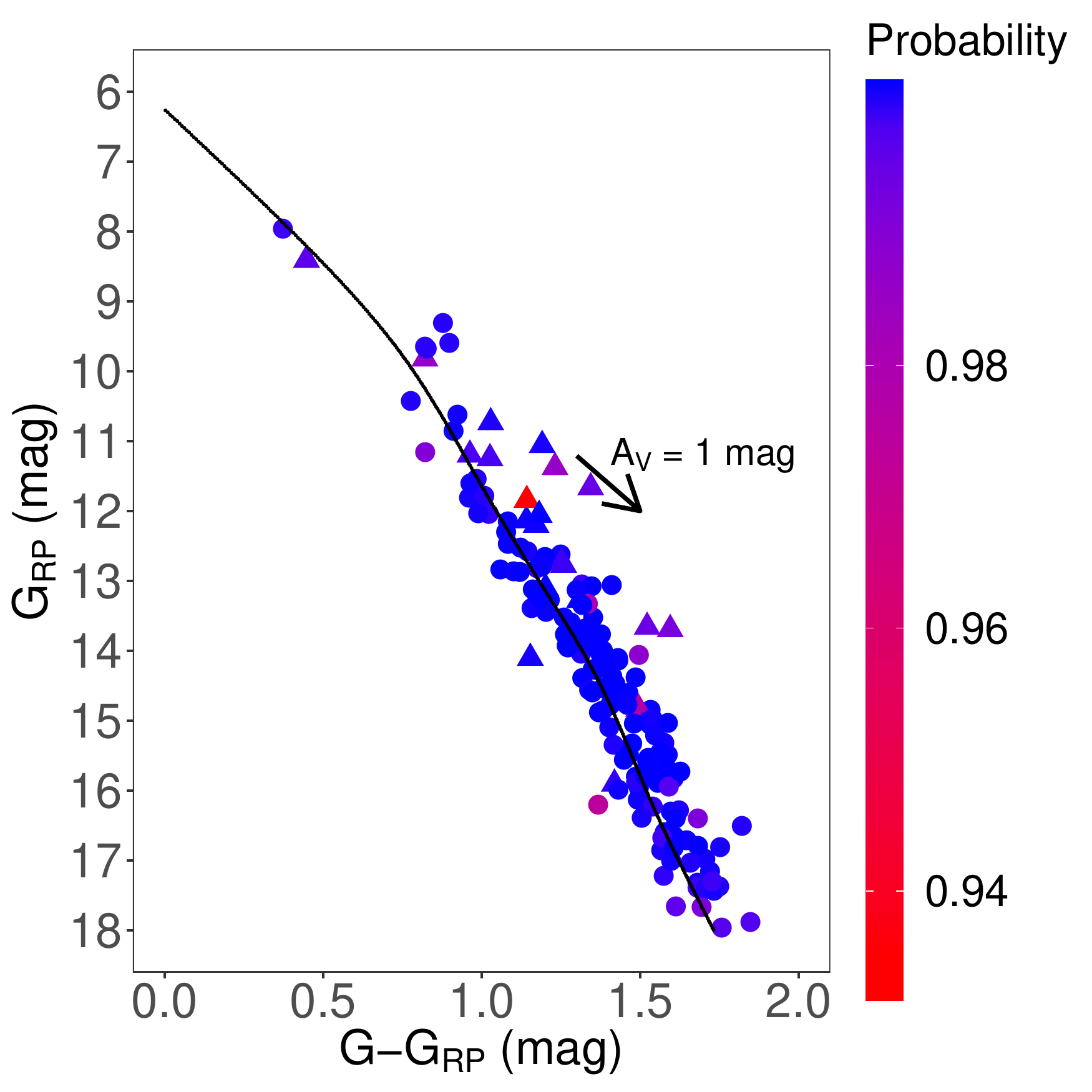}
\includegraphics[width=0.45\textwidth]{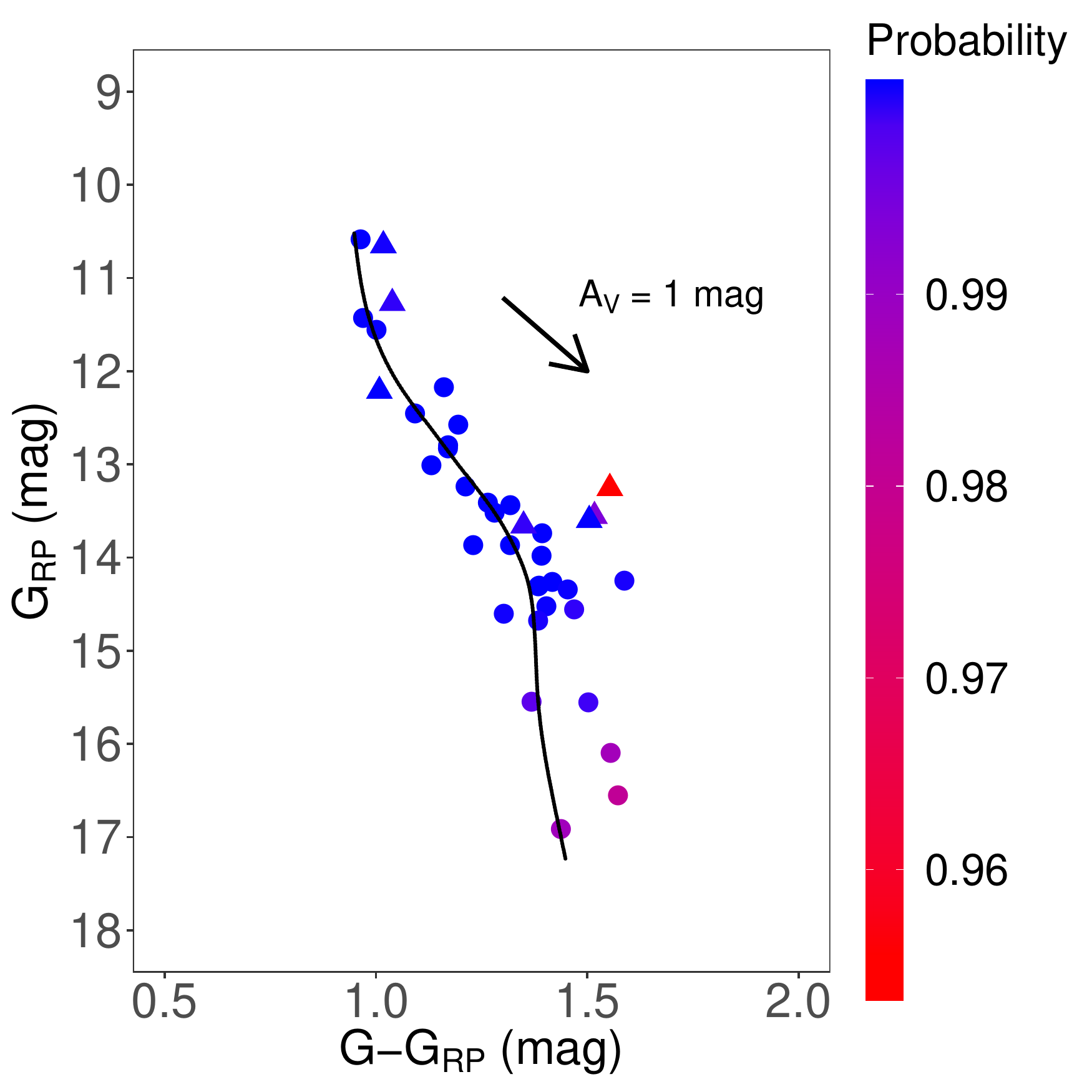}
\caption{Colour-magnitude diagram of the cluster members identified in our membership analysis in Cha~I \textit{(left panel)} and Cha~II \textit{(right panel)}. The black solid line indicates the empirical isochrone derived for each cluster (see Tables~\ref{tab_isochrone_Cha1} and \ref{tab_isochrone_Cha2}). The stars are colour-coded based on their membership probabilities which are scaled from zero to one. Triangles indicate the stars with RUWE $\geq$ 1.4 (see Sect.~\ref{section3.1}). The arrow indicates the extinction
vector of $A_{V}=1$~mag that we converted to the Gaia bands based on the relative extinction values computed by \citet{Wang2019}.
\label{fig_cmd} 
}
\end{center}
\end{figure*}

Figure~\ref{fig_location} confirms the existence of a more dispersed population of cluster members in the immediate vicinity of the Cha~I molecular cloud that is known from pre-\textit{Gaia} studies \citep[see e.g.][]{LopezMarti2013b}. Figures~\ref{fig_pm_plx_cha1} and \ref{fig_pm_plx_cha2} show the distribution of proper motions and parallaxes of the Cha~I and Cha~II members identified in our analysis. As expected, the cluster members in the outskirts of the distributions and with the largest uncertainties have the lowest membership probabilities. The colour-magnitude diagrams of Cha~I and Cha~II in the chosen photometric space are shown in Figure~\ref{fig_cmd}, and the empirical isochrones are given in Tables~\ref{tab_isochrone_Cha1} and \ref{tab_isochrone_Cha2}. Our samples of cluster members cover the magnitude range from about $G=6$ to $G=20$~mag and $G=12$ to $G=18$~mag in Cha~I and Cha~II, respectively, and they represent the most complete censuses of members in the Chamaeleon clouds (with available astrometry in the Gaia-DR2 catalogue) to date.  

\begin{figure*}[!h]
\begin{center}
\includegraphics[width=0.49\textwidth]{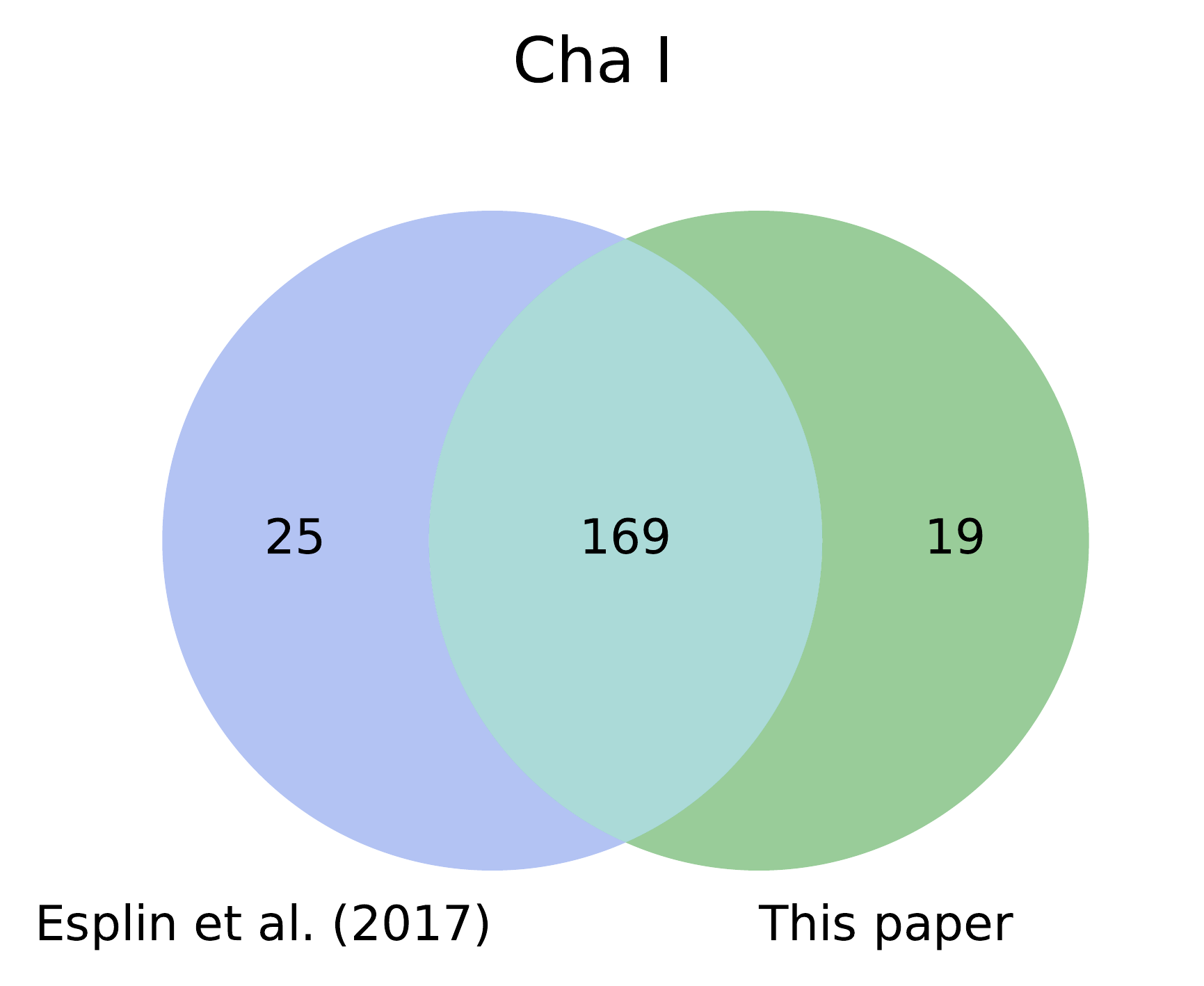}
\includegraphics[width=0.49\textwidth]{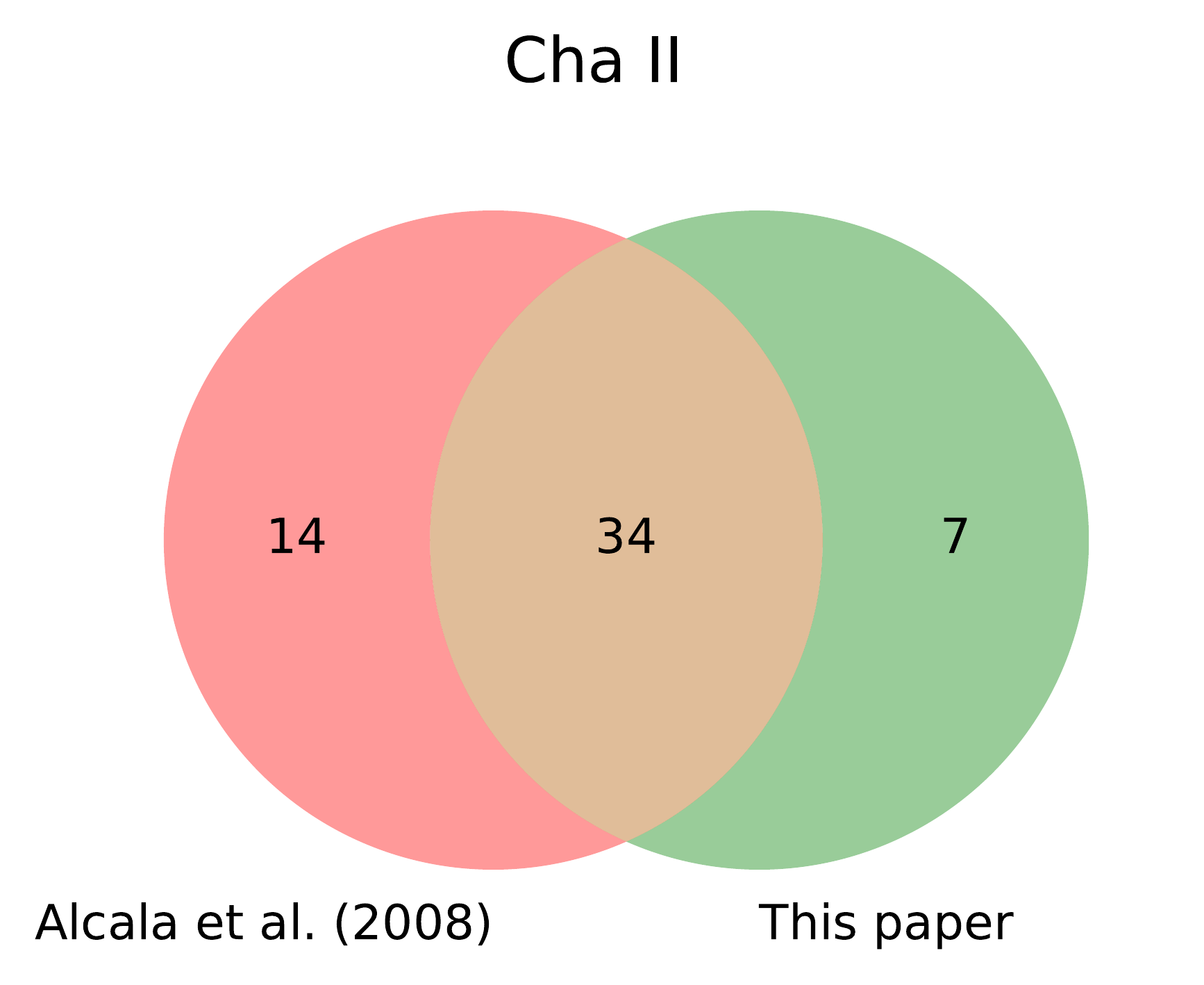}
\caption{Venn diagram comparing the number of members  (with Gaia-DR2 data) identified in previous studies of the literature \citep{Alcala2008,Esplin2017} and the samples of cluster members derived in this paper from our membership analysis. 
\label{fig_venn} 
}
\end{center}
\end{figure*}

In Figure~\ref{fig_venn} we compare the samples of cluster members identified in this paper with the lists of Cha~I and Cha~II stars (with available Gaia-DR2 data) published in other studies. We note that most members in the Cha~I sample (i.e., 169 stars) were also included in the study conducted by \citet{Esplin2017}. However, 25~sources presented in that study could not be recovered by our membership analysis. We verified that they have proper motions and parallaxes that are either not consistent with membership in Cha~I (see e.g. Figure~\ref{fig_pmra_pmdec_init_list}) or result from poor astrometric solutions. For example, if we use the re-normalised unit weight error (RUWE) criterion\footnote{see technical note \href{https://www.cosmos.esa.int/web/gaia/dr2-known-issues}{GAIA-C3-TN-LU-LL-124-01} for more details} to filter the literature sources with poor astrometry (see Sect.~\ref{section3.1}) we note that 17~stars among the rejected sources fail to pass this selection criterion (i.e. RUWE $<$ 1.4). These 17 sources can be considered at this stage as potential candidate members and future data releases of the Gaia space mission will allow us to confirm (or exclude) membership with more precise data. Three stars among the other eight rejected candidate members from the literature lie outside the region surveyed by our analysis. Only one rejected star from the literature among the remaining five sources (namely, Gaia DR2 5201125062389375872) has proper motions and parallaxes consistent with membership in Cha~I. However, it is located below the empirical isochrone of the cluster which explains why it is rejected by our methodology. The same conclusion holds if we try to correct its photometry by the effect of interstellar extinction using the value of $A_{V}\simeq2.5$~mag that is estimated from the \citet{Dobashi2005} maps at the location of the source. 

Similar arguments apply to the analysis in Cha~II. We confirmed 34~members from \citet{Alcala2008} and rejected 14~stars from that study. We note that among the rejected members there are eight sources that are more likely to be background objects ($\varpi<1$~mas) unrelated to the Cha~II population (see Figure~\ref{fig_pmra_pmdec_init_list}). Four stars have poor astrometric solutions based on the RUWE criterion which could possibly explain the discrepant proper motions and parallaxes. The remaining two stars among the other rejected sources, namely Gaia~DR2~5788884464898536960 and Gaia~DR2~5788200298087320320, have proper motions and parallaxes consistent with membership in Cha~II, but they lie below the empirical isochrone defined by the other cluster members. We tentatively corrected their photometry from the interstellar extinction estimated at their position from the \citet{Dobashi2005} extinction maps, but this correction does not place them on the empirical isochrone of the cluster.

The membership analysis conducted in this study allowed us to identify 19 and 7 new members of Cha~I and Cha~II, respectively. This represents an increase of about 11\% and 21\% in the stellar population of Cha~I and Cha~II with respect to the number of confirmed cluster members from previous studies.

\section{Discussion}\label{section3}

\subsection{Refining the sample of cluster members}\label{section3.1}

Let us now refine our sample of cluster members by filtering the stars with the best data available before deriving the overall kinematic properties of the Chamaeleon subgroups. To do so, we use the RUWE criterion as one possible indicator of the goodness of fit of the Gaia-DR2 astrometric solution. The value of 1.4 is often used in the literature to distinguish between the sources with reliable (i.e., RUWE $<$ 1.4) and poor astrometric solutions in the Gaia-DR2 catalogue, but some studies adopt different thresholds \citep[see e.g.][]{Luhman2020}. In this paper we adopt the RUWE threshold of 1.4 to select the sources with good astrometric solutions for consistency with the methodology adopted in other papers of this series \citep[see e.g.][]{Galli2020,Galli2020b}. This step reduces the samples of members to 160 and 31 stars in Cha~I and Cha~II, respectively. The rejected stars exhibit mostly the largest uncertainties in proper motions and parallaxes as shown in Figures~\ref{fig_pm_plx_cha1} and \ref{fig_pm_plx_cha2}. Many of them have been identified as binaries or multiple systems in the literature (see also Figure~\ref{fig_cmd}) which explains the poor astrometry in the Gaia-DR2 catalogue. Future data releases of the Gaia space mission will deliver an improved astrometry for such systems.

We searched the CDS databases \citep{Wenger2000} for RV information of the stars in our sample. Our search for RVs in the literature is based on \citet{Dubath1996}, \citet{Covino1997}, \citet{Joergens2001}, \citet{Torres2006}, \citet{James2006}, \citet{Gontcharov2006}, \citet{Guenther2007}, \citet{Nguyen2012}, \citet{Biazzo2012}, \citet{Sacco2017}, and the Gaia-DR2 catalogue. We found RV measurements for 82 and 19~stars in our samples of cluster members in Cha~I and Cha~II, respectively. Some stars have multiple RV measurements in the literature (collected from different studies) and in such cases we decided to use the most precise value in our analysis.  

We applied the interquartile range (IQR) criterion to identify the stars with discrepant RVs in the sample. This method rejects outliers that lie over $1.5\times$IQR below the first quartile or above the third quartile. Doing so, we found four stars in Cha~I with discrepant RVs as illustrated in Figure~\ref{fig_RV}. We discarded these RV measurements from the analysis presented below, but we still retain the stars in the sample of cluster members. Their proper motions and parallaxes are consistent with membership in Cha~I and the corresponding RVs are more likely to be affected by other reasons (e.g. undetected binaries). This reduces the sample of sources in Cha~I with available RV information to 78~stars. We do not discard any RV measurement in the Cha~II sample based on the IQR criterion. 

Our search for published RV data for Chamaeleon stars in the literature allowed us to retrieve this information for about 49\% and 61\% of the stars in Cha~I and Cha~II, respectively. The main source of RV data in Cha~I is the study conducted by \citet{Sacco2017} as part of the Gaia-ESO survey, and we therefore refer the reader to that paper for further details on these measurements. In the case of the Cha~II molecular cloud, the main source of RV data is the study conducted by \citet{Biazzo2012}. The mean precision of the RVs in our sample is 0.5~km/s and 4.7~km/s in Cha~I and Cha~II, respectively. In the following, we use the precise Gaia-DR2 astrometry combined with published RV data to investigate the distance and spatial velocity of Chamaeleon stars.  

\begin{figure*}[!h]
\begin{center}
\includegraphics[width=0.49\textwidth]{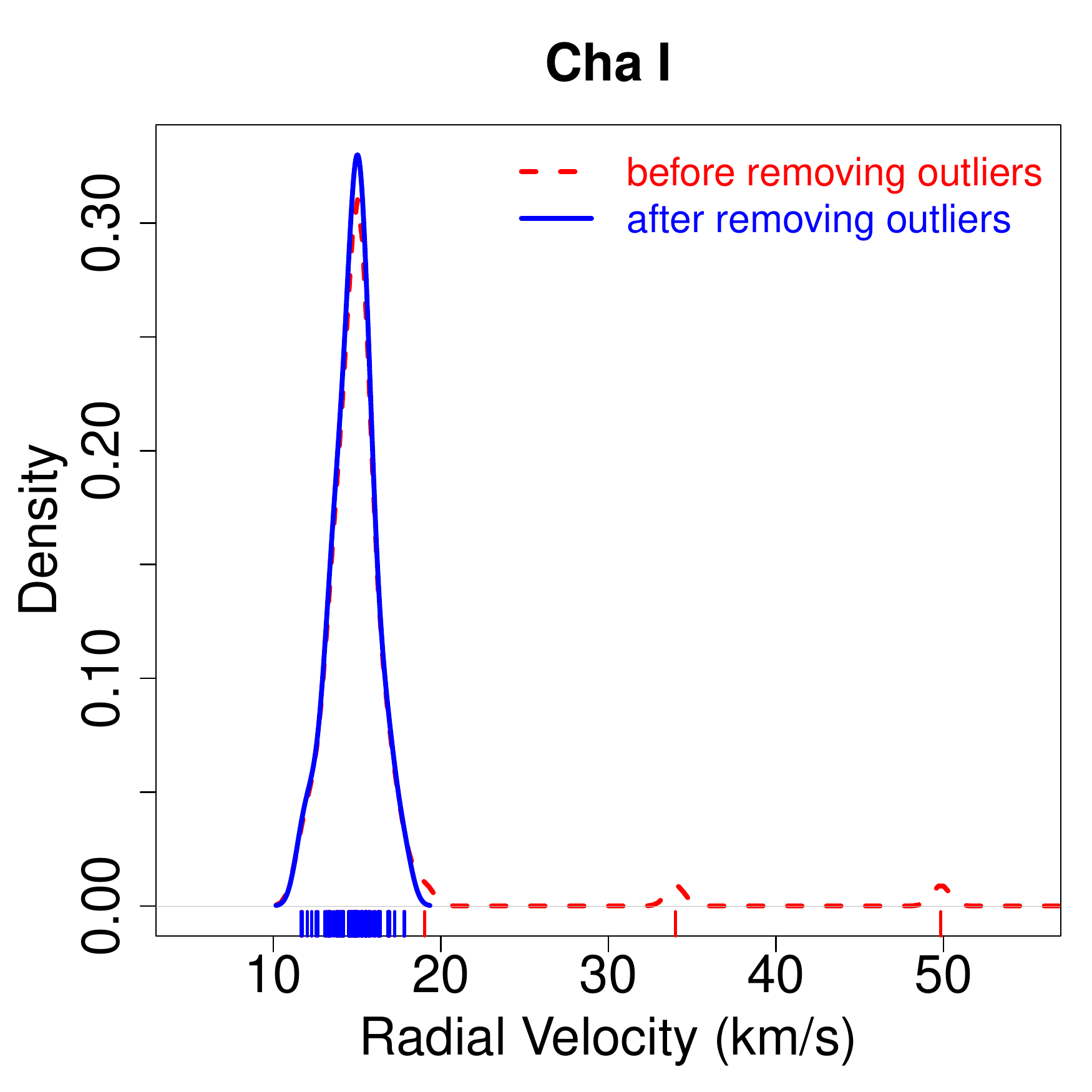}
\includegraphics[width=0.49\textwidth]{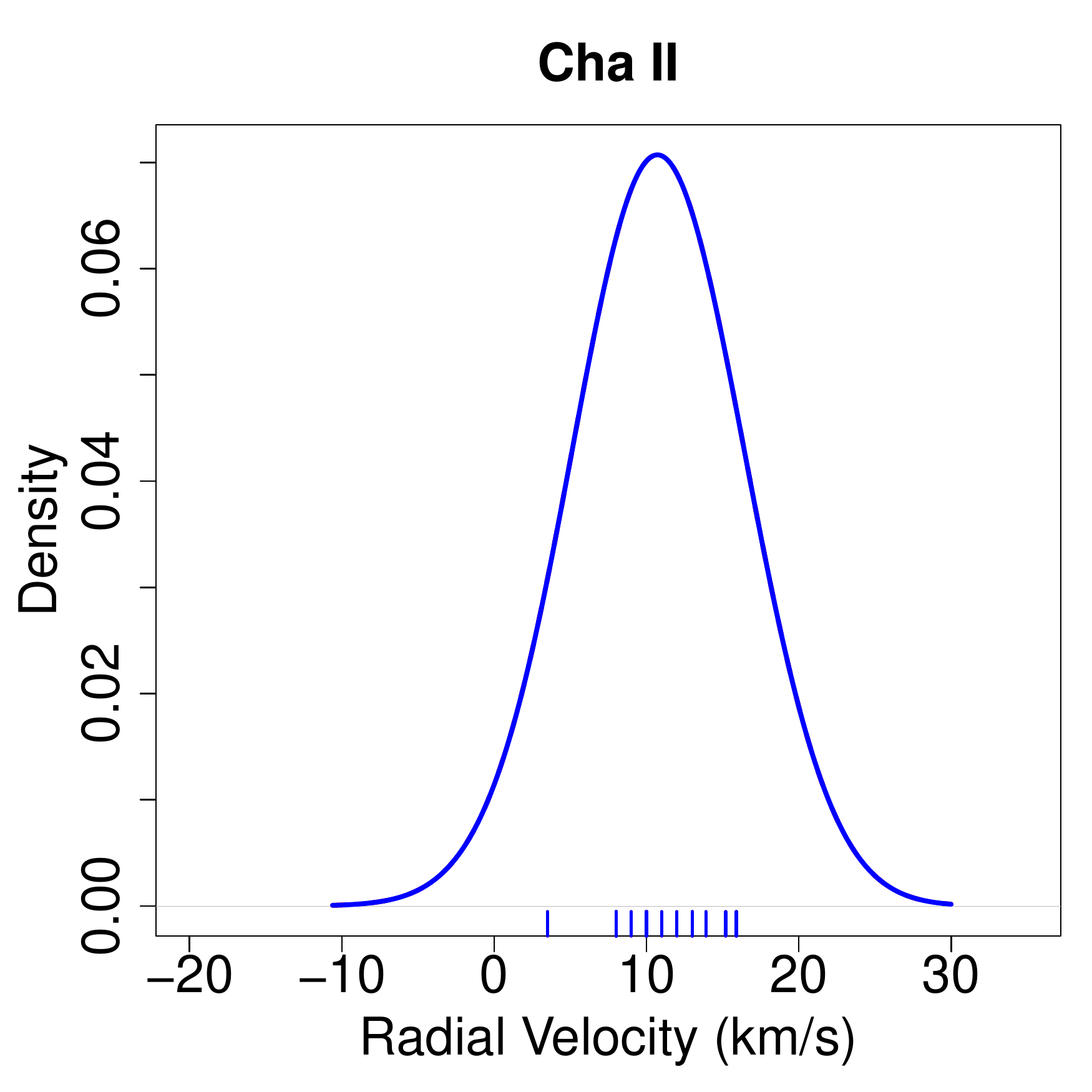}
\caption{Kernel density estimate of the RV distribution for the stars in Cha~I \textit{(left panel)} and Cha~II \textit{(right panel)}. We used the mean precision of the RVs in each sample (see text in Sect.~\ref{section3.1}) as kernel bandwidth. The tick marks in the horizontal axis mark the RVs of individual stars. The star with the most discrepant RV in the Cha~I sample, namely Gaia DR2 5201347408553455616 \citep[$V_{r}=164.11\pm1.61$~km/s, ][]{Sacco2017}, is not shown to improve the visibility of the plot. 
\label{fig_RV} 
}
\end{center}
\end{figure*}

\subsection{Proper motions and parallaxes of the subgroups}\label{section3.2}

The stellar population of Cha~I is composed of two subclusters that have been historically separated based on their position in the sky using the declination of $\delta=-77^{\circ}$ as midpoint to separate the northern and southern subgroups \citep[see e.g.][]{Luhman2007}. However, a recent study conducted by \citet{Roccatagliata2018} shows that the subclusters overlap and that they cannot be strictly divided by their sky positions.

\begin{figure*}[!h]
\begin{center}
\includegraphics[width=0.49\textwidth]{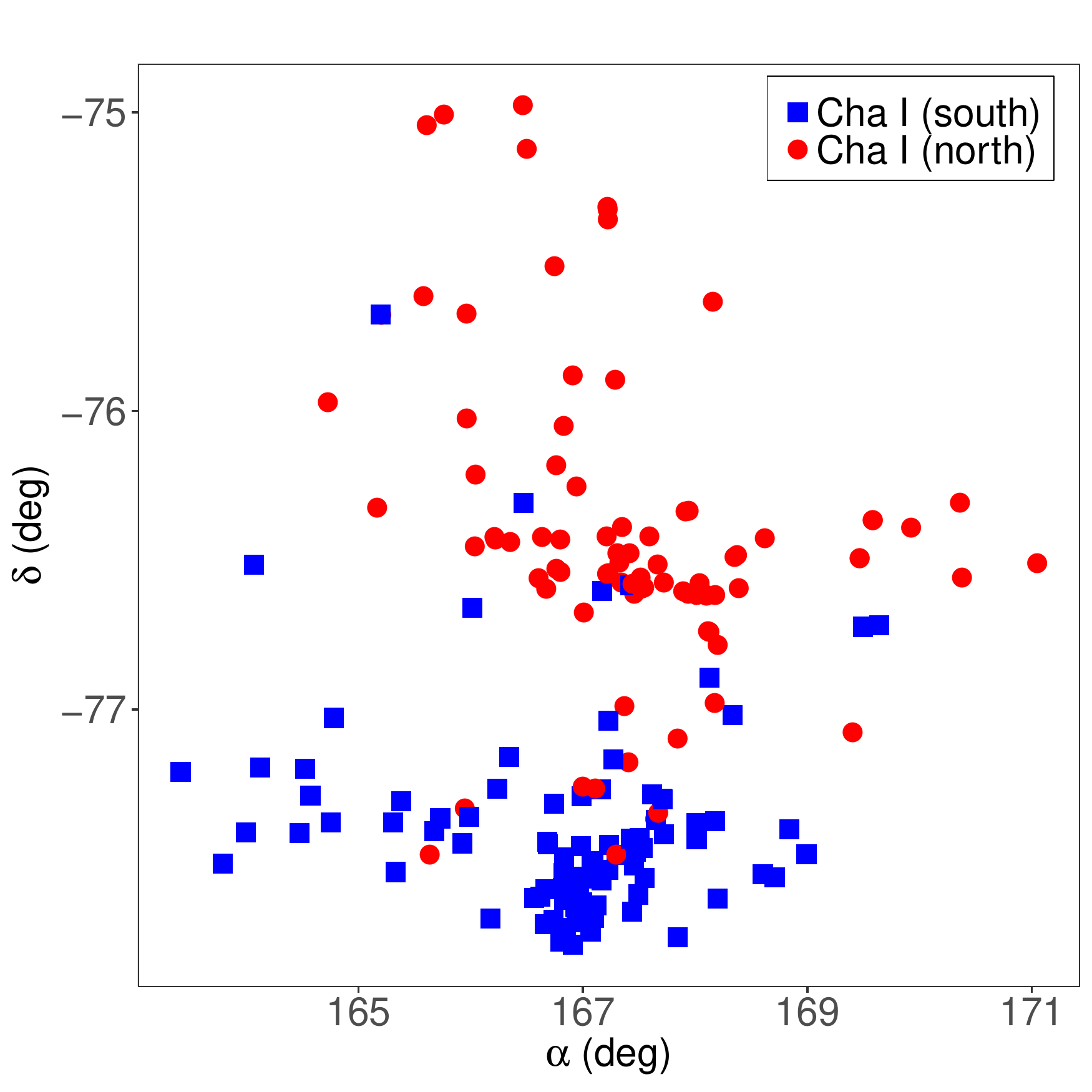}
\includegraphics[width=0.49\textwidth]{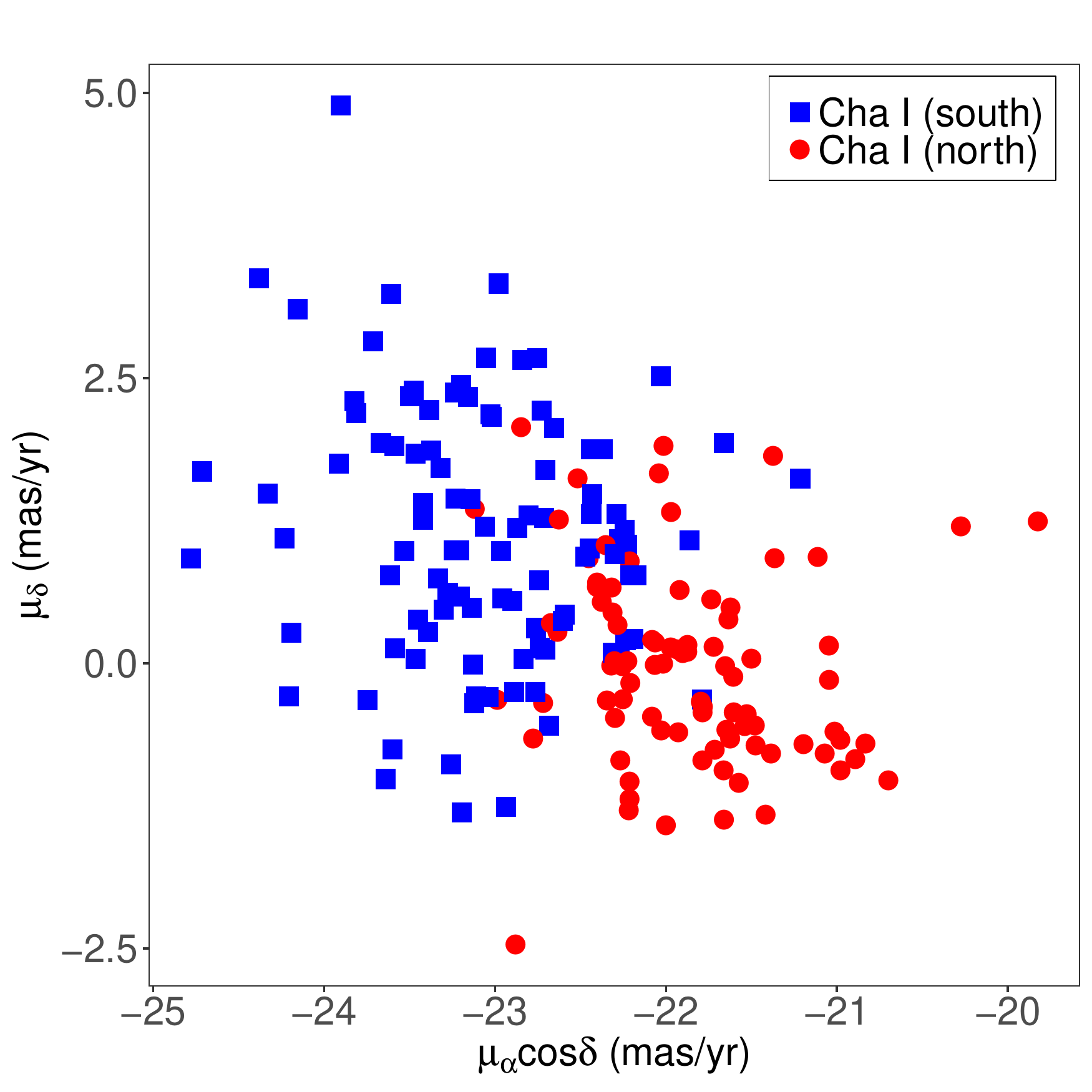}
\includegraphics[width=0.49\textwidth]{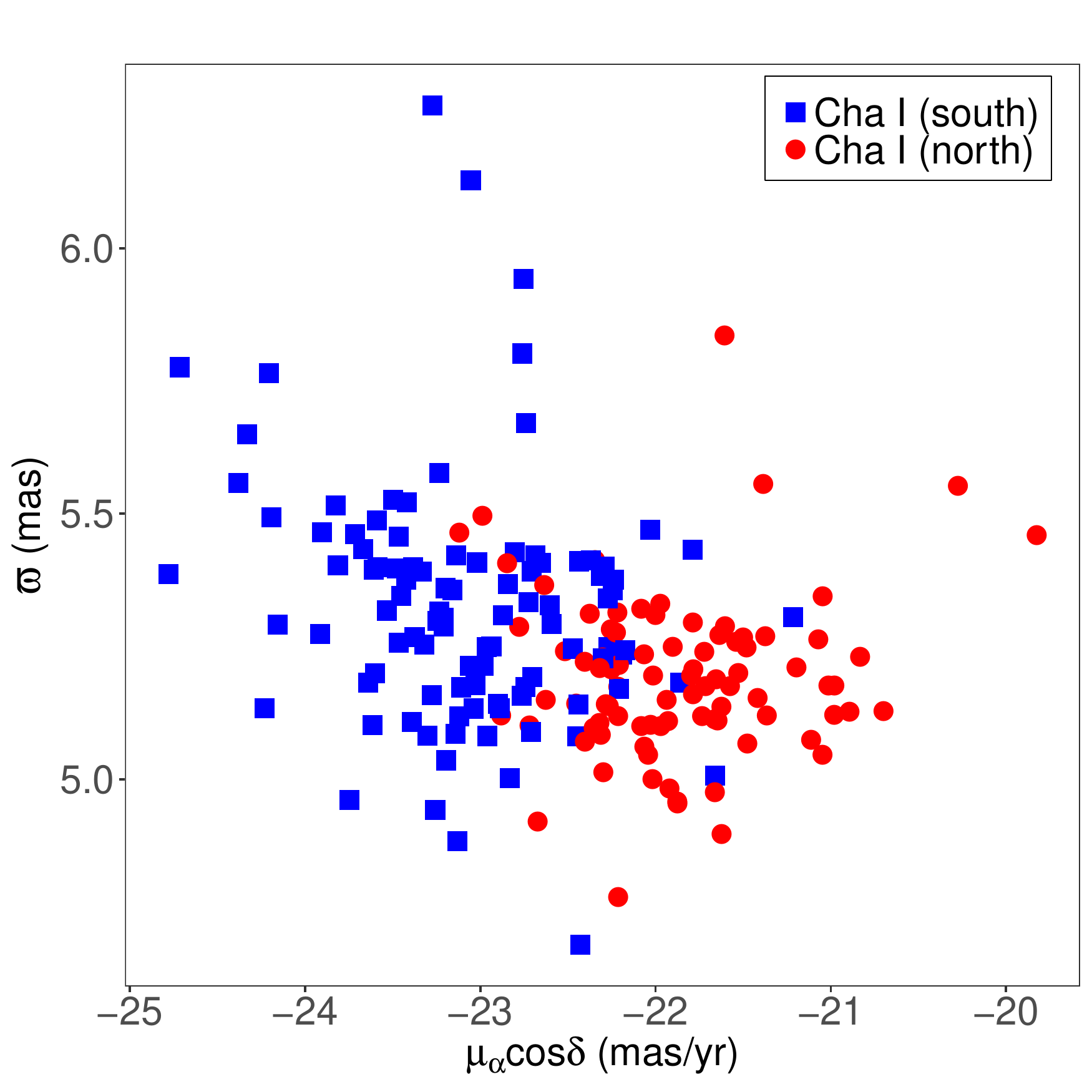}
\includegraphics[width=0.49\textwidth]{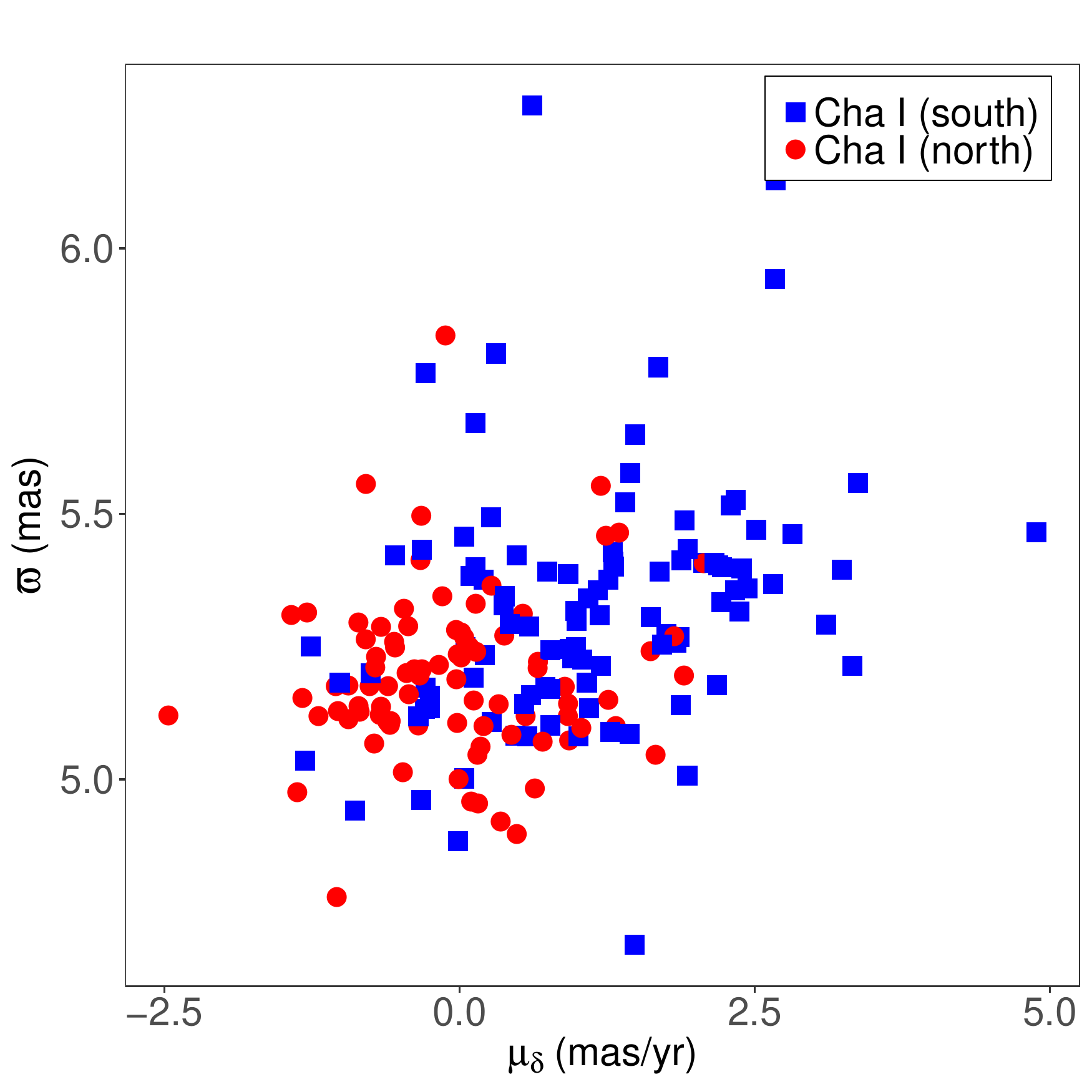}
\caption{Clustering results obtained for Cha~I with the PAM algorithm in the 5D space of position, proper motion, and parallax.
\label{fig_clustering} 
}
\end{center}
\end{figure*}

In this paper, we apply the Partitioning Around Medoids \citep[PAM,][]{PAM} clustering algorithm to separate the two subclusters in Cha~I. The PAM algorithm is an unsupervised machine-learning method for partitioning a $N$-dimensional dataset into $k$ groups (i.e. clusters) that works similarly to the $k$-Means algorithm \citep{kMeans}, but uses medoids to represent the centers of the clusters. We perform the clustering in the 5D space of astrometric observables $(\alpha,\delta,\mu_{\alpha}\cos\delta,\mu_{\delta},\varpi)$ of the stars using the \textit{pam} routine from the \texttt{cluster} package implemented in R programming language. We use the Average Silhouette \citep{ROUSSEEUW198753} and Gap Statistic \citep{GapStatistic} methods to confirm that the optimal number of clusters for our dataset is indeed $k=2$, which we use as input for the clustering analysis. The clustering analysis with PAM divides the sample into 76~stars and 84~stars for Cha~I (north) and Cha~I (south), respectively. Figure~\ref{fig_clustering} confirms that the subclusters in Cha~I overlap in position, proper motion, and parallax in good agreement with the findings reported by \citet{Roccatagliata2018}. 

In Table~\ref{tab_subgroups} we compare the proper motions and parallaxes of the Chamaeleon subgroups in our sample of members. It is apparent from this comparison that the stellar populations in Cha~I and Cha~II exhibit distinct kinematic properties (as already anticipated in Sect.~\ref{section2}). However, we also note a mean difference of about 1~mas/yr in the proper motions of the stars in the two subgroups of Cha~I despite the overlap of the subgroups shown in Figure~\ref{fig_clustering}. \citet{Roccatagliata2018} performed a two-sample Kolmogorov-Smirnov test and concluded that the proper motion distribution of the subclusters is not drawn from the same parent distribution, but the authors do not discuss whether the observed difference is due to projection effects or a different space motion. In Section~\ref{section3.3} we compute the spatial velocity of the stars and investigate whether the two subclusters in Cha~I are indeed kinematically distinct.

\begin{table*}[!h]
\centering
\scriptsize{
\caption{Proper motions and parallaxes of the Chamaeleon subgroups in our sample of cluster members.
\label{tab_subgroups}}
\begin{tabular}{lccccccccccc}
\hline\hline
Sample&$N_{init}$&$N_{RUWE}$&\multicolumn{3}{c}{$\mu_{\alpha}\cos\delta$}&\multicolumn{3}{c}{$\mu_{\delta}$}&\multicolumn{3}{c}{$\varpi$}\\
&&&\multicolumn{3}{c}{(mas/yr)}&\multicolumn{3}{c}{(mas/yr)}&\multicolumn{3}{c}{(mas)}\\
\hline\hline
&&&Mean$\pm$SEM&Median&SD&Mean$\pm$SEM&Median&SD&Mean$\pm$SEM&Median&SD\\
\hline

Cha I (north) & 87 & 76 & $ -21.903 \pm 0.063 $& -21.953 & 0.548 & $ -0.054 \pm 0.099 $& -0.023 & 0.863 & $ 5.178 \pm 0.016 $& 5.175 & 0.138 \\
Cha I (south) & 101 & 84 & $ -23.053 \pm 0.069 $& -23.048 & 0.629 & $ 1.127 \pm 0.118 $& 1.022 & 1.083 & $ 5.314 \pm 0.023 $& 5.322 & 0.211 \\
\hline
Cha I & 188 & 160 & $ -22.507 \pm 0.065 $& -22.403 & 0.824 & $ 0.566 \pm 0.091 $& 0.405 & 1.146 & $ 5.250 \pm 0.015 $& 5.235 & 0.192 \\
Cha II & 41 & 31 & $ -20.207 \pm 0.170 $& -19.954 & 0.945 & $ -7.635 \pm 0.129 $& -7.491 & 0.717 & $ 5.037 \pm 0.021 $& 5.061 & 0.115 \\

\hline
\end{tabular}
\tablefoot{We provide for each subgroup the initial number of stars, final number of stars after the RUWE filtering (see Sect.~\ref{section3.1}), mean, standard error of the mean (SEM), median and standard deviation (SD) of proper motions and parallaxes.}
}
\end{table*}

\subsection{Distance and spatial velocity of Chamaeleon stars}\label{section3.3}

In the following, we convert the stellar proper motions and parallaxes into distances and 2D velocities using Bayesian inference. This analysis uses the \textit{Kalkayotl} package\footnote{The code is available at \href{https://github.com/olivares-j/Kalkayotl}{https://github.com/olivares-j/Kalkayotl}.} \citep{kalkayotl,Olivares2020} in python programming language, which implements a number of priors for the distance. We took advantage of this code to investigate two priors for the distance that are based on purely statistical probability density distributions (hereafter, statistical priors), and another two priors that are based on astrophysical assumptions (hereafter, astrophysical priors). The statistical priors considered in this paper include the Uniform and Gaussian distributions. The two astrophysical priors that we use are based on the surface brightness profile of the Large Magellanic Cloud derived from star counts by \citet{Elson1987} and the King's profile distribution observed in globular clusters \citep{King1962}. More details on the implementation and performance of each prior are given in the \textit{Kalkayotl} paper \citep{Olivares2020}. The prior that we used in the case of the angular velocity (i.e. 2D tangential velocity) is a beta function following the online tutorials available in the \textit{Gaia} archive \citep{Luri2018}. This procedure takes as input the astrometric observables and covariance matrix of the stars as given by the Gaia-DR2 catalogue.

We compared the distances derived from different priors and confirmed that they do not depend on the choice of the prior. We therefore decided to report the distances derived from the uniform prior as our final results, because this is the most simple prior. We then used the resulting distances to compute the $XYZ$ position of the stars in a reference system that has its origin at the Sun where $X$ points to the Galactic centre, $Y$ points in the direction of Galactic rotation, and $Z$ points to the Galactic north pole. We converted the 2D tangential velocities and RVs of the stars into the $UVW$ components of the Galactic velocity (in the same reference system described before) using the transformation outlined by \citet{Johnson1987}. The distance and spatial velocity of the stars are given in Tables~\ref{tab_members_Cha1} and \ref{tab_members_Cha2}. 
 
\begin{table*}[!h]
\centering
\scriptsize{
\caption{Distance and spatial velocity of the Chamaeleon subgroups in our sample of cluster members.
\label{tab_distance_velocity}}
\begin{tabular}{lcccccccccccc}
\hline\hline
Sample&$N_{d}$&$N_{UVW}$&$d$&\multicolumn{3}{c}{$U$}&\multicolumn{3}{c}{$V$}&\multicolumn{3}{c}{$W$}\\
&&&(pc)&\multicolumn{3}{c}{(km/s)}&\multicolumn{3}{c}{(km/s)}&\multicolumn{3}{c}{(km/s)}\\
\hline\hline
&&&&Mean&Median&SD&Mean&Median&SD&Mean&Median&SD\\
\hline
Cha I (north) & 76 & 39 & $ 191.4 _{ -0.8 }^{+ 0.8 }$ &$ -10.6 \pm 0.4 $& -10.6 & 0.7 & $ -19.3 \pm 0.6 $& -19.4 & 1.0 & $ -11.5 \pm 0.2 $& -11.5 & 0.6 \\
Cha I (south) & 84 & 39 & $ 186.7 _{ -1.0 }^{+ 1.0 }$ &$ -11.5 \pm 0.6 $& -11.6 & 0.9 & $ -19.7 \pm 0.9 $& -19.8 & 1.2 & $ -10.9 \pm 0.3 $& -10.8 & 0.8 \\
\hline
Cha I (all stars) & 160 & 78 & $ 189.4 _{ -0.7 }^{+ 0.8 }$ &$ -11.0 \pm 0.5 $& -11.0 & 0.9 & $ -19.5 \pm 0.8 $& -19.6 & 1.1 & $ -11.2 \pm 0.3 $& -11.1 & 0.8 \\
Cha II (all stars) & 31 & 19 & $ 197.5 _{ -0.9 }^{+ 1.0 }$ &$ -11.0 \pm 2.9 $& -11.2 & 1.7 & $ -18.1 \pm 4.2 $& -17.8 & 2.7 & $ -8.5 \pm 1.4 $& -8.6 & 1.1 \\

\hline\hline
\end{tabular}
\tablefoot{We provide for each subgroup the number of stars used to compute the distance (after the RUWE filtering) and spatial velocity, Bayesian distance, mean, median and standard deviation (SD) of the $UVW$ velocity components. The uncertainties in the mean velocities are computed as explained in the text of Sect.~\ref{section3.3}.}
}
\end{table*}

Table~\ref{tab_distance_velocity} lists the distance of the Chamaeleon subgroups in our sample. We confirm with our methodology that the two subgroups of Cha~I (north and south) are located at different distances within the reported uncertainties as previously reported by \citet{Roccatagliata2018}. In addition, we measure the distance variation along the line of sight of $\Delta d=8.1^{+1.3}_{-1.1}$~pc between Cha~I and Cha~II, and find that they are separated by $23.1\pm0.3$~pc in the space of 3D positions.

The distances derived in this paper based on Gaia-DR2 data are more precise than the results of $179^{+11 +11}_{-10 -10}$~pc and $181^{+6 +11}_{-5 -10}$~pc obtained by \citet{Voirin2018} for Cha~I and Cha~II, respectively (the error bars reported in those solutions refer to random and systematic uncertainties in this order). The latter made use of the parallaxes from the Tycho-Gaia Astrometric Solution catalogue \citep[TGAS,][]{Lindegren2016} which were affected by systematic errors of about 0.3~mas, which explains the much larger uncertainties in the distances. Our results are more precise than the distances of $192\pm6$~pc and $198\pm6$ provided by \citet{Dzib2018} for Cha~I and Cha~II, respectively, using different samples of stars. Our distance estimates are also consistent with the results of $192.7^{+0.4}_{-0.4}$~pc and $186.5^{+0.7}_{-0.7}$~pc derived by \citet{Roccatagliata2018} for the northern and southern subclusters of Cha~I, respectively. However, in this case we note that our uncertainties are larger by a factor of almost two. As explained in Sect.~2 of \citet{Roccatagliata2018} the authors did not include the systematic errors on the Gaia-DR2 parallaxes in their analysis and derived the distance from the inverse of the parallax. In this study, we derived the distances from Bayesian inference using the \textit{Kalkayotl} code \citep{Olivares2020} which is designed to deal with some characteristics of the Gaia-DR2 catalogue including the parallax zero-point correction and spatial correlations which are modelled with the covariance function of \citet{Vasiliev2019}. The existence of systematic errors in the Gaia-DR2 catalogue does not compromise the valuable astrometry delivered by the Gaia satellite, but needs to be considered to avoid underestimating uncertainties.  

Before computing the distance and spatial velocity of the stars we corrected the Gaia-DR2 parallaxes by the zero-point shift of -0.030~mas and added the systematic errors of 0.1~mas/yr in quadrature to the formal uncertainties on proper motions \citep[see][for more details]{Lindegren2018}. It is nevertheless important to mention that other zero-point corrections for the Gaia-DR2 parallaxes exist in the literature. They range from $-0.031\pm0.011$~mas \citep{Graczyk2019} to $-0.082\pm0.033$~mas \citep{Stassun2018}. The lowest value is consistent with the zero-point correction provided by the Gaia collaboration \citep{Lindegren2018}, which was adopted throughout this study. We compare in Table~\ref{tab_distance_zeropoint} the minimum and maximum distances of the Chamaeleon subgroups when we consider different zero-point corrections applied to the Gaia-DR2 parallaxes. The largest zero-point correction available in the literature  \citep[$-0.082\pm0.033$~mas,][]{Stassun2018} puts the Chamaeleon subgroups about 2~pc closer to the Sun, but the resulting distances are still consistent with the results obtained in Table~\ref{tab_distance_velocity} within the reported uncertainties. We therefore conclude that our results are not sensitive to the adopted zero-point correction for the parallaxes. 

\begin{table}[!h]
\centering
\footnotesize{
\caption{Distance of the Chamaeleon subgroups using different zero-point corrections for the Gaia-DR2 parallaxes.
\label{tab_distance_zeropoint}}
\begin{tabular}{lccc}
\hline\hline
Sample&$N_{d}$&\multicolumn{2}{c}{distance}\\
&&\multicolumn{2}{c}{(pc)}\\
\hline\hline
&&\citet{Stassun2018}&no correction\\
\hline
Cha~I (north)&76&$189.6^{+0.8}_{-0.8}$&$192.5^{+0.8}_{-0.7}$\\
Cha~I (south)&84&$184.9^{+0.9}_{-0.9}$&$187.8^{+0.9}_{-0.9}$\\
\hline
Cha~I &160&$187.5^{+0.8}_{-0.8}$&$190.5^{+0.8}_{-0.8}$\\
Cha~II &31&$195.5^{+0.9}_{-0.9}$&$198.6^{+1.0}_{-0.9}$\\
\hline\hline
\end{tabular}
\tablefoot{We provide for each subgroup the number of stars and the Bayesian distances derived from the largest zero-point shift reported in the literature \citep[$-0.082\pm0.033$~mas,][]{Stassun2018} and without any zero-point correction applied to the Gaia-DR2 parallaxes.}
}
\end{table}

Although the mean space motion of stellar groups provides useful information, the non-trivial question that arises in our discussion is whether the Chamaeleon subgroups discussed throughout this study are kinematically distinct. In Section~\ref{section3.2} we report a mean difference of about 1~mas/yr in the proper motions of the two subclusters in Cha~I which translates into a relative tangential velocity of the order of 1~km/s. \citet{Sacco2017} also reported a mean difference of about 1~km/s between the RVs of the northern and southern subgroups of Cha~I. This relatively small difference in velocity can result from other effects such as for example intrinsic velocity dispersion, undetected binaries, and correlated noise. For example, the transformation of  parallaxes, proper motions, and RVs to 3D velocities can result in correlated errors between the velocity components even in the absence of correlations between the astrometric observables \citep[see e.g.][]{Brown1997,Perryman1998}. We followed the procedure outlined in Sect.~7.2 of \citet{Perryman1998} to investigate this effect in our results. The mean motion of the Chamaeleon subgroups is obtained by averaging the measured velocities and the uncertainty in the mean is computed from the mean of the covariance matrices of the individual stars. The resulting spatial velocities of the Chamaeleon subgroups are given in Table~\ref{tab_distance_velocity}. Thus, the relative motion between the two subclusters of Cha~I (in the sense `north' minus `south') is $(\Delta U,\Delta V, \Delta W)=(0.9,0.4,-0.6)\pm(0.7,1.1,0.4)$~km/s. Similarly, the relative motion between Cha~I and Cha~II (in the sense `Cha~II' minus `Cha~I') is $(\Delta U,\Delta V, \Delta W)=(0.0,1.4,2.7)\pm(2.9,4.3,1.4)$~km/s. We therefore conclude that the reported differences in the velocity of the subgroups are not significant at the 3$\sigma$ level when we take into account the covariances in our analysis. 

\begin{figure*}[!h]
\begin{center}
\includegraphics[width=1.0\textwidth]{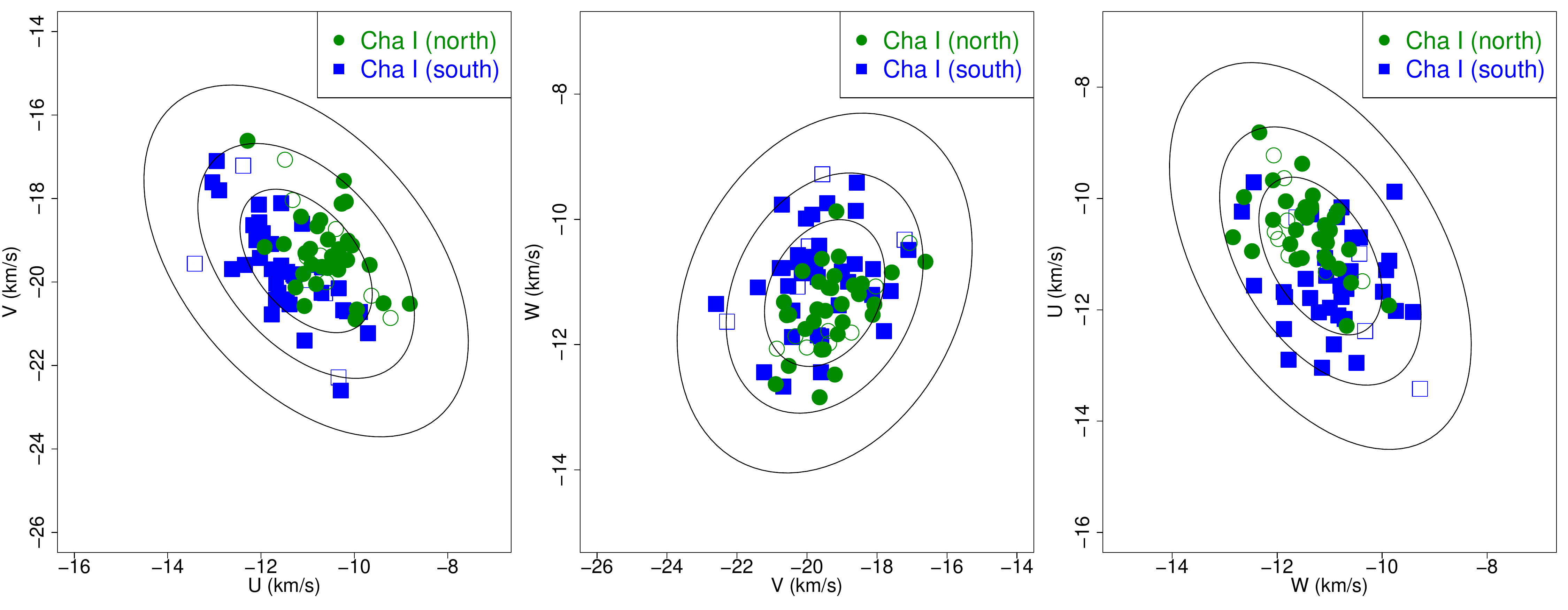}
\includegraphics[width=1.0\textwidth]{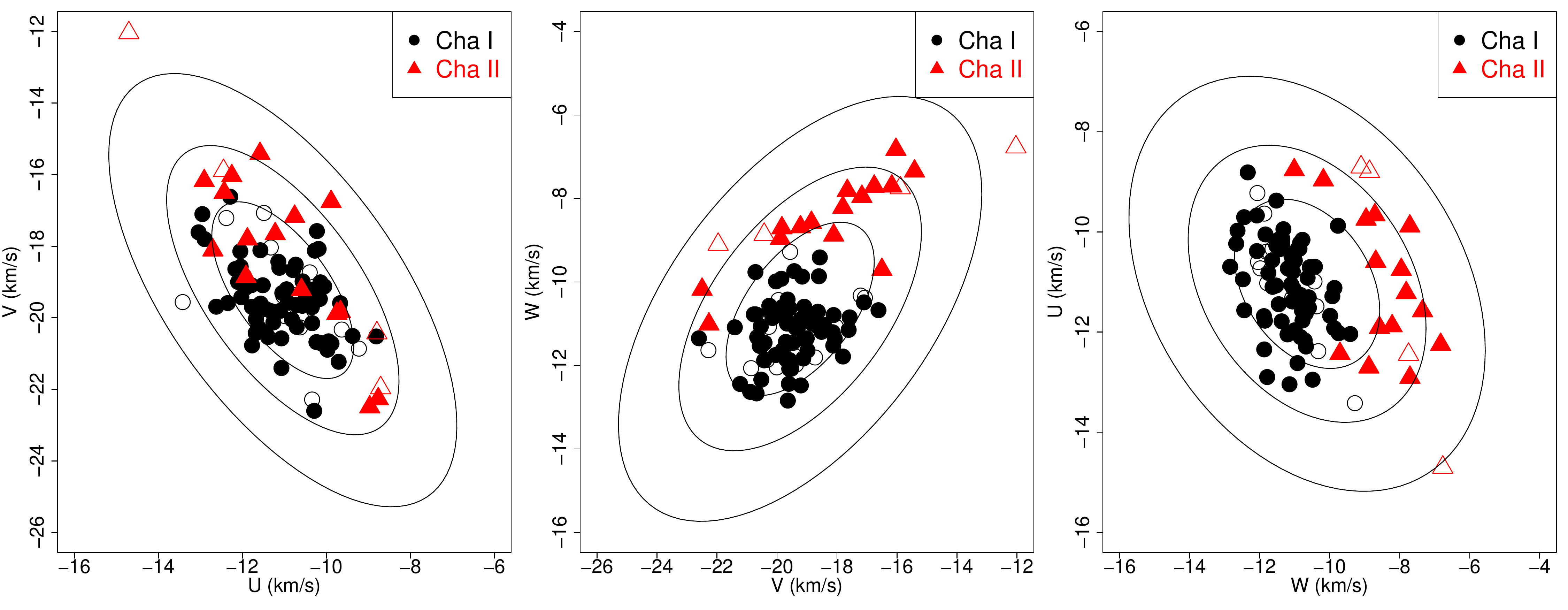}
\caption{Distribution of the 3D spatial velocity of Chamaeleon stars. The upper panels compare the velocity of Cha~I (north) and Cha~I (south), and the lower panels compare the velocity distribution of Cha~I and Cha~II. Filled and open symbols denote single stars and known binaries (or multiple systems), respectively. The contours indicate the  68.3\%, 95.4\%, and 99.9\% confidence levels computed from the covariance matrix of the ensemble of data points illustrated in each panel.
\label{fig_UVW} 
}
\end{center}
\end{figure*}

In addition, we also note from Figure~\ref{fig_UVW} that the space motion of most members in the two subclusters of Cha~I is consistent within $1\sigma$ of the observed velocity dispersion. The observed scatter of the stars in the velocity space results from both measurement errors and the intrinsic velocity dispersion of the cluster. The values listed in Table~\ref{tab_distance_velocity} for the one-dimensional velocity dispersion of Cha~I in each velocity component are in good agreement with the velocity dispersion of $1.10\pm0.15$~km/s reported by \citet{Sacco2017} based on the RV of the stars. The median uncertainties in the UVW velocity components of Cha~I stars are 0.6, 0.5, and 0.4~km/s, suggesting that the internal velocity dispersion is resolved. On the other hand, the median uncertainties in the three velocity components of Cha~II stars are 3.4, 4.7, and 1.8~km/s, implying that the internal velocity dispersion is not resolved. Thus, the observed scatter of Cha~II stars in the space of velocities is most probably due to measurement errors and can be explained by the large uncertainties in the RV of the stars (see Sect.~\ref{section3.1}) which propagate to the 3D spatial velocities. 

We now compare the space motion of the stellar populations associated with the molecular clouds with the other two young stellar groups located in the Chamaeleon star-forming region. The $\epsilon$~Cha and $\eta$~Cha associations are located in the same sky region of the molecular clouds, but they constitute a foreground population of young stars at a distance of about 100~pc \citep{Gagne2018b}. We cross-matched our samples of Cha~I and Cha~II stars identified in this study with the lists of $\epsilon$~Cha and $\eta$~Cha given in the literature \citep[see e.g.][]{Gagne2018a,Gagne2018c}, but we found no sources in common. The relative motion of Cha~I with respect to the space motion of $\epsilon$~Cha and $\eta$~Cha derived by \citet{Murphy2013} is $(\Delta U,\Delta V,\Delta W)=(-0.1,0.9,-1.3)\pm(0.9,1.5,1.4)$~km/s and $(\Delta U,\Delta V,\Delta W)=(-0.8,1.2,0.0)\pm(0.5,0.8,0.3)$~km/s, respectively (in the sense `Cha~I' minus `$\epsilon$~Cha' or `$\eta$~Cha'). Similarly, the relative motion of Cha~II with respect to $\epsilon$~Cha and $\eta$~Cha is $(\Delta U,\Delta V,\Delta W)=(-0.1,2.3,1.4)\pm(3.0,4.4,2.0)$~km/s and  $(\Delta U,\Delta V,\Delta W)=(-0.8,2.6,2.7)\pm(2.9,4.2,1.4)$~km/s, respectively. Therefore, we conclude that the space motion of $\epsilon$~Cha and $\eta$~Cha is consistent with the space motion of the stars in the Cha~I  and Cha~II molecular clouds within 3$\sigma$ of the reported uncertainties in the spatial velocities. The common space motion and similar age of the stars in these stellar groups \citep[see e.g.][]{Luhman2007,Gagne2018b} suggest that they may have formed from the same parent cloud, but further investigation study is warranted to confirm this hypothesis.  

\subsection{Hertzsprung-Russell diagram and relative ages of the subgroups}\label{section3.4}

In this section we use the distances derived from Gaia-DR2 parallaxes (see Sect.~\ref{section3.3}) and spectroscopic data collected from previous studies to construct the most accurate Hertzsprung-Russell diagram (HRD) of the Chamaeleon star-forming region. 

We proceed as follows to place the stars in the HRD. First, we compile the spectral types and extinctions for individual stars  determined from past studies thanks to the numerous spectroscopic surveys performed in this region. We found spectral types and extinctions for 144~stars (out of 160~stars) in Cha~I and 28~stars (out of 31~stars) in Cha~II which are given by \citet{Esplin2017} and \citet{Spezzi2008}, respectively. Second, we compute the photospheric luminosities from the $J$-band given by the 2MASS catalogue \citep{Cutri2003} which is available for all sources with measured spectral type in our sample. We corrected the apparent magnitudes for the extinction values given by \citet{Esplin2017} and \citet{Spezzi2008} using the the extinction relations for various bands given by \citet{Cieza2005} to convert them to the adopted band when necessary. We used the bolometric corrections and effective temperatures for the adopted spectral types given in Table~6 of \citet{Pecaut2013} which is specific for pre-main sequence stars. For a few sources in our sample with spectral types earlier than F0 and later than M5 which are not included in this table we adopt the bolometric corrections and effective temperatures for dwarf stars given in Table~5 of that same study as an approximation. We assumed an uncertainty of half a subclass in the spectral types compiled from the literature which results in an uncertainty of about 20 to 120~K on the effective temperatures for most stars. The resulting stellar luminosities are presented in Table~\ref{tab_HRD}.

\begin{figure*}[!h]
\begin{center}
\includegraphics[width=0.49\textwidth]{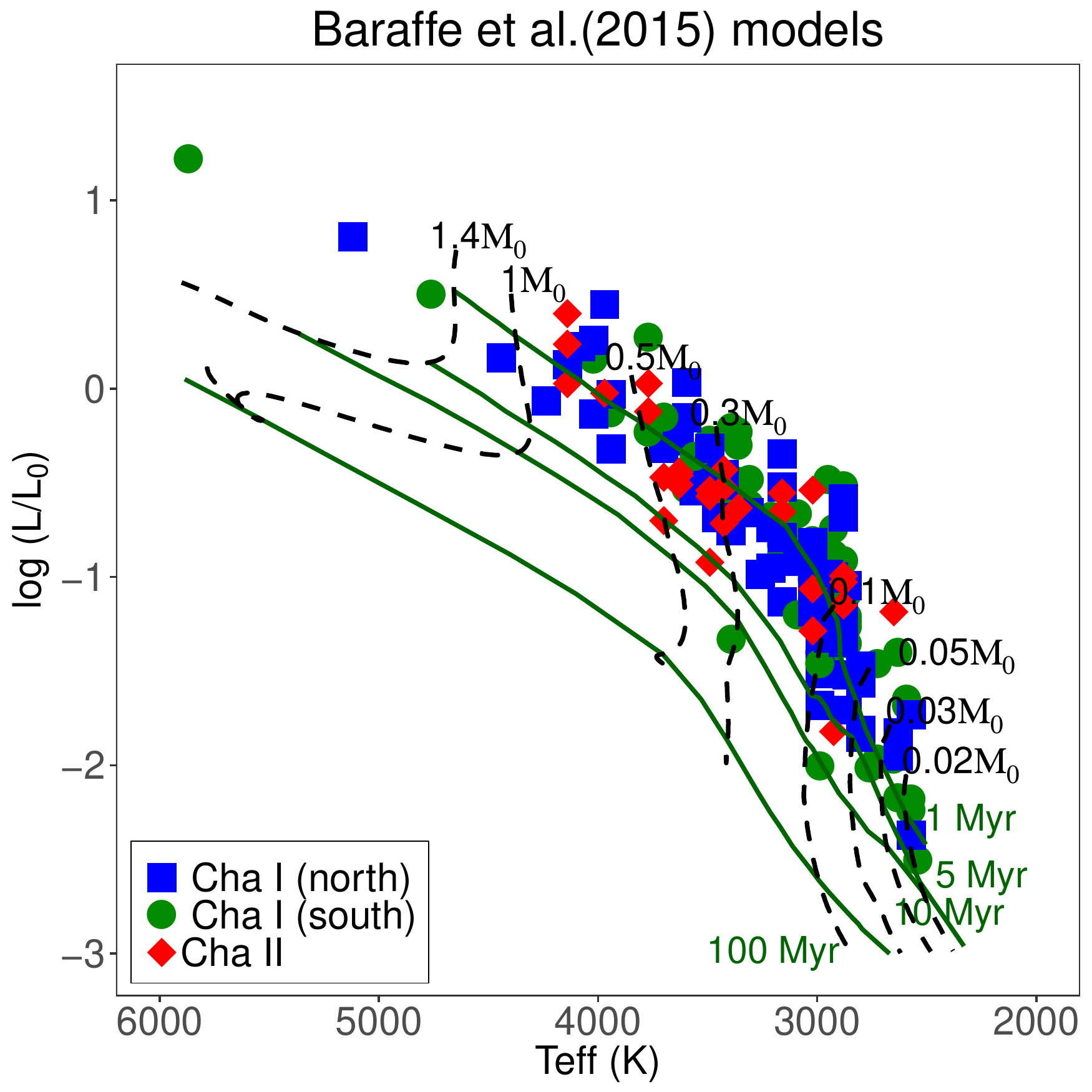}
\includegraphics[width=0.49\textwidth]{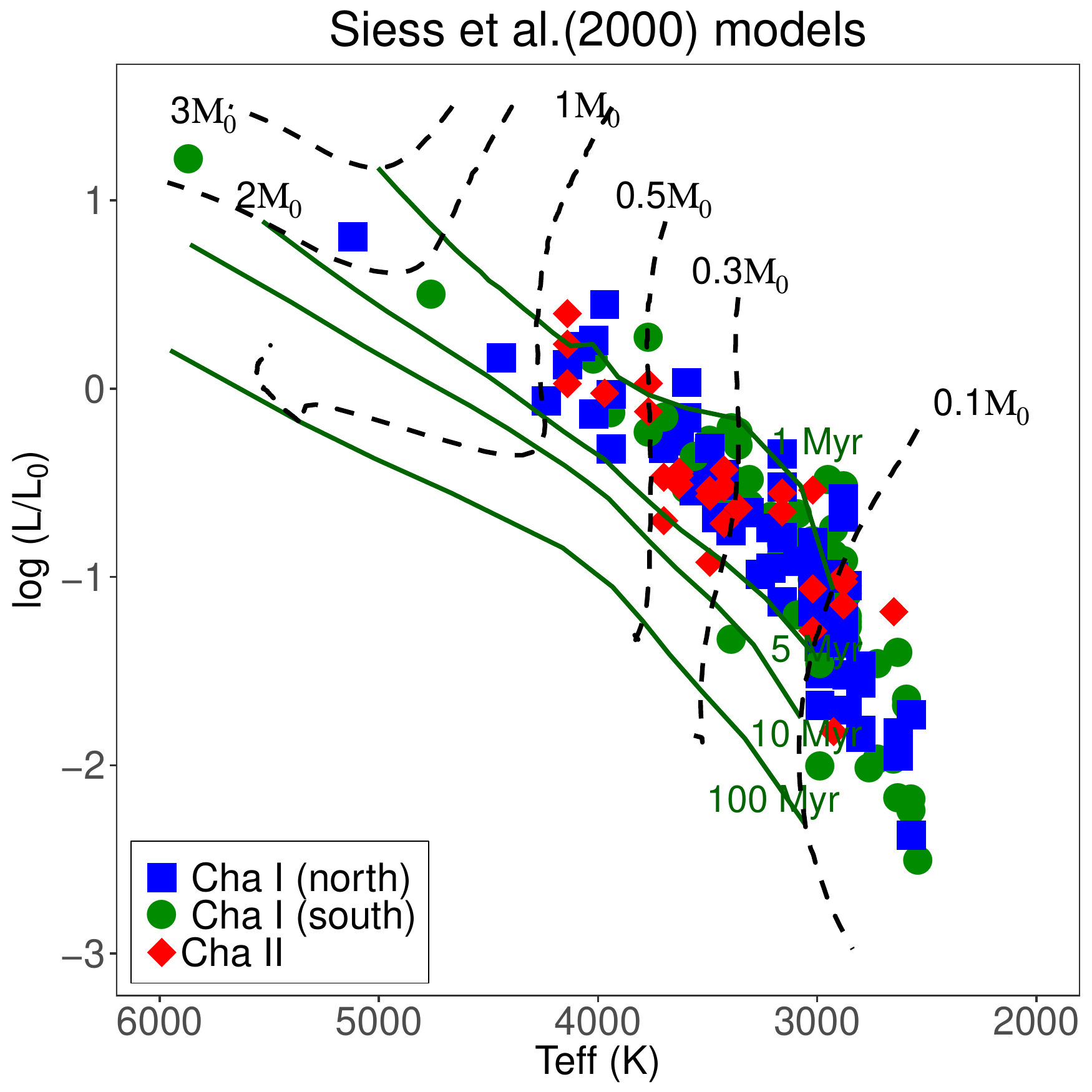}
\caption{HRD of the Chamaeleon star-forming region compared to the grid of isochrones and tracks from pre-main sequence stars models. The green solid and black dashed lines denote the isochrones and tracks for each model, respectively, with the ages (in Myr) and masses (in M$_{\odot}$) indicated in the panels. The different colours and symbols represent the Chamaeleon subgroups discussed throughout this paper. The most massive star in our sample, namely Gaia DR2 5201128124701636864, is not shown here to improve the visibility of the low-mass stars that largely dominate our sample.
\label{fig_HRD} 
}
\end{center}
\end{figure*}

Figure~\ref{fig_HRD} shows the HRD with the pre-main sequence star models of \citet{BHAC15} and \citet{Siess2000}. The two grids of models combined together cover the entire mass domain of our sample which ranges from about 0.02 to 3~M$_{\odot}$. HD~97048 (Gaia~DR2~5201128124701636864) in Cha~I is the most massive cluster member in our sample and a A0V Herbig Ae/Be star \citep[see e.g.][]{Chen2016}. Although age determination at these early stages of stellar evolution is rather uncertain, the HRD analysis suggests that the stars in our sample are mostly younger than 5~Myr. In particular, we note that most sources in the HRD are distributed along the 1~Myr isochrone. Some of them are probably binaries or high-order multiple systems that will require further investigation in future studies. However, it is interesting to note that the Chamaeleon stars appear to be younger as compared to previous studies in light of our new analysis based on the stellar distances derived from the Gaia-DR2 parallaxes. For example, the bolometric luminosities derived in this paper are systematically higher than the values obtained by \citet{Luhman2007} who derived the ages of 3-4~Myr and 5-6~Myr for the northern and southern subclusters of Cha~I, respectively. The latter study adopted a distance modulus of 6.05 which defines a (common) distance of 162~pc for all stars in the region (i.e., about 30~pc closer to the Sun than the value reported in Table~\ref{tab_distance_velocity}) and leads to overestimated ages in the HRD. A similar argument also applies to the results obtained by \citet{Spezzi2008} in Cha~II who adopted the distance of 178$\pm$18~pc derived by \citet{Whittet1997} to this cloud and reported ages of 3-4~Myr for most sources in their sample. We computed the isochronal age of the stars in our sample by interpolating between the isochrones of the \citet{BHAC15} and
\citet{Siess2000} models (see Table~\ref{tab_HRD}). This analysis is restricted to the stars covered by these grids of models (see Figure~\ref{fig_HRD}). The median age of the Cha~I and Cha~II subgroups is 1-2~Myr (see Table~\ref{tab_YSO} and discussion below) which suggests that they are indeed younger than previously thought. 

\begin{figure}[!h]
\begin{center}
\includegraphics[width=0.49\textwidth]{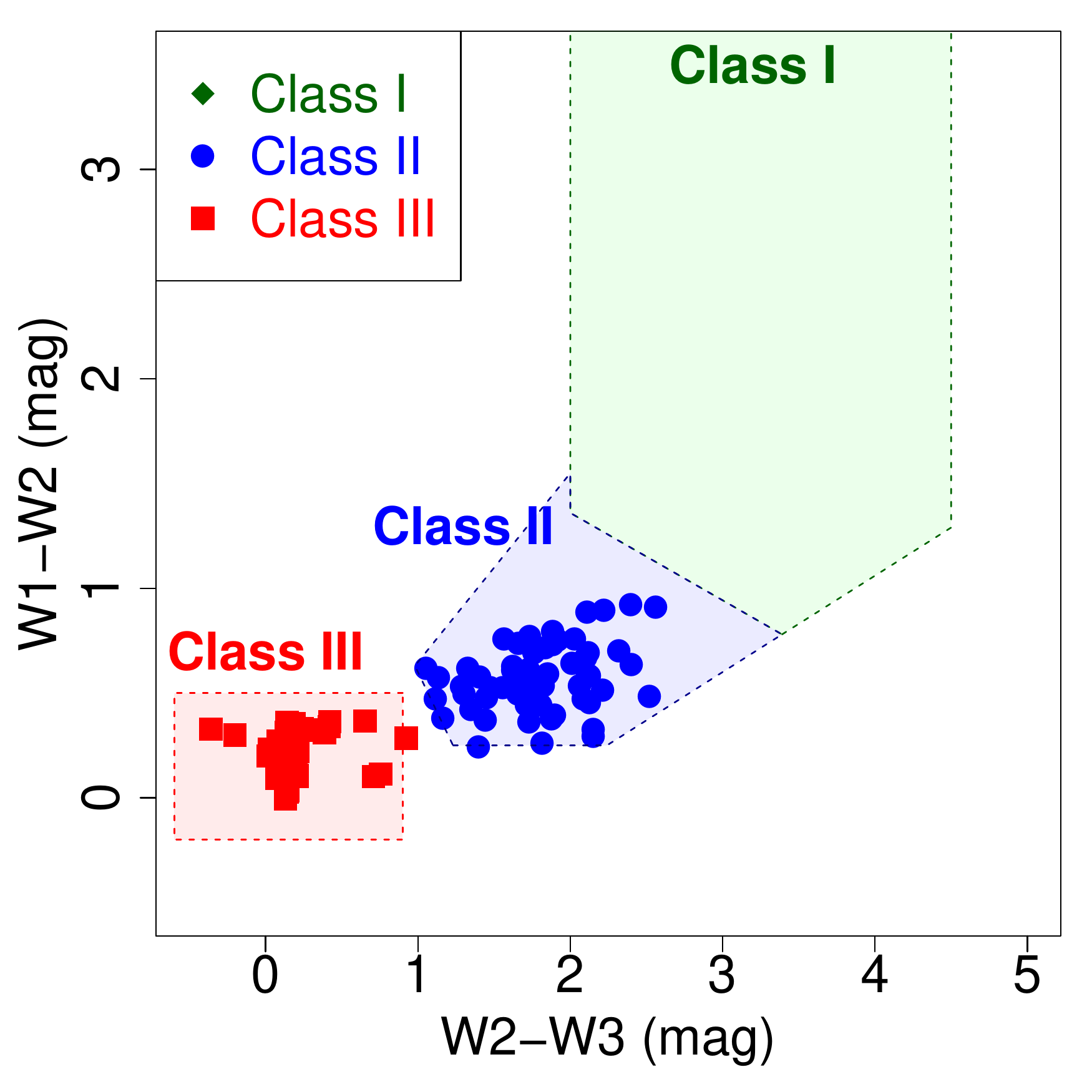}
\caption{Colour-colour diagram constructed from infrared colours for the Chamaeleon stars identified in our membership analysis. This is one of the diagrams used in the classification scheme proposed by \citet{Koenig2014}.
\label{fig_Koenig2014} 
}
\end{center}
\end{figure}

Let us now compare the relative ages of the Chamaeleon subgroups based on the fraction of disc-bearing stars in each sample. \citet{Luhman2008} classified the spectral energy distribution (SED) of 122~stars in our sample of 160~cluster members in Cha~I based on the spectral index $\alpha$ \citep{Lada1987} computed from infrared photometry. We compared these results with the ones derived from the classification scheme developed by \citet{Koenig2014} based on infrared colours from the AllWISE catalogue \citep{WISE}. The latter method was originally designed to identify Class~I and Class~II sources which exhibit strong infrared excess emission as illustrated in Figure~\ref{fig_Koenig2014}. We note that many sources in our sample fall into the region between $W2-W3<1.0$ and $W1-W2<0.5$ where Class~III and asymptotic giant branch (AGB) stars reside \citep[see e.g.][]{Koenig2014}. However, given the very young ages of the stars in our sample (as derived from the HRD) it seems unlikely that these sources are AGB stars and we therefore classify them as Class~III stars. When we compare the SED classification derived by \citet{Luhman2008} with the methodology proposed by \citet{Koenig2014} we find a perfect match for all sources in common between the two methods. By combining the two methodologies we were able to classify the SED of 142~stars in Cha~I. We proceeded in a similar manner for the Cha~II sample. We compiled the SED classification for 28~stars from the study of \citet{Alcala2008}, and derived the SED subclass for one additional star based on the method of \citet{Koenig2014}. 

\begin{table}[!h]
\centering
\scriptsize{
\caption{Fraction of disc-bearing stars and age estimates of the Chamaeleon subgroups. 
\label{tab_YSO}}
\begin{tabular}{lccrrcccc}
\hline\hline
&&&&\multicolumn{2}{c}{BHAC15}&\multicolumn{2}{c}{SDF00}\\
&&&&\multicolumn{2}{c}{models}&\multicolumn{2}{c}{models}\\
\hline
Sample&$N_{\star}$&Class~II&Class~III&$N_{\star}$&Age&$N_{\star}$&Age \\
&&&&&(Myr)&&(Myr)\\
\hline\hline

Cha~I (north)&70&33 (47\%)&37 (53\%)&33&1.9&49&2.4\\
Cha~I (south)&72&30 (42\%)&42 (58\%)&30&1.4&53&2.4\\
\hline
Cha~I&142&63 (44\%)&79 (56\%)&63&1.7&102&2.4\\
Cha~II&29&22 (76\%)&7 (24\%)&15&1.7&23&2.3\\
\hline\hline
\end{tabular}
\tablefoot{We provide the number of stars and relative fraction of the SED subclasses (in parenthesis), number of stars and median age inferred from \citet[][BHAC15]{BHAC15} and \citet[][SDF00]{Siess2000} models. }
}
\end{table}

\begin{figure*}[!h]
\begin{center}
\includegraphics[width=0.33\textwidth]{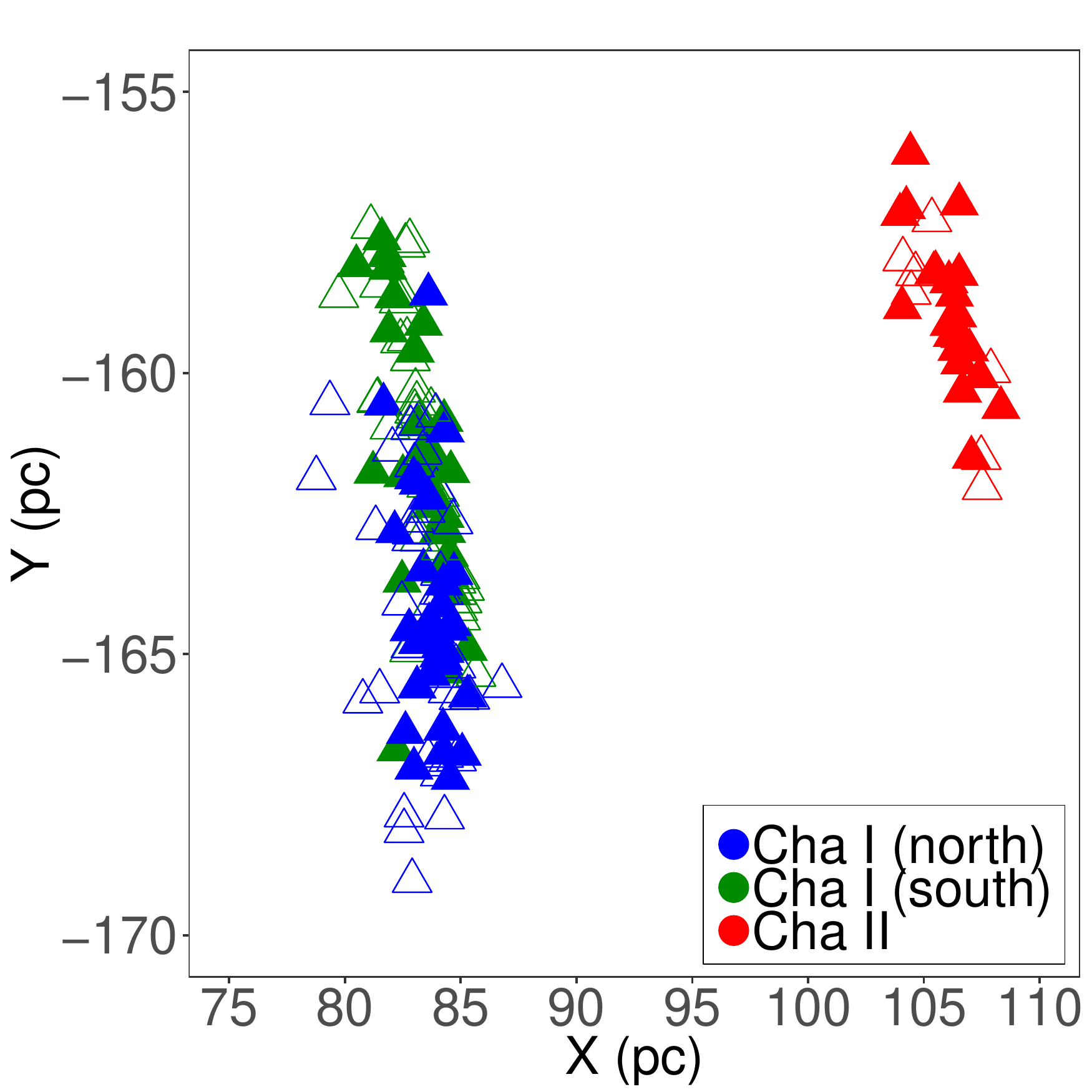}
\includegraphics[width=0.33\textwidth]{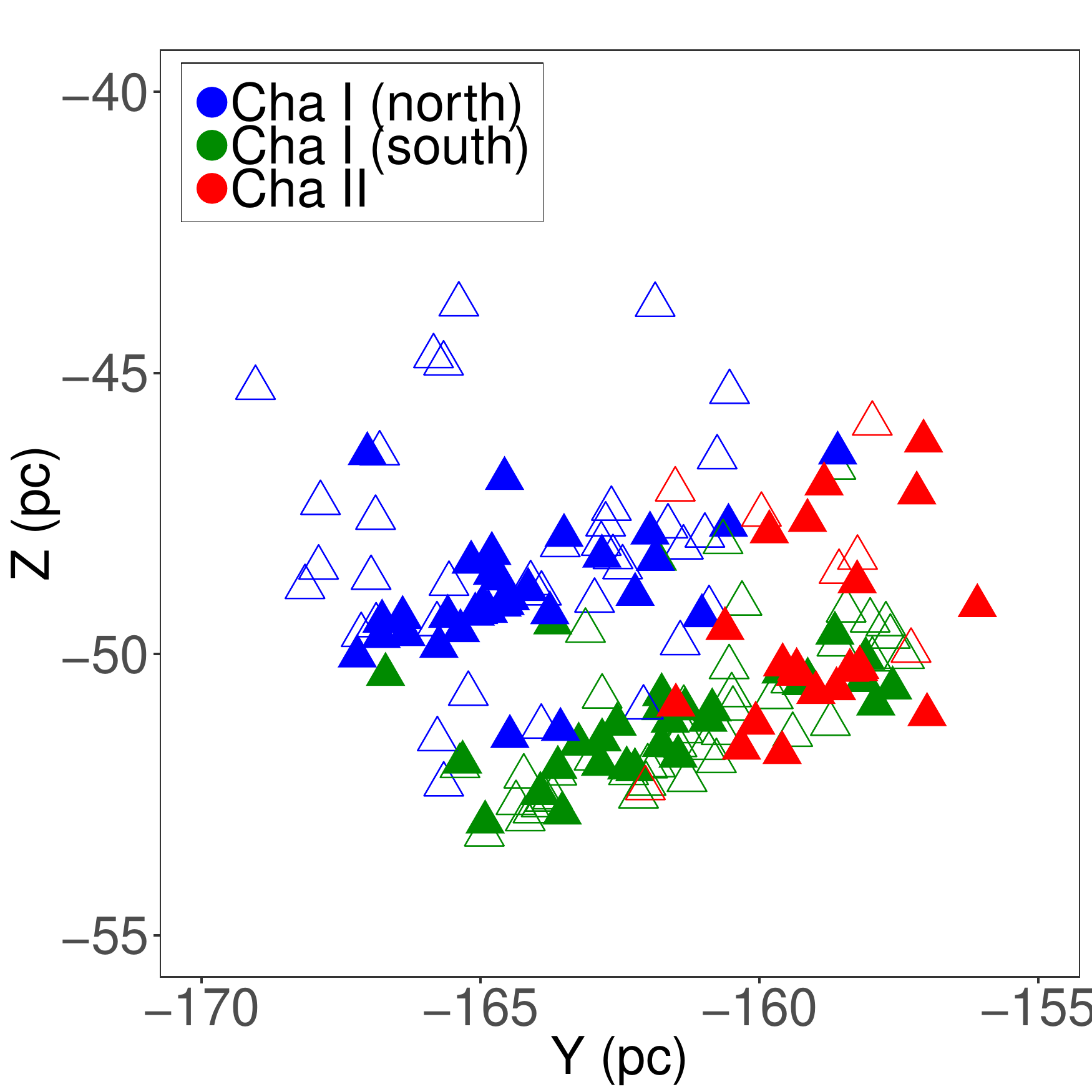}
\includegraphics[width=0.33\textwidth]{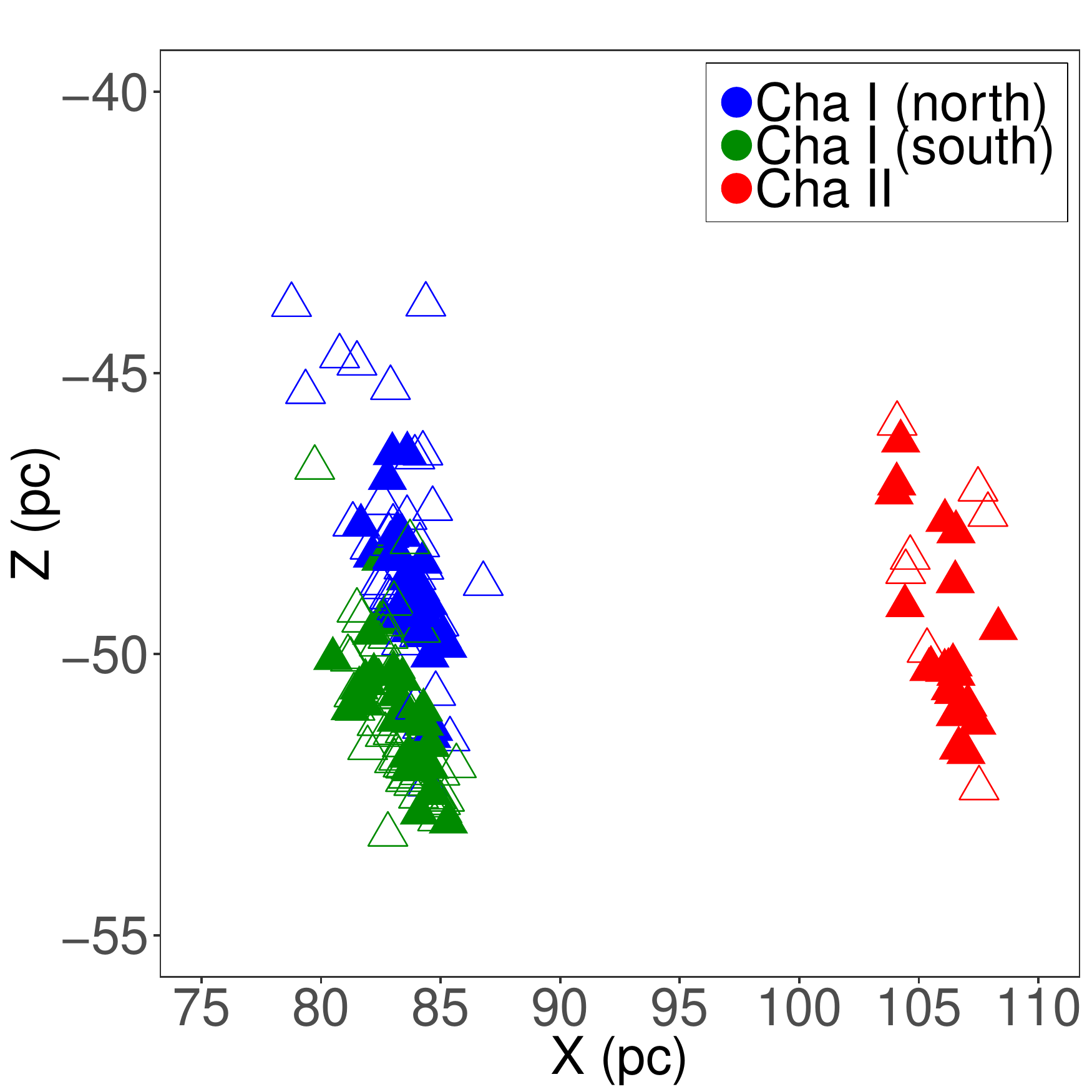}
\caption{Spatial distribution of Chamaeleon stars. The different colours denote the subgroups of stars in our sample. Filled and open symbols indicate Class~II and Class~III stars, respectively.
\label{fig_XYZ} 
}
\end{center}
\end{figure*}

In Table~\ref{tab_YSO} we compare the fractions of SED classes and age estimates in the Chamaeleon subgroups. This comparison confirms that the two subclusters in Cha~I exhibit approximately the same number of Class~II and Class~III stars as shown previously by \citet{Luhman2008}. However, our analysis suggests that the fraction of disc-bearing stars is these subgroups is somewhat lower than the results reported in that study which is also related to the different samples of stars used in each study. On the other hand, when we compare Cha~I and Cha~II we see a different situation. The fraction of disc-bearing stars in Cha~II is almost twice as large as in Cha~I, although the two populations appear to have similar ages as inferred from the HRD. This suggests that the survival time of circumstellar discs in the Cha~I subgroup is somewhat shorter. Similarly, we do not detect significant age differences between the northern and southern subclusters of Cha~I. The two SED classes overlap in the 3D space of stellar positions, but the most dispersed cluster members are mostly Class~III stars as shown in Figure~\ref{fig_XYZ}. We find that the two subgroups of Cha~I define two parallel filamentary structures with lengths of about 10~pc in the $Y$ direction (see middle panel of Figure~\ref{fig_XYZ}). We investigated the 3D position of Chamaeleon stars using the distances derived from the different priors discussed in Sect.~\ref{section3.3} and confirmed that the existence of this structure is not an artifact caused by our choice of prior. The median uncertainties in the XYZ positions of Cha~I stars are 1.1, 2.1 and 0.6~pc. The larger uncertainties in Y can be explained by the higher correlation of the distance with Y ($\rho_{Y}=-0.95$) as compared to the X and Z coordinates ($\rho_{X}=0.73$ and $\rho_{Z}=-0.09$). Thus, the filamentary structure observed for Cha~I is more likely to be explained by the large uncertainties in Y.

We now compare the results obtained in this paper with the ones derived by our team for the Corona-Australis \citep{Galli2020} and Lupus \citep{Galli2020b} star-forming regions using the same methodology applied in this paper. The two age indicators used in this study (isochronal ages and fraction of disc-bearing stars) suggest that the Chamaeleon stars are younger than the stellar population in the Corona-Australis star-forming region. On the other hand, the Chamaeleon subgroups appear to be coeval with the Lupus association and we do not detect important age differences among the various subgroups in these two star-forming regions.

\section{Conclusions}\label{section4}

In this paper we revise the census of stars with available astrometry in the Gaia-DR2 catalogue that are associated to the molecular clouds of the Chamaeleon star-forming region. We applied a probabilistic method to infer membership probabilities and our  analysis allowed us to identify 188~members and 41~members in Cha~I and Cha~II, respectively. We confirm most of the historical members from the literature (with available Gaia-DR2 astrometry) and increase the samples of cluster members by 11\% and 21\% in Cha~I and Cha~II, respectively.

We combined the Gaia-DR2 astrometry with ancillary RV data from previous studies to investigate the 3D spatial distribution and 3D space motion of Chamaeleon stars. We confirm that Cha~I and Cha~II are located at different distances and are separated by about 23~pc in the 3D space of positions. The two subclusters of Cha~I (north and south) are also located at different distances and the observed difference in their proper motion distributions is more likely to be due to projection effects. Our analysis shows that Cha~I (north) and Cha~I (south) have consistent space motions within the reported uncertainties. The HRD analysis reveals that the stars in our sample are mostly younger than 5~Myr and cover the mass range from 0.02 to 3~$M_{\odot}$. The median age of the stars is about 1-2~Myr. We detect no significant age differences between Cha~I and Cha~II, but show that these stellar populations exhibit different fractions of disc-bearing stars. When we compare the results obtained in the present study with those of previous studies conducted by our team in other star-forming regions, we conclude that the Chamaeleon subgroups appear to be coeval with the Lupus association and are younger than the stellar population in the Corona-Australis star-forming region. 

In this study we restricted our analysis of the stellar population of the Chamaeleon clouds to the sources with available data in the Gaia-DR2 catalogue. Our team is measuring precise proper motions of faint sources (beyond the sensitivity limit of the \textit{Gaia} satellite) as part of the DANCe project and we will soon be able to expand the current census of Chamaeleon stars based on ancillary data and spectroscopic observations to accurately derive the initial mass function of this stellar group.  

\begin{acknowledgements}
We thank the referee for constructive criticism that improved the manuscript. This research has received funding from the European Research Council (ERC) under the European Union’s Horizon 2020 research and innovation programme (grant agreement No 682903, P.I. H. Bouy), and from the French State in the framework of the ``Investments for the future” Program, IdEx Bordeaux, reference ANR-10-IDEX-03-02. This research has made use of the SIMBAD database, operated at CDS, Strasbourg, France. This work has made use of data from the European Space Agency (ESA) mission {\it Gaia} (\url{https://www.cosmos.esa.int/gaia}), processed by the {\it Gaia} Data Processing and Analysis Consortium (DPAC, \url{https://www.cosmos.esa.int/web/gaia/dpac/consortium}). Funding for the DPAC has been provided by national institutions, in particular the institutions participating in the {\it Gaia} Multilateral Agreement. This publication makes use of data products from the Wide-field Infrared Survey Explorer, which is a joint project of the University of California, Los Angeles, and the Jet Propulsion Laboratory/California Institute of Technology, funded by the National Aeronautics and Space Administration. 
\end{acknowledgements}

\bibliographystyle{aa} 
\bibliography{references} 

\begin{thebibliography}{105}
\expandafter\ifx\csname natexlab\endcsname\relax\def\natexlab#1{#1}\fi

\bibitem[{{Alcal{\'a}} {et~al.}(2000){Alcal{\'a}}, {Covino}, {Sterzik},
  {Schmitt}, {Krautter}, \& {Neuh{\"a}user}}]{Alcala2000}
{Alcal{\'a}}, J.~M., {Covino}, E., {Sterzik}, M.~F., {et~al.} 2000, \aap, 355,
  629

\bibitem[{{Alcala} {et~al.}(1995){Alcala}, {Krautter}, {Schmitt}, {Covino},
  {Wichmann}, \& {Mundt}}]{Alcala1995}
{Alcala}, J.~M., {Krautter}, J., {Schmitt}, J.~H.~M.~M., {et~al.} 1995, \aaps,
  114, 109

\bibitem[{{Alcal{\'a}} {et~al.}(2008){Alcal{\'a}}, {Spezzi}, {Chapman},
  {Evans}, {Huard}, {J{\o}rgensen}, {Mer{\'\i}n}, {Stapelfeldt}, {Covino},
  {Frasca}, {Gandolfi}, \& {Oliveira}}]{Alcala2008}
{Alcal{\'a}}, J.~M., {Spezzi}, L., {Chapman}, N., {et~al.} 2008, \apj, 676, 427

\bibitem[{{Allers} {et~al.}(2007){Allers}, {Jaffe}, {Luhman}, {Liu}, {Wilson},
  {Skrutskie}, {Nelson}, {Peterson}, {Smith}, \& {Cushing}}]{Allers2007}
{Allers}, K.~N., {Jaffe}, D.~T., {Luhman}, K.~L., {et~al.} 2007, \apj, 657, 511

\bibitem[{{Baraffe} {et~al.}(2015){Baraffe}, {Homeier}, {Allard}, \&
  {Chabrier}}]{BHAC15}
{Baraffe}, I., {Homeier}, D., {Allard}, F., \& {Chabrier}, G. 2015, \aap, 577,
  A42

\bibitem[{{Barrado y Navascu{\'e}s} \& {Jayawardhana}(2004)}]{Barrado2004}
{Barrado y Navascu{\'e}s}, D. \& {Jayawardhana}, R. 2004, \apj, 615, 840

\bibitem[{{Bertout} {et~al.}(1999){Bertout}, {Robichon}, \&
  {Arenou}}]{Bertout1999}
{Bertout}, C., {Robichon}, N., \& {Arenou}, F. 1999, \aap, 352, 574

\bibitem[{{Biazzo} {et~al.}(2012){Biazzo}, {Alcal{\'a}}, {Covino}, {Frasca},
  {Getman}, \& {Spezzi}}]{Biazzo2012}
{Biazzo}, K., {Alcal{\'a}}, J.~M., {Covino}, E., {et~al.} 2012, \aap, 547, A104

\bibitem[{{Bouy} {et~al.}(2013){Bouy}, {Bertin}, {Moraux}, {Cuillandre},
  {Bouvier}, {Barrado}, {Solano}, \& {Bayo}}]{Bouy2013}
{Bouy}, H., {Bertin}, E., {Moraux}, E., {et~al.} 2013, \aap, 554, A101

\bibitem[{{Brown} {et~al.}(1997){Brown}, {Perryman}, {Kovalevsky}, {Robichon},
  {Turon}, \& {Mermilliod}}]{Brown1997}
{Brown}, A.~G.~A., {Perryman}, M.~A.~C., {Kovalevsky}, J., {et~al.} 1997, in
  ESA Special Publication, Vol. 402, Hipparcos - Venice '97, ed. R.~M.
  {Bonnet}, E.~{H{\o}g}, P.~L. {Bernacca}, L.~{Emiliani}, A.~{Blaauw},
  C.~{Turon}, J.~{Kovalevsky}, L.~{Lindegren}, H.~{Hassan}, M.~{Bouffard},
  B.~{Strim}, D.~{Heger}, M.~A.~C. {Perryman}, \& L.~{Woltjer}, 681--686

\bibitem[{{Cambr{\'e}sy}(1999)}]{Cambresy1999}
{Cambr{\'e}sy}, L. 1999, \aap, 345, 965

\bibitem[{{Cambresy} {et~al.}(1998){Cambresy}, {Copet}, {Epchtein}, {de Batz},
  {Borsenberger}, {Fouque}, {Kimeswenger}, \& {Tiphene}}]{Cambresy1998}
{Cambresy}, L., {Copet}, E., {Epchtein}, N., {et~al.} 1998, \aap, 338, 977

\bibitem[{{Cambresy} {et~al.}(1997){Cambresy}, {Epchtein}, {Copet}, {de Batz},
  {Kimeswenger}, {Le Bertre}, {Rouan}, \& {Tiphene}}]{Cambresy1997}
{Cambresy}, L., {Epchtein}, N., {Copet}, E., {et~al.} 1997, \aap, 324, L5

\bibitem[{{Carpenter} {et~al.}(2002){Carpenter}, {Hillenbrand}, {Skrutskie}, \&
  {Meyer}}]{Carpenter2002}
{Carpenter}, J.~M., {Hillenbrand}, L.~A., {Skrutskie}, M.~F., \& {Meyer}, M.~R.
  2002, \aj, 124, 1001

\bibitem[{{Chen} {et~al.}(2016){Chen}, {Shan}, \& {Zhang}}]{Chen2016}
{Chen}, P.~S., {Shan}, H.~G., \& {Zhang}, P. 2016, \na, 44, 1

\bibitem[{{Cieza} {et~al.}(2005){Cieza}, {Kessler-Silacci}, {Jaffe}, {Harvey},
  \& {Evans}}]{Cieza2005}
{Cieza}, L.~A., {Kessler-Silacci}, J.~E., {Jaffe}, D.~T., {Harvey}, P.~M., \&
  {Evans}, Neal~J., I. 2005, \apj, 635, 422

\bibitem[{{Comer{\'o}n} \& {Claes}(2004)}]{Comeron2004}
{Comer{\'o}n}, F. \& {Claes}, P. 2004, \apj, 602, 298

\bibitem[{{Comer{\'o}n} {et~al.}(2000){Comer{\'o}n}, {Neuh{\"a}user}, \&
  {Kaas}}]{Comeron2000}
{Comer{\'o}n}, F., {Neuh{\"a}user}, R., \& {Kaas}, A.~A. 2000, \aap, 359, 269

\bibitem[{{Comer{\'o}n} {et~al.}(1999){Comer{\'o}n}, {Rieke}, \&
  {Neuh{\"a}user}}]{Comeron1999}
{Comer{\'o}n}, F., {Rieke}, G.~H., \& {Neuh{\"a}user}, R. 1999, \aap, 343, 477

\bibitem[{{Covino} {et~al.}(1997){Covino}, {Alcala}, {Allain}, {Bouvier},
  {Terranegra}, \& {Krautter}}]{Covino1997}
{Covino}, E., {Alcala}, J.~M., {Allain}, S., {et~al.} 1997, \aap, 328, 187

\bibitem[{{Cutri} {et~al.}(2003){Cutri}, {Skrutskie}, {van Dyk}, {Beichman},
  {Carpenter}, {Chester}, {Cambresy}, {Evans}, {Fowler}, {Gizis}, {Howard},
  {Huchra}, {Jarrett}, {Kopan}, {Kirkpatrick}, {Light}, {Marsh}, {McCallon},
  {Schneider}, {Stiening}, {Sykes}, {Weinberg}, {Wheaton}, {Wheelock}, \&
  {Zacarias}}]{Cutri2003}
{Cutri}, R.~M., {Skrutskie}, M.~F., {van Dyk}, S., {et~al.} 2003, VizieR Online
  Data Catalog, II/246

\bibitem[{{Dobashi} {et~al.}(2005){Dobashi}, {Uehara}, {Kandori}, {Sakurai},
  {Kaiden}, {Umemoto}, \& {Sato}}]{Dobashi2005}
{Dobashi}, K., {Uehara}, H., {Kandori}, R., {et~al.} 2005, \pasj, 57, S1

\bibitem[{{Dubath} {et~al.}(1996){Dubath}, {Reipurth}, \& {Mayor}}]{Dubath1996}
{Dubath}, P., {Reipurth}, B., \& {Mayor}, M. 1996, \aap, 308, 107

\bibitem[{{Dzib} {et~al.}(2018){Dzib}, {Loinard}, {Ortiz-Le{\'o}n},
  {Rodr{\'\i}guez}, \& {Galli}}]{Dzib2018}
{Dzib}, S.~A., {Loinard}, L., {Ortiz-Le{\'o}n}, G.~N., {Rodr{\'\i}guez}, L.~F.,
  \& {Galli}, P. A.~B. 2018, \apj, 867, 151

\bibitem[{{Elson} {et~al.}(1987){Elson}, {Fall}, \& {Freeman}}]{Elson1987}
{Elson}, R. A.~W., {Fall}, S.~M., \& {Freeman}, K.~C. 1987, \apj, 323, 54

\bibitem[{{ESA}(1997)}]{Hipparcos}
{ESA}, . 1997, VizieR Online Data Catalog, I/239

\bibitem[{{Esplin} {et~al.}(2017){Esplin}, {Luhman}, {Faherty}, {Mamajek}, \&
  {Bochanski}}]{Esplin2017}
{Esplin}, T.~L., {Luhman}, K.~L., {Faherty}, J.~K., {Mamajek}, E.~E., \&
  {Bochanski}, J.~J. 2017, \aj, 154, 46

\bibitem[{{Feigelson} {et~al.}(1993){Feigelson}, {Casanova}, {Montmerle}, \&
  {Guibert}}]{Feigelson1993}
{Feigelson}, E.~D., {Casanova}, S., {Montmerle}, T., \& {Guibert}, J. 1993,
  \apj, 416, 623

\bibitem[{{Feigelson} \& {Lawson}(2004)}]{Feigelson2004}
{Feigelson}, E.~D. \& {Lawson}, W.~A. 2004, \apj, 614, 267

\bibitem[{{Fitzgerald} {et~al.}(1976){Fitzgerald}, {Stephens}, \&
  {Witt}}]{Fitzgerald1976}
{Fitzgerald}, M.~P., {Stephens}, T.~C., \& {Witt}, A.~N. 1976, \apj, 208, 709

\bibitem[{{Franco}(1991)}]{Franco1991}
{Franco}, G.~A.~P. 1991, \aap, 251, 581

\bibitem[{{Gagn{\'e}} \& {Faherty}(2018)}]{Gagne2018c}
{Gagn{\'e}}, J. \& {Faherty}, J.~K. 2018, \apj, 862, 138

\bibitem[{{Gagn{\'e}} {et~al.}(2018{\natexlab{a}}){Gagn{\'e}}, {Mamajek},
  {Malo}, {Riedel}, {Rodriguez}, {Lafreni{\`e}re}, {Faherty}, {Roy-Loubier},
  {Pueyo}, {Robin}, \& {Doyon}}]{Gagne2018a}
{Gagn{\'e}}, J., {Mamajek}, E.~E., {Malo}, L., {et~al.} 2018{\natexlab{a}},
  \apj, 856, 23

\bibitem[{{Gagn{\'e}} {et~al.}(2018{\natexlab{b}}){Gagn{\'e}}, {Roy-Loubier},
  {Faherty}, {Doyon}, \& {Malo}}]{Gagne2018b}
{Gagn{\'e}}, J., {Roy-Loubier}, O., {Faherty}, J.~K., {Doyon}, R., \& {Malo},
  L. 2018{\natexlab{b}}, \apj, 860, 43

\bibitem[{{Gaia Collaboration} {et~al.}(2018){Gaia Collaboration}, {Brown},
  {Vallenari}, {Prusti}, {de Bruijne}, {Babusiaux}, {Bailer-Jones}, {Biermann},
  {Evans}, {Eyer}, {Jansen}, {Jordi}, {Klioner}, {Lammers}, {Lindegren},
  {Luri}, {Mignard}, {Panem}, {Pourbaix}, {Randich}, {Sartoretti}, {Siddiqui},
  {Soubiran}, {van Leeuwen}, {Walton}, {Arenou}, {Bastian}, {Cropper},
  {Drimmel}, {Katz}, {Lattanzi}, {Bakker}, {Cacciari}, {Casta{\~n}eda},
  {Chaoul}, {Cheek}, {De Angeli}, {Fabricius}, {Guerra}, {Holl}, {Masana},
  {Messineo}, {Mowlavi}, {Nienartowicz}, {Panuzzo}, {Portell}, {Riello},
  {Seabroke}, {Tanga}, {Th{\'e}venin}, {Gracia-Abril}, {Comoretto},
  {Garcia-Reinaldos}, {Teyssier}, {Altmann}, {Andrae}, {Audard},
  {Bellas-Velidis}, {Benson}, {Berthier}, {Blomme}, {Burgess}, {Busso},
  {Carry}, {Cellino}, {Clementini}, {Clotet}, {Creevey}, {Davidson}, {De
  Ridder}, {Delchambre}, {Dell'Oro}, {Ducourant},
  {Fern{\'a}ndez-Hern{\'a}ndez}, {Fouesneau}, {Fr{\'e}mat}, {Galluccio},
  {Garc{\'\i}a-Torres}, {Gonz{\'a}lez-N{\'u}{\~n}ez}, {Gonz{\'a}lez-Vidal},
  {Gosset}, {Guy}, {Halbwachs}, {Hambly}, {Harrison}, {Hern{\'a}ndez},
  {Hestroffer}, {Hodgkin}, {Hutton}, {Jasniewicz}, {Jean-Antoine-Piccolo},
  {Jordan}, {Korn}, {Krone-Martins}, {Lanzafame}, {Lebzelter}, {L{\"o}ffler},
  {Manteiga}, {Marrese}, {Mart{\'\i}n-Fleitas}, {Moitinho}, {Mora}, {Muinonen},
  {Osinde}, {Pancino}, {Pauwels}, {Petit}, {Recio-Blanco}, {Richards},
  {Rimoldini}, {Robin}, {Sarro}, {Siopis}, {Smith}, {Sozzetti}, {S{\"u}veges},
  {Torra}, {van Reeven}, {Abbas}, {Abreu Aramburu}, {Accart}, {Aerts},
  {Altavilla}, {{\'A}lvarez}, {Alvarez}, {Alves}, {Anderson}, {Andrei},
  {Anglada Varela}, {Antiche}, {Antoja}, {Arcay}, {Astraatmadja}, {Bach},
  {Baker}, {Balaguer-N{\'u}{\~n}ez}, {Balm}, {Barache}, {Barata}, {Barbato},
  {Barblan}, {Barklem}, {Barrado}, {Barros}, {Barstow}, {Bartholom{\'e}
  Mu{\~n}oz}, {Bassilana}, {Becciani}, {Bellazzini}, {Berihuete}, {Bertone},
  {Bianchi}, {Bienaym{\'e}}, {Blanco-Cuaresma}, {Boch}, {Boeche}, {Bombrun},
  {Borrachero}, {Bossini}, {Bouquillon}, {Bourda}, {Bragaglia}, {Bramante},
  {Breddels}, {Bressan}, {Brouillet}, {Br{\"u}semeister}, {Brugaletta},
  {Bucciarelli}, {Burlacu}, {Busonero}, {Butkevich}, {Buzzi}, {Caffau},
  {Cancelliere}, {Cannizzaro}, {Cantat-Gaudin}, {Carballo}, {Carlucci},
  {Carrasco}, {Casamiquela}, {Castellani}, {Castro-Ginard}, {Charlot},
  {Chemin}, {Chiavassa}, {Cocozza}, {Costigan}, {Cowell}, {Crifo}, {Crosta},
  {Crowley}, {Cuypers}, {Dafonte}, {Damerdji}, {Dapergolas}, {David}, {David},
  {de Laverny}, {De Luise}, {De March}, {de Martino}, {de Souza}, {de Torres},
  {Debosscher}, {del Pozo}, {Delbo}, {Delgado}, {Delgado}, {Di Matteo},
  {Diakite}, {Diener}, {Distefano}, {Dolding}, {Drazinos}, {Dur{\'a}n},
  {Edvardsson}, {Enke}, {Eriksson}, {Esquej}, {Eynard Bontemps}, {Fabre},
  {Fabrizio}, {Faigler}, {Falc{\~a}o}, {Farr{\`a}s Casas}, {Federici},
  {Fedorets}, {Fernique}, {Figueras}, {Filippi}, {Findeisen}, {Fonti},
  {Fraile}, {Fraser}, {Fr{\'e}zouls}, {Gai}, {Galleti}, {Garabato},
  {Garc{\'\i}a-Sedano}, {Garofalo}, {Garralda}, {Gavel}, {Gavras}, {Gerssen},
  {Geyer}, {Giacobbe}, {Gilmore}, {Girona}, {Giuffrida}, {Glass}, {Gomes},
  {Granvik}, {Gueguen}, {Guerrier}, {Guiraud}, {Guti{\'e}rrez-S{\'a}nchez},
  {Haigron}, {Hatzidimitriou}, {Hauser}, {Haywood}, {Heiter}, {Helmi}, {Heu},
  {Hilger}, {Hobbs}, {Hofmann}, {Holland}, {Huckle}, {Hypki}, {Icardi},
  {Jan{\ss}en}, {Jevardat de Fombelle}, {Jonker}, {Juh{\'a}sz}, {Julbe},
  {Karampelas}, {Kewley}, {Klar}, {Kochoska}, {Kohley}, {Kolenberg},
  {Kontizas}, {Kontizas}, {Koposov}, {Kordopatis}, {Kostrzewa-Rutkowska},
  {Koubsky}, {Lambert}, {Lanza}, {Lasne}, {Lavigne}, {Le Fustec}, {Le
  Poncin-Lafitte}, {Lebreton}, {Leccia}, {Leclerc}, {Lecoeur-Taibi},
  {Lenhardt}, {Leroux}, {Liao}, {Licata}, {Lindstr{\o}m}, {Lister}, {Livanou},
  {Lobel}, {L{\'o}pez}, {Managau}, {Mann}, {Mantelet}, {Marchal}, {Marchant},
  {Marconi}, {Marinoni}, {Marschalk{\'o}}, {Marshall}, {Martino}, {Marton},
  {Mary}, {Massari}, {Matijevi{\v{c}}}, {Mazeh}, {McMillan}, {Messina},
  {Michalik}, {Millar}, {Molina}, {Molinaro}, {Moln{\'a}r}, {Montegriffo},
  {Mor}, {Morbidelli}, {Morel}, {Morris}, {Mulone}, {Muraveva}, {Musella},
  {Nelemans}, {Nicastro}, {Noval}, {O'Mullane}, {Ord{\'e}novic},
  {Ord{\'o}{\~n}ez-Blanco}, {Osborne}, {Pagani}, {Pagano}, {Pailler},
  {Palacin}, {Palaversa}, {Panahi}, {Pawlak}, {Piersimoni}, {Pineau}, {Plachy},
  {Plum}, {Poggio}, {Poujoulet}, {Pr{\v{s}}a}, {Pulone}, {Racero}, {Ragaini},
  {Rambaux}, {Ramos-Lerate}, {Regibo}, {Reyl{\'e}}, {Riclet}, {Ripepi}, {Riva},
  {Rivard}, {Rixon}, {Roegiers}, {Roelens}, {Romero-G{\'o}mez}, {Rowell},
  {Royer}, {Ruiz-Dern}, {Sadowski}, {Sagrist{\`a} Sell{\'e}s}, {Sahlmann},
  {Salgado}, {Salguero}, {Sanna}, {Santana-Ros}, {Sarasso}, {Savietto},
  {Schultheis}, {Sciacca}, {Segol}, {Segovia}, {S{\'e}gransan}, {Shih},
  {Siltala}, {Silva}, {Smart}, {Smith}, {Solano}, {Solitro}, {Sordo}, {Soria
  Nieto}, {Souchay}, {Spagna}, {Spoto}, {Stampa}, {Steele},
  {Steidelm{\"u}ller}, {Stephenson}, {Stoev}, {Suess}, {Surdej}, {Szabados},
  {Szegedi-Elek}, {Tapiador}, {Taris}, {Tauran}, {Taylor}, {Teixeira},
  {Terrett}, {Teyssand ier}, {Thuillot}, {Titarenko}, {Torra Clotet}, {Turon},
  {Ulla}, {Utrilla}, {Uzzi}, {Vaillant}, {Valentini}, {Valette}, {van Elteren},
  {Van Hemelryck}, {van Leeuwen}, {Vaschetto}, {Vecchiato}, {Veljanoski},
  {Viala}, {Vicente}, {Vogt}, {von Essen}, {Voss}, {Votruba}, {Voutsinas},
  {Walmsley}, {Weiler}, {Wertz}, {Wevers}, {Wyrzykowski}, {Yoldas},
  {{\v{Z}}erjal}, {Ziaeepour}, {Zorec}, {Zschocke}, {Zucker}, {Zurbach}, \&
  {Zwitter}}]{GaiaDR2}
{Gaia Collaboration}, {Brown}, A.~G.~A., {Vallenari}, A., {et~al.} 2018, \aap,
  616, A1

\bibitem[{{Gaia Collaboration} {et~al.}(2016){Gaia Collaboration}, {Brown},
  {Vallenari}, {Prusti}, {de Bruijne}, {Mignard}, {Drimmel}, {Babusiaux},
  {Bailer-Jones}, {Bastian}, {Biermann}, {Evans}, {Eyer}, {Jansen}, {Jordi},
  {Katz}, {Klioner}, {Lammers}, {Lindegren}, {Luri}, {O'Mullane}, {Panem},
  {Pourbaix}, {Randich}, {Sartoretti}, {Siddiqui}, {Soubiran}, {Valette}, {van
  Leeuwen}, {Walton}, {Aerts}, {Arenou}, {Cropper}, {H{\o}g}, {Lattanzi},
  {Grebel}, {Holland}, {Huc}, {Passot}, {Perryman}, {Bramante}, {Cacciari},
  {Casta{\~n}eda}, {Chaoul}, {Cheek}, {De Angeli}, {Fabricius}, {Guerra},
  {Hern{\'a}ndez}, {Jean-Antoine-Piccolo}, {Masana}, {Messineo}, {Mowlavi},
  {Nienartowicz}, {Ord{\'o}{\~n}ez-Blanco}, {Panuzzo}, {Portell}, {Richards},
  {Riello}, {Seabroke}, {Tanga}, {Th{\'e}venin}, {Torra}, {Els},
  {Gracia-Abril}, {Comoretto}, {Garcia-Reinaldos}, {Lock}, {Mercier},
  {Altmann}, {Andrae}, {Astraatmadja}, {Bellas-Velidis}, {Benson}, {Berthier},
  {Blomme}, {Busso}, {Carry}, {Cellino}, {Clementini}, {Cowell}, {Creevey},
  {Cuypers}, {Davidson}, {De Ridder}, {de Torres}, {Delchambre}, {Dell'Oro},
  {Ducourant}, {Fr{\'e}mat}, {Garc{\'\i}a-Torres}, {Gosset}, {Halbwachs},
  {Hambly}, {Harrison}, {Hauser}, {Hestroffer}, {Hodgkin}, {Huckle}, {Hutton},
  {Jasniewicz}, {Jordan}, {Kontizas}, {Korn}, {Lanzafame}, {Manteiga},
  {Moitinho}, {Muinonen}, {Osinde}, {Pancino}, {Pauwels}, {Petit},
  {Recio-Blanco}, {Robin}, {Sarro}, {Siopis}, {Smith}, {Smith}, {Sozzetti},
  {Thuillot}, {van Reeven}, {Viala}, {Abbas}, {Abreu Aramburu}, {Accart},
  {Aguado}, {Allan}, {Allasia}, {Altavilla}, {{\'A}lvarez}, {Alves},
  {Anderson}, {Andrei}, {Anglada Varela}, {Antiche}, {Antoja}, {Ant{\'o}n},
  {Arcay}, {Bach}, {Baker}, {Balaguer-N{\'u}{\~n}ez}, {Barache}, {Barata},
  {Barbier}, {Barblan}, {Barrado y Navascu{\'e}s}, {Barros}, {Barstow},
  {Becciani}, {Bellazzini}, {Bello Garc{\'\i}a}, {Belokurov}, {Bendjoya},
  {Berihuete}, {Bianchi}, {Bienaym{\'e}}, {Billebaud}, {Blagorodnova},
  {Blanco-Cuaresma}, {Boch}, {Bombrun}, {Borrachero}, {Bouquillon}, {Bourda},
  {Bouy}, {Bragaglia}, {Breddels}, {Brouillet}, {Br{\"u}semeister},
  {Bucciarelli}, {Burgess}, {Burgon}, {Burlacu}, {Busonero}, {Buzzi}, {Caffau},
  {Cambras}, {Campbell}, {Cancelliere}, {Cantat-Gaudin}, {Carlucci},
  {Carrasco}, {Castellani}, {Charlot}, {Charnas}, {Chiavassa}, {Clotet},
  {Cocozza}, {Collins}, {Costigan}, {Crifo}, {Cross}, {Crosta}, {Crowley},
  {Dafonte}, {Damerdji}, {Dapergolas}, {David}, {David}, {De Cat}, {de Felice},
  {de Laverny}, {De Luise}, {De March}, {de Martino}, {de Souza}, {Debosscher},
  {del Pozo}, {Delbo}, {Delgado}, {Delgado}, {Di Matteo}, {Diakite},
  {Distefano}, {Dolding}, {Dos Anjos}, {Drazinos}, {Duran}, {Dzigan},
  {Edvardsson}, {Enke}, {Evans}, {Eynard Bontemps}, {Fabre}, {Fabrizio},
  {Faigler}, {Falc{\~a}o}, {Farr{\`a}s Casas}, {Federici}, {Fedorets},
  {Fern{\'a}ndez-Hern{\'a}ndez}, {Fernique}, {Fienga}, {Figueras}, {Filippi},
  {Findeisen}, {Fonti}, {Fouesneau}, {Fraile}, {Fraser}, {Fuchs}, {Gai},
  {Galleti}, {Galluccio}, {Garabato}, {Garc{\'\i}a-Sedano}, {Garofalo},
  {Garralda}, {Gavras}, {Gerssen}, {Geyer}, {Gilmore}, {Girona}, {Giuffrida},
  {Gomes}, {Gonz{\'a}lez-Marcos}, {Gonz{\'a}lez-N{\'u}{\~n}ez},
  {Gonz{\'a}lez-Vidal}, {Granvik}, {Guerrier}, {Guillout}, {Guiraud},
  {G{\'u}rpide}, {Guti{\'e}rrez-S{\'a}nchez}, {Guy}, {Haigron},
  {Hatzidimitriou}, {Haywood}, {Heiter}, {Helmi}, {Hobbs}, {Hofmann}, {Holl},
  {Holland }, {Hunt}, {Hypki}, {Icardi}, {Irwin}, {Jevardat de Fombelle},
  {Jofr{\'e}}, {Jonker}, {Jorissen}, {Julbe}, {Karampelas}, {Kochoska},
  {Kohley}, {Kolenberg}, {Kontizas}, {Koposov}, {Kordopatis}, {Koubsky},
  {Krone-Martins}, {Kudryashova}, {Kull}, {Bachchan}, {Lacoste-Seris}, {Lanza},
  {Lavigne}, {Le Poncin-Lafitte}, {Lebreton}, {Lebzelter}, {Leccia}, {Leclerc},
  {Lecoeur-Taibi}, {Lemaitre}, {Lenhardt}, {Leroux}, {Liao}, {Licata},
  {Lindstr{\o}m}, {Lister}, {Livanou}, {Lobel}, {L{\"o}ffler}, {L{\'o}pez},
  {Lorenz}, {MacDonald}, {Magalh{\~a}es Fernandes}, {Managau}, {Mann},
  {Mantelet}, {Marchal}, {Marchant}, {Marconi}, {Marinoni}, {Marrese},
  {Marschalk{\'o}}, {Marshall}, {Mart{\'\i}n-Fleitas}, {Martino}, {Mary},
  {Matijevi{\v{c}}}, {Mazeh}, {McMillan}, {Messina}, {Michalik}, {Millar},
  {Mirand a}, {Molina}, {Molinaro}, {Molinaro}, {Moln{\'a}r}, {Moniez},
  {Montegriffo}, {Mor}, {Mora}, {Morbidelli}, {Morel}, {Morgenthaler},
  {Morris}, {Mulone}, {Muraveva}, {Musella}, {Narbonne}, {Nelemans},
  {Nicastro}, {Noval}, {Ord{\'e}novic}, {Ordieres-Mer{\'e}}, {Osborne},
  {Pagani}, {Pagano}, {Pailler}, {Palacin}, {Palaversa}, {Parsons}, {Pecoraro},
  {Pedrosa}, {Pentik{\"a}inen}, {Pichon}, {Piersimoni}, {Pineau}, {Plachy},
  {Plum}, {Poujoulet}, {Pr{\v{s}}a}, {Pulone}, {Ragaini}, {Rago}, {Rambaux},
  {Ramos-Lerate}, {Ranalli}, {Rauw}, {Read}, {Regibo}, {Reyl{\'e}}, {Ribeiro},
  {Rimoldini}, {Ripepi}, {Riva}, {Rixon}, {Roelens}, {Romero-G{\'o}mez},
  {Rowell}, {Royer}, {Ruiz-Dern}, {Sadowski}, {Sagrist{\`a} Sell{\'e}s},
  {Sahlmann}, {Salgado}, {Salguero}, {Sarasso}, {Savietto}, {Schultheis},
  {Sciacca}, {Segol}, {Segovia}, {Segransan}, {Shih}, {Smareglia}, {Smart},
  {Solano}, {Solitro}, {Sordo}, {Soria Nieto}, {Souchay}, {Spagna}, {Spoto},
  {Stampa}, {Steele}, {Steidelm{\"u}ller}, {Stephenson}, {Stoev}, {Suess},
  {S{\"u}veges}, {Surdej}, {Szabados}, {Szegedi-Elek}, {Tapiador}, {Taris},
  {Tauran}, {Taylor}, {Teixeira}, {Terrett}, {Tingley}, {Trager}, {Turon},
  {Ulla}, {Utrilla}, {Valentini}, {van Elteren}, {Van Hemelryck}, {van
  Leeuwen}, {Varadi}, {Vecchiato}, {Veljanoski}, {Via}, {Vicente}, {Vogt},
  {Voss}, {Votruba}, {Voutsinas}, {Walmsley}, {Weiler}, {Weingrill}, {Wevers},
  {Wyrzykowski}, {Yoldas}, {{\v{Z}}erjal}, {Zucker}, {Zurbach}, {Zwitter},
  {Alecu}, {Allen}, {Allende Prieto}, {Amorim}, {Anglada-Escud{\'e}},
  {Arsenijevic}, {Azaz}, {Balm}, {Beck}, {Bernstein}, {Bigot}, {Bijaoui},
  {Blasco}, {Bonfigli}, {Bono}, {Boudreault}, {Bressan}, {Brown}, {Brunet},
  {Bunclark}, {Buonanno}, {Butkevich}, {Carret}, {Carrion}, {Chemin},
  {Ch{\'e}reau}, {Corcione}, {Darmigny}, {de Boer}, {de Teodoro}, {de Zeeuw},
  {Delle Luche}, {Domingues}, {Dubath}, {Fodor}, {Fr{\'e}zouls}, {Fries},
  {Fustes}, {Fyfe}, {Gallardo}, {Gallegos}, {Gardiol}, {Gebran}, {Gomboc},
  {G{\'o}mez}, {Grux}, {Gueguen}, {Heyrovsky}, {Hoar}, {Iannicola}, {Isasi
  Parache}, {Janotto}, {Joliet}, {Jonckheere}, {Keil}, {Kim}, {Klagyivik},
  {Klar}, {Knude}, {Kochukhov}, {Kolka}, {Kos}, {Kutka}, {Lainey}, {LeBouquin},
  {Liu}, {Loreggia}, {Makarov}, {Marseille}, {Martayan}, {Martinez-Rubi},
  {Massart}, {Meynadier}, {Mignot}, {Munari}, {Nguyen}, {Nordlander}, {Ocvirk},
  {O'Flaherty}, {Olias Sanz}, {Ortiz}, {Osorio}, {Oszkiewicz}, {Ouzounis},
  {Palmer}, {Park}, {Pasquato}, {Peltzer}, {Peralta}, {P{\'e}turaud},
  {Pieniluoma}, {Pigozzi}, {Poels}, {Prat}, {Prod'homme}, {Raison}, {Rebordao},
  {Risquez}, {Rocca-Volmerange}, {Rosen}, {Ruiz-Fuertes}, {Russo}, {Sembay},
  {Serraller Vizcaino}, {Short}, {Siebert}, {Silva}, {Sinachopoulos}, {Slezak},
  {Soffel}, {Sosnowska}, {Strai{\v{z}}ys}, {ter Linden}, {Terrell}, {Theil},
  {Tiede}, {Troisi}, {Tsalmantza}, {Tur}, {Vaccari}, {Vachier}, {Valles}, {Van
  Hamme}, {Veltz}, {Virtanen}, {Wallut}, {Wichmann}, {Wilkinson}, {Ziaeepour},
  \& {Zschocke}}]{GaiaDR1}
{Gaia Collaboration}, {Brown}, A.~G.~A., {Vallenari}, A., {et~al.} 2016, \aap,
  595, A2

\bibitem[{{Galli} {et~al.}(2020{\natexlab{a}}){Galli}, {Bouy}, {Olivares},
  {Miret-Roig}, {Sarro}, {Barrado}, {Berihuete}, \& {Brandner}}]{Galli2020}
{Galli}, P.~A.~B., {Bouy}, H., {Olivares}, J., {et~al.} 2020{\natexlab{a}},
  \aap, 634, A98

\bibitem[{{Galli} {et~al.}(2020{\natexlab{b}}){Galli}, {Bouy}, {Olivares},
  {Miret-Roig}, {Vieira}, {Sarro}, {Barrado}, {Berihuete}, {Bertout}, {Bertin},
  \& {Cuillandre}}]{Galli2020b}
{Galli}, P.~A.~B., {Bouy}, H., {Olivares}, J., {et~al.} 2020{\natexlab{b}},
  arXiv e-prints, arXiv:2010.00233

\bibitem[{{Gilmore} {et~al.}(2012){Gilmore}, {Randich}, {Asplund}, {Binney},
  {Bonifacio}, {Drew}, {Feltzing}, {Ferguson}, {Jeffries}, {Micela},
  {Negueruela}, {Prusti}, {Rix}, {Vallenari}, {Alfaro}, {Allende-Prieto},
  {Babusiaux}, {Bensby}, {Blomme}, {Bragaglia}, {Flaccomio}, {Fran{\c{c}}ois},
  {Irwin}, {Koposov}, {Korn}, {Lanzafame}, {Pancino}, {Paunzen},
  {Recio-Blanco}, {Sacco}, {Smiljanic}, {Van Eck}, {Walton}, {Aden}, {Aerts},
  {Affer}, {Alcala}, {Altavilla}, {Alves}, {Antoja}, {Arenou}, {Argiroffi},
  {Asensio Ramos}, {Bailer-Jones}, {Balaguer-Nunez}, {Bayo}, {Barbuy},
  {Barisevicius}, {Barrado y Navascues}, {Battistini}, {Bellas Velidis},
  {Bellazzini}, {Belokurov}, {Bergemann}, {Bertelli}, {Biazzo}, {Bienayme},
  {Bland-Hawthorn}, {Boeche}, {Bonito}, {Boudreault}, {Bouvier}, {Brandao},
  {Brown}, {de Bruijne}, {Burleigh}, {Caballero}, {Caffau}, {Calura},
  {Capuzzo-Dolcetta}, {Caramazza}, {Carraro}, {Casagrande}, {Casewell},
  {Chapman}, {Chiappini}, {Chorniy}, {Christlieb}, {Cignoni}, {Cocozza},
  {Colless}, {Collet}, {Collins}, {Correnti}, {Covino}, {Crnojevic}, {Cropper},
  {Cunha}, {Damiani}, {David}, {Delgado}, {Duffau}, {Edvardsson}, {Eldridge},
  {Enke}, {Eriksson}, {Evans}, {Eyer}, {Famaey}, {Fellhauer}, {Ferreras},
  {Figueras}, {Fiorentino}, {Flynn}, {Folha}, {Franciosini}, {Frasca},
  {Freeman}, {Fremat}, {Friel}, {Gaensicke}, {Gameiro}, {Garzon}, {Geier},
  {Geisler}, {Gerhard}, {Gibson}, {Gomboc}, {Gomez}, {Gonzalez-Fernandez},
  {Gonzalez Hernandez}, {Gosset}, {Grebel}, {Greimel}, {Groenewegen},
  {Grundahl}, {Guarcello}, {Gustafsson}, {Hadrava}, {Hatzidimitriou}, {Hambly},
  {Hammersley}, {Hansen}, {Haywood}, {Heber}, {Heiter}, {Held}, {Helmi},
  {Hensler}, {Herrero}, {Hill}, {Hodgkin}, {Huelamo}, {Huxor}, {Ibata},
  {Jackson}, {de Jong}, {Jonker}, {Jordan}, {Jordi}, {Jorissen}, {Katz},
  {Kawata}, {Keller}, {Kharchenko}, {Klement}, {Klutsch}, {Knude}, {Koch},
  {Kochukhov}, {Kontizas}, {Koubsky}, {Lallement}, {de Laverny}, {van Leeuwen},
  {Lemasle}, {Lewis}, {Lind}, {Lindstrom}, {Lobel}, {Lopez Santiago}, {Lucas},
  {Ludwig}, {Lueftinger}, {Magrini}, {Maiz Apellaniz}, {Maldonado}, {Marconi},
  {Marino}, {Martayan}, {Martinez-Valpuesta}, {Matijevic}, {McMahon},
  {Messina}, {Meyer}, {Miglio}, {Mikolaitis}, {Minchev}, {Minniti}, {Moitinho},
  {Momany}, {Monaco}, {Montalto}, {Monteiro}, {Monier}, {Montes}, {Mora},
  {Moraux}, {Morel}, {Mowlavi}, {Mucciarelli}, {Munari}, {Napiwotzki},
  {Nardetto}, {Naylor}, {Naze}, {Nelemans}, {Okamoto}, {Ortolani}, {Pace},
  {Palla}, {Palous}, {Parker}, {Penarrubia}, {Pillitteri}, {Piotto}, {Posbic},
  {Prisinzano}, {Puzeras}, {Quirrenbach}, {Ragaini}, {Read}, {Read}, {Reyle},
  {De Ridder}, {Robichon}, {Robin}, {Roeser}, {Romano}, {Royer}, {Ruchti},
  {Ruzicka}, {Ryan}, {Ryde}, {Santos}, {Sanz Forcada}, {Sarro Baro},
  {Sbordone}, {Schilbach}, {Schmeja}, {Schnurr}, {Schoenrich}, {Scholz},
  {Seabroke}, {Sharma}, {De Silva}, {Smith}, {Solano}, {Sordo}, {Soubiran},
  {Sousa}, {Spagna}, {Steffen}, {Steinmetz}, {Stelzer}, {Stempels},
  {Tabernero}, {Tautvaisiene}, {Thevenin}, {Torra}, {Tosi}, {Tolstoy}, {Turon},
  {Walker}, {Wambsganss}, {Worley}, {Venn}, {Vink}, {Wyse}, {Zaggia},
  {Zeilinger}, {Zoccali}, {Zorec}, {Zucker}, {Zwitter}, \& {Gaia-ESO Survey
  Team}}]{Gilmore2012}
{Gilmore}, G., {Randich}, S., {Asplund}, M., {et~al.} 2012, The Messenger, 147,
  25

\bibitem[{{Gontcharov}(2006)}]{Gontcharov2006}
{Gontcharov}, G.~A. 2006, Astronomy Letters, 32, 759

\bibitem[{{Graczyk} {et~al.}(2019){Graczyk}, {Pietrzy{\'n}ski}, {Gieren},
  {Storm}, {Nardetto}, {Gallenne}, {Maxted}, {Kervella}, {Ko{\l}aczkowski},
  {Konorski}, {Pilecki}, {Zgirski}, {G{\'o}rski}, {Suchomska}, {Karczmarek},
  {Taormina}, {Wielg{\'o}rski}, {Narloch}, {Smolec}, {Chini}, \&
  {Breuval}}]{Graczyk2019}
{Graczyk}, D., {Pietrzy{\'n}ski}, G., {Gieren}, W., {et~al.} 2019, \apj, 872,
  85

\bibitem[{{Graham} \& {Hartigan}(1988)}]{Graham1988}
{Graham}, J.~A. \& {Hartigan}, P. 1988, \aj, 95, 1197

\bibitem[{{Grasdalen} {et~al.}(1975){Grasdalen}, {Joyce}, {Knacke}, {Strom}, \&
  {Strom}}]{Grasdalen1975}
{Grasdalen}, G., {Joyce}, R., {Knacke}, R.~F., {Strom}, S.~E., \& {Strom},
  K.~M. 1975, \aj, 80, 117

\bibitem[{{Gregorio Hetem} {et~al.}(1988){Gregorio Hetem}, {Sanzovo}, \&
  {Lepine}}]{Gregorio-Hetem1988}
{Gregorio Hetem}, J.~C., {Sanzovo}, G.~C., \& {Lepine}, J.~R.~D. 1988, \aaps,
  76, 347

\bibitem[{{Guenther} {et~al.}(2007){Guenther}, {Esposito}, {Mundt}, {Covino},
  {Alcal{\'a}}, {Cusano}, \& {Stecklum}}]{Guenther2007}
{Guenther}, E.~W., {Esposito}, M., {Mundt}, R., {et~al.} 2007, \aap, 467, 1147

\bibitem[{{Hartigan}(1993)}]{Hartigan1993}
{Hartigan}, P. 1993, \aj, 105, 1511

\bibitem[{{Henize} \& {Mendoza}(1973)}]{Henize1973}
{Henize}, K.~G. \& {Mendoza}, E.~E. 1973, \apj, 180, 115

\bibitem[{{Hughes} \& {Hartigan}(1992)}]{Hughes1992}
{Hughes}, J. \& {Hartigan}, P. 1992, \aj, 104, 680

\bibitem[{{Hyland} {et~al.}(1982){Hyland}, {Jones}, \& {Mitchell}}]{Hyland1982}
{Hyland}, A.~R., {Jones}, T.~J., \& {Mitchell}, R.~M. 1982, \mnras, 201, 1095

\bibitem[{{James} {et~al.}(2006){James}, {Melo}, {Santos}, \&
  {Bouvier}}]{James2006}
{James}, D.~J., {Melo}, C., {Santos}, N.~C., \& {Bouvier}, J. 2006, \aap, 446,
  971

\bibitem[{{Joergens} \& {Guenther}(2001)}]{Joergens2001}
{Joergens}, V. \& {Guenther}, E. 2001, \aap, 379, L9

\bibitem[{{Johnson} \& {Soderblom}(1987)}]{Johnson1987}
{Johnson}, D. R.~H. \& {Soderblom}, D.~R. 1987, \aj, 93, 864

\bibitem[{Kaufmann \& Rousseeuw(1987)}]{PAM}
Kaufmann, L. \& Rousseeuw, P. 1987, Data Analysis based on the L1-Norm and
  Related Methods, 405

\bibitem[{{King}(1962)}]{King1962}
{King}, I. 1962, \aj, 67, 471

\bibitem[{{Koenig} \& {Leisawitz}(2014)}]{Koenig2014}
{Koenig}, X.~P. \& {Leisawitz}, D.~T. 2014, \apj, 791, 131

\bibitem[{{Lada}(1987)}]{Lada1987}
{Lada}, C.~J. 1987, in IAU Symposium, Vol. 115, Star Forming Regions, ed.
  M.~{Peimbert} \& J.~{Jugaku}, 1

\bibitem[{{Lindegren} {et~al.}(2018){Lindegren}, {Hern{\'a}ndez}, {Bombrun},
  {Klioner}, {Bastian}, {Ramos-Lerate}, {de Torres}, {Steidelm{\"u}ller},
  {Stephenson}, {Hobbs}, {Lammers}, {Biermann}, {Geyer}, {Hilger}, {Michalik},
  {Stampa}, {McMillan}, {Casta{\~n}eda}, {Clotet}, {Comoretto}, {Davidson},
  {Fabricius}, {Gracia}, {Hambly}, {Hutton}, {Mora}, {Portell}, {van Leeuwen},
  {Abbas}, {Abreu}, {Altmann}, {Andrei}, {Anglada}, {Balaguer-N{\'u}{\~n}ez},
  {Barache}, {Becciani}, {Bertone}, {Bianchi}, {Bouquillon}, {Bourda},
  {Br{\"u}semeister}, {Bucciarelli}, {Busonero}, {Buzzi}, {Cancelliere},
  {Carlucci}, {Charlot}, {Cheek}, {Crosta}, {Crowley}, {de Bruijne}, {de
  Felice}, {Drimmel}, {Esquej}, {Fienga}, {Fraile}, {Gai}, {Garralda},
  {Gonz{\'a}lez-Vidal}, {Guerra}, {Hauser}, {Hofmann}, {Holl}, {Jordan},
  {Lattanzi}, {Lenhardt}, {Liao}, {Licata}, {Lister}, {L{\"o}ffler},
  {Marchant}, {Martin-Fleitas}, {Messineo}, {Mignard}, {Morbidelli}, {Poggio},
  {Riva}, {Rowell}, {Salguero}, {Sarasso}, {Sciacca}, {Siddiqui}, {Smart},
  {Spagna}, {Steele}, {Taris}, {Torra}, {van Elteren}, {van Reeven}, \&
  {Vecchiato}}]{Lindegren2018}
{Lindegren}, L., {Hern{\'a}ndez}, J., {Bombrun}, A., {et~al.} 2018, \aap, 616,
  A2

\bibitem[{{Lindegren} {et~al.}(2016){Lindegren}, {Lammers}, {Bastian},
  {Hern{\'a}ndez}, {Klioner}, {Hobbs}, {Bombrun}, {Michalik}, {Ramos-Lerate},
  {Butkevich}, {Comoretto}, {Joliet}, {Holl}, {Hutton}, {Parsons},
  {Steidelm{\"u}ller}, {Abbas}, {Altmann}, {Andrei}, {Anton}, {Bach},
  {Barache}, {Becciani}, {Berthier}, {Bianchi}, {Biermann}, {Bouquillon},
  {Bourda}, {Br{\"u}semeister}, {Bucciarelli}, {Busonero}, {Carlucci},
  {Casta{\~n}eda}, {Charlot}, {Clotet}, {Crosta}, {Davidson}, {de Felice},
  {Drimmel}, {Fabricius}, {Fienga}, {Figueras}, {Fraile}, {Gai}, {Garralda},
  {Geyer}, {Gonz{\'a}lez-Vidal}, {Guerra}, {Hambly}, {Hauser}, {Jordan},
  {Lattanzi}, {Lenhardt}, {Liao}, {L{\"o}ffler}, {McMillan}, {Mignard}, {Mora},
  {Morbidelli}, {Portell}, {Riva}, {Sarasso}, {Serraller}, {Siddiqui}, {Smart},
  {Spagna}, {Stampa}, {Steele}, {Taris}, {Torra}, {van Reeven}, {Vecchiato},
  {Zschocke}, {de Bruijne}, {Gracia}, {Raison}, {Lister}, {Marchant},
  {Messineo}, {Soffel}, {Osorio}, {de Torres}, \& {O'Mullane}}]{Lindegren2016}
{Lindegren}, L., {Lammers}, U., {Bastian}, U., {et~al.} 2016, \aap, 595, A4

\bibitem[{{L{\'o}pez Mart{\'\i}} {et~al.}(2005){L{\'o}pez Mart{\'\i}},
  {Eisl{\"o}ffel}, \& {Mundt}}]{LopezMarti2005}
{L{\'o}pez Mart{\'\i}}, B., {Eisl{\"o}ffel}, J., \& {Mundt}, R. 2005, \aap,
  444, 175

\bibitem[{{L{\'o}pez Mart{\'\i}} {et~al.}(2004){L{\'o}pez Mart{\'\i}},
  {Eisl{\"o}ffel}, {Scholz}, \& {Mundt}}]{LopezMarti2004}
{L{\'o}pez Mart{\'\i}}, B., {Eisl{\"o}ffel}, J., {Scholz}, A., \& {Mundt}, R.
  2004, \aap, 416, 555

\bibitem[{{L{\'o}pez Mart{\'\i}} {et~al.}(2013){L{\'o}pez Mart{\'\i}},
  {Jim{\'e}nez-Esteban}, {Bayo}, {Barrado}, {Solano}, {Bouy}, \&
  {Rodrigo}}]{LopezMarti2013b}
{L{\'o}pez Mart{\'\i}}, B., {Jim{\'e}nez-Esteban}, F., {Bayo}, A., {et~al.}
  2013, \aap, 556, A144

\bibitem[{{Lopez Mart{\'\i}} {et~al.}(2013){Lopez Mart{\'\i}}, {Jimenez
  Esteban}, {Bayo}, {Barrado}, {Solano}, \& {Rodrigo}}]{LopezMarti2013a}
{Lopez Mart{\'\i}}, B., {Jimenez Esteban}, F., {Bayo}, A., {et~al.} 2013, \aap,
  551, A46

\bibitem[{{Luhman}(2004)}]{Luhman2004a}
{Luhman}, K.~L. 2004, \apj, 602, 816

\bibitem[{{Luhman}(2007)}]{Luhman2007}
{Luhman}, K.~L. 2007, \apjs, 173, 104

\bibitem[{{Luhman}(2008)}]{Luhman2008_review}
{Luhman}, K.~L. 2008, {Chamaeleon}, ed. B.~{Reipurth}, Vol.~5, 169

\bibitem[{{Luhman} {et~al.}(2008){Luhman}, {Allen}, {Allen}, {Gutermuth},
  {Hartmann}, {Mamajek}, {Megeath}, {Myers}, \& {Fazio}}]{Luhman2008}
{Luhman}, K.~L., {Allen}, L.~E., {Allen}, P.~R., {et~al.} 2008, \apj, 675, 1375

\bibitem[{{Luhman} \& {Esplin}(2020)}]{Luhman2020}
{Luhman}, K.~L. \& {Esplin}, T.~L. 2020, \aj, 160, 44

\bibitem[{{Luhman} {et~al.}(2004){Luhman}, {Peterson}, \&
  {Megeath}}]{Luhman2004b}
{Luhman}, K.~L., {Peterson}, D.~E., \& {Megeath}, S.~T. 2004, \apj, 617, 565

\bibitem[{{Luri} {et~al.}(2018){Luri}, {Brown}, {Sarro}, {Arenou},
  {Bailer-Jones}, {Castro-Ginard}, {de Bruijne}, {Prusti}, {Babusiaux}, \&
  {Delgado}}]{Luri2018}
{Luri}, X., {Brown}, A.~G.~A., {Sarro}, L.~M., {et~al.} 2018, \aap, 616, A9

\bibitem[{MacQueen(1967)}]{kMeans}
MacQueen, J. 1967, in Proceedings of the Fifth Berkeley Symposium on
  Mathematical Statistics and Probability, Volume 1: Statistics (Berkeley,
  Calif.: University of California Press), 281--297

\bibitem[{{Ma{\'\i}z Apell{\'a}niz} \& {Weiler}(2018)}]{Apellaniz2018}
{Ma{\'\i}z Apell{\'a}niz}, J. \& {Weiler}, M. 2018, \aap, 619, A180

\bibitem[{{Mizuno} {et~al.}(2001){Mizuno}, {Yamaguchi}, {Tachihara}, {Toyoda},
  {Aoyama}, {Yamamoto}, {Onishi}, \& {Fukui}}]{Mizuno2001}
{Mizuno}, A., {Yamaguchi}, R., {Tachihara}, K., {et~al.} 2001, \pasj, 53, 1071

\bibitem[{{Murphy} {et~al.}(2013){Murphy}, {Lawson}, \& {Bessell}}]{Murphy2013}
{Murphy}, S.~J., {Lawson}, W.~A., \& {Bessell}, M.~S. 2013, \mnras, 435, 1325

\bibitem[{{Neuhauser} \& {Comeron}(1998)}]{Neuhauser1998}
{Neuhauser}, R. \& {Comeron}, F. 1998, Science, 282, 83

\bibitem[{{Neuh{\"a}user} \& {Comer{\'o}n}(1999)}]{Neuhauser1999}
{Neuh{\"a}user}, R. \& {Comer{\'o}n}, F. 1999, \aap, 350, 612

\bibitem[{{Nguyen} {et~al.}(2012){Nguyen}, {Brandeker}, {van Kerkwijk}, \&
  {Jayawardhana}}]{Nguyen2012}
{Nguyen}, D.~C., {Brandeker}, A., {van Kerkwijk}, M.~H., \& {Jayawardhana}, R.
  2012, \apj, 745, 119

\bibitem[{Olivares(2019)}]{kalkayotl}
Olivares, J. 2019, olivares-j/Kalkayotl: Basic functionality

\bibitem[{{Olivares} {et~al.}(2019){Olivares}, {Bouy}, {Sarro}, {Miret-Roig},
  {Berihuete}, {Bertin}, {Barrado}, {Hu{\'e}lamo}, {Tamura}, {Allen},
  {Beletsky}, {Serre}, \& {Cuillandre}}]{Olivares2019}
{Olivares}, J., {Bouy}, H., {Sarro}, L.~M., {et~al.} 2019, \aap, 625, A115

\bibitem[{{Olivares} {et~al.}(2020){Olivares}, {Sarro}, {Bouy}, {Miret-Roig},
  {Casamiquela}, {Galli}, {Berihuete}, \& {Tarricq}}]{Olivares2020}
{Olivares}, J., {Sarro}, L.~M., {Bouy}, H., {et~al.} 2020, arXiv e-prints,
  arXiv:2010.00272

\bibitem[{{Pecaut} \& {Mamajek}(2013)}]{Pecaut2013}
{Pecaut}, M.~J. \& {Mamajek}, E.~E. 2013, \apjs, 208, 9

\bibitem[{{Perryman} {et~al.}(1998){Perryman}, {Brown}, {Lebreton}, {Gomez},
  {Turon}, {Cayrel de Strobel}, {Mermilliod}, {Robichon}, {Kovalevsky}, \&
  {Crifo}}]{Perryman1998}
{Perryman}, M.~A.~C., {Brown}, A.~G.~A., {Lebreton}, Y., {et~al.} 1998, \aap,
  331, 81

\bibitem[{{Persi} {et~al.}(2000){Persi}, {Marenzi}, {Olofsson}, {Kaas},
  {Nordh}, {Huldtgren}, {Abergel}, {Andr{\'e}}, {Bontemps}, {Boulanger},
  {Burggdorf}, {Casali}, {Cesarsky}, {Copet}, {Davies}, {Falgarone},
  {Montmerle}, {Perault}, {Prusti}, {Puget}, \& {Sibille}}]{Persi2000}
{Persi}, P., {Marenzi}, A.~R., {Olofsson}, G., {et~al.} 2000, \aap, 357, 219

\bibitem[{{Prusti} {et~al.}(1992){Prusti}, {Whittet}, {Assendorp}, \&
  {Wesselius}}]{Prusti1992}
{Prusti}, T., {Whittet}, D.~C.~B., {Assendorp}, R., \& {Wesselius}, P.~R. 1992,
  \aap, 260, 151

\bibitem[{{Robrade} \& {Schmitt}(2007)}]{Robrade2007}
{Robrade}, J. \& {Schmitt}, J.~H.~M.~M. 2007, \aap, 461, 669

\bibitem[{{Roccatagliata} {et~al.}(2018){Roccatagliata}, {Sacco},
  {Franciosini}, \& {Rand ich}}]{Roccatagliata2018}
{Roccatagliata}, V., {Sacco}, G.~G., {Franciosini}, E., \& {Rand ich}, S. 2018,
  \aap, 617, L4

\bibitem[{Rousseeuw(1987)}]{ROUSSEEUW198753}
Rousseeuw, P.~J. 1987, Journal of Computational and Applied Mathematics, 20, 53

\bibitem[{Rousseeuw \& Driessen(1999)}]{Rousseeuw1999}
Rousseeuw, P.~J. \& Driessen, K.~V. 1999, Technometrics, 41, 212

\bibitem[{{Rydgren}(1980)}]{Rydgren1980}
{Rydgren}, A.~E. 1980, \aj, 85, 444

\bibitem[{{Sacco} {et~al.}(2017){Sacco}, {Spina}, {Randich}, {Palla}, {Parker},
  {Jeffries}, {Jackson}, {Meyer}, {Mapelli}, {Lanzafame}, {Bonito}, {Damiani},
  {Franciosini}, {Frasca}, {Klutsch}, {Prisinzano}, {Tognelli},
  {Degl'Innocenti}, {Prada Moroni}, {Alfaro}, {Micela}, {Prusti}, {Barrado},
  {Biazzo}, {Bouy}, {Bravi}, {Lopez-Santiago}, {Wright}, {Bayo}, {Gilmore},
  {Bragaglia}, {Flaccomio}, {Koposov}, {Pancino}, {Casey}, {Costado}, {Donati},
  {Hourihane}, {Jofr{\'e}}, {Lardo}, {Lewis}, {Magrini}, {Monaco},
  {Morbidelli}, {Sousa}, {Worley}, \& {Zaggia}}]{Sacco2017}
{Sacco}, G.~G., {Spina}, L., {Randich}, S., {et~al.} 2017, \aap, 601, A97

\bibitem[{{Sarro} {et~al.}(2014){Sarro}, {Bouy}, {Berihuete}, {Bertin},
  {Moraux}, {Bouvier}, {Cuillandre}, {Barrado}, \& {Solano}}]{Sarro2014}
{Sarro}, L.~M., {Bouy}, H., {Berihuete}, A., {et~al.} 2014, \aap, 563, A45

\bibitem[{{Schwartz}(1977)}]{Schwartz1977}
{Schwartz}, R.~D. 1977, \apjs, 35, 161

\bibitem[{{Schwartz}(1992)}]{Schwartz1992}
{Schwartz}, R.~D. 1992, {The Chamaeleon Dark Clouds and T-Associations}, ed.
  B.~{Reipurth}, 93

\bibitem[{{Siess} {et~al.}(2000){Siess}, {Dufour}, \& {Forestini}}]{Siess2000}
{Siess}, L., {Dufour}, E., \& {Forestini}, M. 2000, \aap, 358, 593

\bibitem[{{Spezzi} {et~al.}(2008){Spezzi}, {Alcal{\'a}}, {Covino}, {Frasca},
  {Gandolfi}, {Oliveira}, {Chapman}, {Evans}, {Huard}, {J{\o}rgensen},
  {Mer{\'\i}n}, \& {Stapelfeldt}}]{Spezzi2008}
{Spezzi}, L., {Alcal{\'a}}, J.~M., {Covino}, E., {et~al.} 2008, \apj, 680, 1295

\bibitem[{{Stassun} \& {Torres}(2018)}]{Stassun2018}
{Stassun}, K.~G. \& {Torres}, G. 2018, \apj, 862, 61

\bibitem[{{Stelzer} {et~al.}(2004){Stelzer}, {Micela}, \&
  {Neuh{\"a}user}}]{Stelzer2004}
{Stelzer}, B., {Micela}, G., \& {Neuh{\"a}user}, R. 2004, \aap, 423, 1029

\bibitem[{Tibshirani {et~al.}(2001)Tibshirani, Walther, \&
  Hastie}]{GapStatistic}
Tibshirani, R., Walther, G., \& Hastie, T. 2001, Journal of the Royal
  Statistical Society: Series B (Statistical Methodology), 63, 411

\bibitem[{{Torres} {et~al.}(2006){Torres}, {Quast}, {da Silva}, {de La Reza},
  {Melo}, \& {Sterzik}}]{Torres2006}
{Torres}, C.~A.~O., {Quast}, G.~R., {da Silva}, L., {et~al.} 2006, \aap, 460,
  695

\bibitem[{{Vasiliev}(2019)}]{Vasiliev2019}
{Vasiliev}, E. 2019, \mnras, 489, 623

\bibitem[{{Voirin} {et~al.}(2018){Voirin}, {Manara}, \& {Prusti}}]{Voirin2018}
{Voirin}, J., {Manara}, C.~F., \& {Prusti}, T. 2018, \aap, 610, A64

\bibitem[{{Vuong} {et~al.}(2001){Vuong}, {Cambr{\'e}sy}, \&
  {Epchtein}}]{Vuong2001}
{Vuong}, M.~H., {Cambr{\'e}sy}, L., \& {Epchtein}, N. 2001, \aap, 379, 208

\bibitem[{{Wang} \& {Chen}(2019)}]{Wang2019}
{Wang}, S. \& {Chen}, X. 2019, \apj, 877, 116

\bibitem[{{Wenger} {et~al.}(2000){Wenger}, {Ochsenbein}, {Egret}, {Dubois},
  {Bonnarel}, {Borde}, {Genova}, {Jasniewicz}, {Lalo{\"e}}, {Lesteven}, \&
  {Monier}}]{Wenger2000}
{Wenger}, M., {Ochsenbein}, F., {Egret}, D., {et~al.} 2000, \aaps, 143, 9

\bibitem[{{Whittet} {et~al.}(1997){Whittet}, {Prusti}, {Franco}, {Gerakines},
  {Kilkenny}, {Larson}, \& {Wesselius}}]{Whittet1997}
{Whittet}, D.~C.~B., {Prusti}, T., {Franco}, G.~A.~P., {et~al.} 1997, \aap,
  327, 1194

\bibitem[{{Wright} {et~al.}(2010){Wright}, {Eisenhardt}, {Mainzer}, {Ressler},
  {Cutri}, {Jarrett}, {Kirkpatrick}, {Padgett}, {McMillan}, {Skrutskie},
  {Stanford}, {Cohen}, {Walker}, {Mather}, {Leisawitz}, {Gautier}, {McLean},
  {Benford}, {Lonsdale}, {Blain}, {Mendez}, {Irace}, {Duval}, {Liu}, {Royer},
  {Heinrichsen}, {Howard}, {Shannon}, {Kendall}, {Walsh}, {Larsen}, {Cardon},
  {Schick}, {Schwalm}, {Abid}, {Fabinsky}, {Naes}, \& {Tsai}}]{WISE}
{Wright}, E.~L., {Eisenhardt}, P. R.~M., {Mainzer}, A.~K., {et~al.} 2010, \aj,
  140, 1868

\end{thebibliography}

\begin{appendix}
\section{Tables (online material)}\label{appendix_tables}

\begin{landscape}
\begin{table}
\caption{Properties of the 188 cluster members selected from our membership analysis in Cha~I. (This table will be available in its entirety in machine-readable form.) }\label{tab_members_Cha1}
\scriptsize{
\begin{tabular}{lcccccccccccccccc}
\hline\hline
Source Identifier&$\alpha$&$\delta$&$\mu_{\alpha}\cos\delta$&$\mu_{\delta}$&$\varpi$&RUWE&Prob.&$V_{r}$&Ref&$d$&$U$&$V$&$W$&Cloud&SED\\
&(h:m:s) &($^{\circ}$ $^\prime$ $^\prime$$^\prime$)&(mas/yr)&(mas/yr)&(mas)&&&(km/s)&&(pc)&(km/s)&(km/s)&(km/s)&&\\
\hline\hline

Gaia DR2 5201384581492347904 & 10 53 39.65 & -77 12 33.9 & $ -22.980 \pm 0.172 $& $ 3.329 \pm 0.154 $& $ 5.213 \pm 0.103 $& $ 1.08 $& $ 0.9723 $& \nodata & \nodata & $ 190.8 ^{+ 2.9 }_{ -3.5 } $& \nodata & \nodata & \nodata & Cha I (south) & \nodata \\
Gaia DR2 5201001852664912512 & 10 55 09.54 & -77 30 54.2 & $ -22.210 \pm 0.136 $& $ 0.771 \pm 0.120 $& $ 5.170 \pm 0.071 $& $ 1.32 $& $ 1.0000 $& $ 16.83 \pm 0.90 $& 1& $ 192.1 ^{+ 2.1 }_{ -2.5 } $& $ -10.2 ^{+ 0.9 }_{ -0.9 } $& $ -20.7 ^{+ 1.0 }_{ -1.0 } $& $ -12.7 ^{+ 0.6 }_{ -0.6 } $& Cha I (south) & Class~III \\
Gaia DR2 5201378641555926656 & 10 55 59.05 & -77 24 39.4 & $ -23.904 \pm 0.077 $& $ 4.889 \pm 0.072 $& $ 5.466 \pm 0.041 $& $ 1.31 $& $ 0.9794 $& \nodata & \nodata & $ 184.3 ^{+ 1.3 }_{ -1.2 } $& \nodata & \nodata & \nodata & Cha I (south) & \nodata \\
Gaia DR2 5201378641555926784 & 10 55 59.69 & -77 24 40.1 & $ -23.813 \pm 0.059 $& $ 2.193 \pm 0.053 $& $ 5.403 \pm 0.031 $& $ 1.21 $& $ 0.9998 $& \nodata & \nodata & $ 185.6 ^{+ 1.6 }_{ -1.4 } $& \nodata & \nodata & \nodata & Cha I (south) & \nodata \\
Gaia DR2 5201536661993294720 & 10 56 16.28 & -76 30 53.1 & $ -22.712 \pm 0.180 $& $ 1.275 \pm 0.160 $& $ 5.089 \pm 0.106 $& $ 1.22 $& $ 0.9999 $& $ 12.56 \pm 2.01 $& 1& $ 192.6 ^{+ 2.0 }_{ -2.9 } $& $ -13.0 ^{+ 1.5 }_{ -1.5 } $& $ -17.6 ^{+ 2.0 }_{ -2.0 } $& $ -11.1 ^{+ 1.0 }_{ -1.0 } $& Cha I (south) & Class~II \\
Gaia DR2 5201388292344009984 & 10 56 30.28 & -77 11 39.4 & $ -23.714 \pm 0.037 $& $ 2.822 \pm 0.035 $& $ 5.462 \pm 0.019 $& $ 1.07 $& $ 0.9993 $& $ 15.30 \pm 0.03 $& 2& $ 184.1 ^{+ 1.2 }_{ -1.0 } $& $ -11.6 ^{+ 0.4 }_{ -0.4 } $& $ -20.3 ^{+ 0.2 }_{ -0.2 } $& $ -10.7 ^{+ 0.3 }_{ -0.3 } $& Cha I (south) & \nodata \\
Gaia DR2 5201191037382571904 & 10 57 53.66 & -77 24 49.7 & $ -23.477 \pm 0.092 $& $ 2.389 \pm 0.080 $& $ 5.397 \pm 0.049 $& $ 1.20 $& $ 0.9999 $& $ 15.71 \pm 0.32 $& 1& $ 186.1 ^{+ 2.1 }_{ -1.8 } $& $ -11.4 ^{+ 0.6 }_{ -0.6 } $& $ -20.5 ^{+ 0.5 }_{ -0.5 } $& $ -11.1 ^{+ 0.4 }_{ -0.4 } $& Cha I (south) & \nodata \\
Gaia DR2 5201387776948008320 & 10 58 05.86 & -77 11 50.1 & $ -23.199 \pm 0.231 $& $ 2.441 \pm 0.204 $& $ 5.360 \pm 0.104 $& $ 1.00 $& $ 0.9999 $& \nodata & \nodata & $ 188.0 ^{+ 3.3 }_{ -3.0 } $& \nodata & \nodata & \nodata & Cha I (south) & Class~II \\
Gaia DR2 5201199348142249216 & 10 58 16.64 & -77 17 17.2 & $ -23.375 \pm 0.081 $& $ 1.865 \pm 0.070 $& $ 5.267 \pm 0.042 $& $ 2.17 $& $ 0.9993 $& \nodata & \nodata & $ 191.1 ^{+ 1.9 }_{ -1.9 } $& \nodata & \nodata & \nodata & Cha I (south) & \nodata \\
Gaia DR2 5201199352439663872 & 10 58 17.95 & -77 17 19.9 & $ -22.369 \pm 0.166 $& $ 1.878 \pm 0.192 $& $ 5.412 \pm 0.091 $& $ 1.22 $& $ 0.9999 $& \nodata & \nodata & $ 186.7 ^{+ 3.1 }_{ -2.3 } $& \nodata & \nodata & \nodata & Cha I (south) & \nodata \\
Gaia DR2 5201568891428175872 & 10 58 54.74 & -75 58 19.6 & $ -22.403 \pm 0.111 $& $ 0.704 \pm 0.083 $& $ 5.070 \pm 0.058 $& $ 1.17 $& $ 0.9999 $& \nodata & \nodata & $ 193.6 ^{+ 1.4 }_{ -2.0 } $& \nodata & \nodata & \nodata & Cha I (north) & Class~III \\
Gaia DR2 5201196599363180928 & 10 59 00.95 & -77 22 40.9 & $ -23.385 \pm 0.036 $& $ 2.220 \pm 0.033 $& $ 5.399 \pm 0.019 $& $ 1.07 $& $ 0.9998 $& $ 17.80 \pm 0.10 $& 3& $ 186.2 ^{+ 1.4 }_{ -1.4 } $& $ -10.3 ^{+ 0.5 }_{ -0.5 } $& $ -22.3 ^{+ 0.3 }_{ -0.3 } $& $ -11.6 ^{+ 0.3 }_{ -0.3 } $& Cha I (south) & \nodata \\
Gaia DR2 5201226389256838528 & 10 59 06.87 & -77  01 40.3 & $ -22.726 \pm 0.050 $& $ 2.217 \pm 0.044 $& $ 5.334 \pm 0.023 $& $ 1.10 $& $ 0.9989 $& $ 14.99 \pm 0.06 $& 2& $ 187.1 ^{+ 1.6 }_{ -1.5 } $& $ -11.3 ^{+ 0.4 }_{ -0.4 } $& $ -19.8 ^{+ 0.2 }_{ -0.2 } $& $ -10.6 ^{+ 0.3 }_{ -0.3 } $& Cha I (south) & \nodata \\
Gaia DR2 5201553704423697792 & 11  00 40.14 & -76 19 28.0 & $ -22.404 \pm 0.076 $& $ 0.664 \pm 0.064 $& $ 5.221 \pm 0.043 $& $ 1.09 $& $ 1.0000 $& $ 15.69 \pm 0.03 $& 2& $ 192.2 ^{+ 1.8 }_{ -1.9 } $& $ -10.8 ^{+ 0.4 }_{ -0.4 } $& $ -20.0 ^{+ 0.2 }_{ -0.2 } $& $ -11.8 ^{+ 0.3 }_{ -0.3 } $& Cha I (north) & Class~II \\
Gaia DR2 5225597893420402176 & 11  00 46.99 & -75 40 36.4 & $ -22.753 \pm 0.981 $& $ 2.673 \pm 0.680 $& $ 5.943 \pm 0.450 $& $ 4.90 $& $ 0.9985 $& \nodata & \nodata & $ 187.5 ^{+ 4.7 }_{ -3.2 } $& \nodata & \nodata & \nodata & Cha I (south) & \nodata \\
Gaia DR2 5225597889120103424 & 11  00 49.18 & -75 40 41.4 & $ -22.042 \pm 0.139 $& $ 1.660 \pm 0.098 $& $ 5.045 \pm 0.063 $& $ 1.10 $& $ 0.9994 $& \nodata & \nodata & $ 193.6 ^{+ 1.4 }_{ -1.8 } $& \nodata & \nodata & \nodata & Cha I (north) & \nodata \\
Gaia DR2 5201194816953813504 & 11  01 13.58 & -77 22 38.7 & $ -23.609 \pm 0.193 $& $ 3.238 \pm 0.182 $& $ 5.395 \pm 0.086 $& $ 1.08 $& $ 0.9993 $& \nodata & \nodata & $ 187.0 ^{+ 3.0 }_{ -2.5 } $& \nodata & \nodata & \nodata & Cha I (south) & Class~III \\
Gaia DR2 5201185574182015744 & 11  01 19.09 & -77 32 38.7 & $ -22.653 \pm 0.435 $& $ 2.062 \pm 0.397 $& $ 5.408 \pm 0.188 $& $ 0.98 $& $ 0.9992 $& \nodata & \nodata & $ 187.9 ^{+ 4.1 }_{ -3.4 } $& \nodata & \nodata & \nodata & Cha I (south) & Class~III \\
Gaia DR2 5201185574184171776 & 11  01 19.31 & -77 32 37.5 & $ -23.668 \pm 0.748 $& $ 1.931 \pm 0.723 $& $ 5.433 \pm 0.337 $& $ 1.24 $& $ 0.9951 $& \nodata & \nodata & $ 188.5 ^{+ 4.2 }_{ -3.7 } $& \nodata & \nodata & \nodata & Cha I (south) & \nodata \\
Gaia DR2 5201218864474119808 & 11  01 31.93 & -77 18 25.0 & $ -24.715 \pm 0.578 $& $ 1.682 \pm 0.515 $& $ 5.776 \pm 0.301 $& $ 1.04 $& $ 0.9903 $& \nodata & \nodata & $ 187.0 ^{+ 4.4 }_{ -2.8 } $& \nodata & \nodata & \nodata & Cha I (south) & Class~III \\

\hline
\end{tabular}
\tablefoot{For each star, we provide the Gaia-DR2 identifier, position, proper motion and parallax (not corrected for zero-point offset) from the Gaia-DR2 catalogue, RUWE value, membership probability, RV with reference, distance derived from Bayesian inference, UVW spatial velocity, molecular cloud, and SED class. References for radial velocities are: (1)~\citet{Sacco2017}, (2)~\citet{Nguyen2012}, (3)~\citet{Guenther2007}, and (4)~\citet{Joergens2001}.}
}
\end{table}
\end{landscape}
\clearpage

\begin{landscape}
\begin{table}
\caption{Properties of the 41 cluster members selected from our membership analysis in Cha~II. (This table will be available in its entirety in machine-readable form.) }\label{tab_members_Cha2}
\scriptsize{
\begin{tabular}{lcccccccccccccccc}
\hline\hline
Source Identifier&$\alpha$&$\delta$&$\mu_{\alpha}\cos\delta$&$\mu_{\delta}$&$\varpi$&RUWE&Prob.&$V_{r}$&Ref&$d$&$U$&$V$&$W$&Cloud&SED\\
&(h:m:s) &($^{\circ}$ $^\prime$ $^\prime$$^\prime$)&(mas/yr)&(mas/yr)&(mas)&&&(km/s)&&(pc)&(km/s)&(km/s)&(km/s)&&\\
\hline\hline

Gaia DR2 5789146187323908736 & 12 56 33.55 & -76 45 45.6 & $ -21.265 \pm 0.113 $& $ -7.648 \pm 0.077 $& $ 5.106 \pm 0.052 $& $ 1.14 $& $ 1.0000 $& $ 15.90 \pm 0.50 $& 1& $ 195.6 ^{+ 2.3 }_{ -2.0 } $& $ -9.0 ^{+ 0.7 }_{ -0.7 } $& $ -22.5 ^{+ 0.7 }_{ -0.7 } $& $ -10.2 ^{+ 0.4 }_{ -0.4 } $& Cha~II & Class~II \\
Gaia DR2 5789258681110332416 & 12 58 56.05 & -76 30 10.7 & $ -21.082 \pm 0.117 $& $ -9.394 \pm 0.112 $& $ 5.189 \pm 0.067 $& $ 1.01 $& $ 0.9999 $& $ 9.00 \pm 5.40 $& 1& $ 194.7 ^{+ 2.1 }_{ -1.9 } $& $ -12.4 ^{+ 3.4 }_{ -3.4 } $& $ -16.5 ^{+ 4.7 }_{ -4.7 } $& $ -9.7 ^{+ 1.6 }_{ -1.6 } $& Cha~II & Class~III \\
Gaia DR2 5788954902364046848 & 12 59 09.78 & -76 51 03.9 & $ -21.189 \pm 0.276 $& $ -8.357 \pm 0.234 $& $ 5.242 \pm 0.147 $& $ 1.01 $& $ 0.9872 $& \nodata & \nodata & $ 195.5 ^{+ 3.2 }_{ -2.5 } $& \nodata & \nodata & \nodata & Cha~II & Class~II \\
Gaia DR2 5788930820483641088 & 12 59 10.05 & -77 12 14.0 & $ -22.775 \pm 0.146 $& $ -8.302 \pm 0.122 $& $ 5.096 \pm 0.072 $& $ 1.01 $& $ 0.9999 $& \nodata & \nodata & $ 196.0 ^{+ 2.8 }_{ -2.3 } $& \nodata & \nodata & \nodata & Cha~II & Class~III \\
Gaia DR2 5788932607187147136 & 13  00 53.14 & -77  09 09.4 & $ -20.751 \pm 0.130 $& $ -7.734 \pm 0.138 $& $ 5.002 \pm 0.087 $& $ 1.62 $& $ 0.9995 $& \nodata & \nodata & $ 198.6 ^{+ 2.6 }_{ -3.0 } $& \nodata & \nodata & \nodata & Cha~II & \nodata \\
Gaia DR2 5789044482500658560 & 13  00 53.15 & -76 54 15.3 & $ -20.653 \pm 0.076 $& $ -8.238 \pm 0.062 $& $ 5.137 \pm 0.044 $& $ 1.20 $& $ 1.0000 $& \nodata & \nodata & $ 194.3 ^{+ 1.8 }_{ -1.6 } $& \nodata & \nodata & \nodata & Cha~II & Class~II \\
Gaia DR2 5788932607187147264 & 13  00 53.49 & -77  09 08.5 & $ -21.401 \pm 0.110 $& $ -8.245 \pm 0.099 $& $ 5.039 \pm 0.067 $& $ 1.05 $& $ 0.9997 $& \nodata & \nodata & $ 197.8 ^{+ 2.6 }_{ -2.8 } $& \nodata & \nodata & \nodata & Cha~II & \nodata \\
Gaia DR2 5788932778988729856 & 13  00 55.24 & -77  08 29.8 & $ -21.916 \pm 0.085 $& $ -8.171 \pm 0.068 $& $ 5.098 \pm 0.048 $& $ 1.00 $& $ 1.0000 $& $ 10.00 \pm 6.00 $& 1& $ 195.8 ^{+ 2.2 }_{ -2.0 } $& $ -12.7 ^{+ 3.7 }_{ -3.7 } $& $ -18.1 ^{+ 5.2 }_{ -5.2 } $& $ -8.9 ^{+ 1.8 }_{ -1.8 } $& Cha~II & Class~III \\
Gaia DR2 5789044478200012160 & 13  00 56.14 & -76 54 02.5 & $ -19.570 \pm 0.078 $& $ -7.927 \pm 0.063 $& $ 4.876 \pm 0.046 $& $ 1.66 $& $ 0.9998 $& \nodata & \nodata & $ 201.4 ^{+ 1.6 }_{ -1.6 } $& \nodata & \nodata & \nodata & Cha~II & \nodata \\
Gaia DR2 5788090759241383680 & 13  01 58.80 & -77 51 22.0 & $ -20.869 \pm 0.038 $& $ -7.759 \pm 0.031 $& $ 5.030 \pm 0.020 $& $ 1.44 $& $ 1.0000 $& \nodata & \nodata & $ 195.9 ^{+ 1.7 }_{ -1.6 } $& \nodata & \nodata & \nodata & Cha~II & \nodata \\
Gaia DR2 5789239989411572608 & 13  02 13.43 & -76 37 58.0 & $ -20.990 \pm 0.040 $& $ -9.144 \pm 0.041 $& $ 5.152 \pm 0.024 $& $ 1.15 $& $ 0.9999 $& $ 15.90 \pm 1.10 $& 2& $ 194.1 ^{+ 1.5 }_{ -1.5 } $& $ -8.8 ^{+ 1.0 }_{ -1.0 } $& $ -22.3 ^{+ 1.2 }_{ -1.2 } $& $ -11.0 ^{+ 0.5 }_{ -0.5 } $& Cha~II & Class~II \\
Gaia DR2 5788861173293937792 & 13  02 22.75 & -77 34 49.6 & $ -19.786 \pm 0.165 $& $ -7.035 \pm 0.139 $& $ 4.884 \pm 0.097 $& $ 1.01 $& $ 0.9984 $& $ 12.00 \pm 7.20 $& 1& $ 200.3 ^{+ 2.1 }_{ -2.6 } $& $ -10.6 ^{+ 4.4 }_{ -4.6 } $& $ -19.2 ^{+ 6.2 }_{ -6.3 } $& $ -8.7 ^{+ 2.2 }_{ -2.3 } $& Cha~II & Class~II \\
Gaia DR2 5789026065680783104 & 13  03 04.36 & -77  07 02.9 & $ -20.011 \pm 0.361 $& $ -8.649 \pm 0.290 $& $ 5.367 \pm 0.201 $& $ 6.78 $& $ 0.9991 $& \nodata & \nodata & $ 195.6 ^{+ 3.3 }_{ -2.5 } $& \nodata & \nodata & \nodata & Cha~II & \nodata \\
Gaia DR2 5788087666864871168 & 13  03 08.96 & -77 55 59.7 & $ -19.791 \pm 0.181 $& $ -6.943 \pm 0.138 $& $ 4.677 \pm 0.095 $& $ 1.22 $& $ 0.9990 $& $ 11.00 \pm 6.60 $& 1& $ 201.4 ^{+ 1.8 }_{ -2.0 } $& $ -11.9 ^{+ 4.2 }_{ -4.3 } $& $ -18.9 ^{+ 5.7 }_{ -5.8 } $& $ -8.6 ^{+ 2.1 }_{ -2.2 } $& Cha~II & Class~III \\
Gaia DR2 5789245005933444096 & 13  03 16.08 & -76 29 38.2 & $ -19.677 \pm 0.127 $& $ -5.945 \pm 0.105 $& $ 4.965 \pm 0.077 $& $ 1.03 $& $ 0.9997 $& \nodata & \nodata & $ 199.6 ^{+ 2.3 }_{ -2.7 } $& \nodata & \nodata & \nodata & Cha~II & Class~III \\
Gaia DR2 5789045856890263936 & 13  04 22.76 & -76 50 05.6 & $ -19.954 \pm 0.054 $& $ -7.434 \pm 0.046 $& $ 5.025 \pm 0.029 $& $ 1.02 $& $ 1.0000 $& \nodata & \nodata & $ 197.1 ^{+ 1.9 }_{ -1.8 } $& \nodata & \nodata & \nodata & Cha~II & Class~II \\
Gaia DR2 5789045856890265216 & 13 04 24.02 & -76 50 01.3 & $ -20.258 \pm 0.050 $& $ -7.117 \pm 0.044 $& $ 5.003 \pm 0.026 $& $ 1.03 $& $ 0.9999 $& $ 15.20 \pm 0.20 $& 1& $ 198.0 ^{+ 1.8 }_{ -1.8 } $& $ -8.7 ^{+ 0.5 }_{ -0.5 } $& $ -22.0 ^{+ 0.5 }_{ -0.4 } $& $ -9.1 ^{+ 0.3 }_{ -0.3 } $& Cha~II & Class~II \\
Gaia DR2 5788088457138909056 & 13  04 24.77 & -77 52 30.4 & $ -19.406 \pm 0.158 $& $ -6.898 \pm 0.119 $& $ 4.911 \pm 0.080 $& $ 1.21 $& $ 0.9963 $& $ 13.00 \pm 7.80 $& 1& $ 199.4 ^{+ 2.4 }_{ -2.9 } $& $ -9.7 ^{+ 4.8 }_{ -4.7 } $& $ -19.8 ^{+ 6.7 }_{ -6.6 } $& $ -8.7 ^{+ 2.4 }_{ -2.4 } $& Cha~II & Class~II \\
Gaia DR2 5788199129851179392 & 13  04 55.62 & -77 39 51.0 & $ -19.568 \pm 0.427 $& $ -8.617 \pm 0.319 $& $ 5.003 \pm 0.203 $& $ 1.98 $& $ 0.9889 $& \nodata & \nodata & $ 197.9 ^{+ 3.2 }_{ -3.3 } $& \nodata & \nodata & \nodata & Cha~II & \nodata \\
Gaia DR2 5788199129856209792 & 13  04 55.62 & -77 39 49.3 & $ -20.629 \pm 0.054 $& $ -7.880 \pm 0.046 $& $ 5.071 \pm 0.033 $& $ 1.13 $& $ 1.0000 $& \nodata & \nodata & $ 196.7 ^{+ 1.9 }_{ -1.9 } $& \nodata & \nodata & \nodata & Cha~II & Class~II \\

\hline
\end{tabular}
\tablefoot{For each star, we provide the Gaia-DR2 identifier, position, proper motion and parallax (not corrected for zero-point offset) from the Gaia-DR2 catalogue, RUWE value, membership probability, RV with reference, distance derived from Bayesian inference, UVW spatial velocity, molecular cloud, and SED class. References for radial velocities are: (1)~\citet{Biazzo2012}, (2)~\citet{Torres2006}, and (3)~Gaia-DR2.}
}
\end{table}
\end{landscape}
\clearpage

\begin{table*}
\centering
\caption{Membership probability of all sources in the field derived independently using different $p_{in}$ values in the membership analysis conducted for Cha~I. (This table will be available in its entirety in machine-readable form.)
\label{tab_prob_Cha1}}
\begin{tabular}{cccccc}
\hline\hline
Source Identifier&probability&probability&probability&probability&probability\\
&($p_{in}=0.5$)&($p_{in}=0.6$)&($p_{in}=0.7$)&($p_{in}=0.8$)&($p_{in}=0.9$)\\
\hline\hline

Gaia DR2 5201074042471837568 &		1.4349E-147  &		2.7899E-149 &		5.4960E-151 &		7.4803E-152 &		5.5727E-157\\
Gaia DR2 5201073836316342528 &		8.6941E-281  &		2.9651E-282 &		4.4077E-284 &		5.4432E-292 &		1.6288E-294\\
Gaia DR2 5201073836313415424 &		4.3543E-109 &		1.1548E-109 &		3.3785E-112 &		6.4841E-113 &		1.9674E-118\\
Gaia DR2 5201073935099216512 &		1.0370E-187 &		4.3339E-190 &		3.6931E-193 &		8.3850E-197 &		9.3870E-203\\
Gaia DR2 5201073836316344320 &		2.3984E-298 &		2.3984E-298 &		2.3984E-298 &		2.3983E-298 &		2.3983E-298\\
Gaia DR2 5201073935099219072 &		3.6232E-116 &		2.6937E-117 &		5.4370E-119 &		1.8111E-119 &		1.0402E-123\\
Gaia DR2 5201073832019988480 &		9.1292E-259 &		9.9697E-261 &		1.8108E-264 &		2.4042E-269 &		2.1896E-278\\
Gaia DR2 5201073939393526528 &		1.5073E-17 &		1.4872E-17 &		9.2370E-18 &		6.6952E-18 &		1.7472E-18\\
Gaia DR2 5201073935099210240 &		1.9355E-106 &		2.9213E-107 &		6.1859E-109 &		1.8016E-109 &		9.9123E-114\\
Gaia DR2 5201073935099216128 &		1.4987E-168 &		2.5022E-169 &		1.7000E-172 &		7.1575E-174 &		1.7152E-181\\
Gaia DR2 5201073832019985536 &		5.7921E-186 &		8.4142E-187 &		6.1339E-190 &		3.3510E-192 &		1.2762E-199\\
Gaia DR2 5201073973755611904 &		2.3469E-76 &		1.3902E-76 &		1.5145E-78 &		7.1632E-79 &		2.7975E-83\\
Gaia DR2 5201073939393029632 &		6.6388E-126 &		6.1494E-126 &		3.7579E-127 &		8.4288E-128 &		4.6618E-131\\
Gaia DR2 5201074351710395392 &		1.6818E-83 &		1.6836E-83 &		1.0307E-84 &		3.9460E-85 &		1.5117E-88\\
Gaia DR2 5201073935102778624 &		1.4768E-297 &		1.4768E-297 &		1.4768E-297 &		1.4767E-297 &		1.4767E-297\\
Gaia DR2 5201073969459128320 &		6.1980E-30 &		6.3911E-30 &		1.8565E-30 &		1.3947E-30 &		5.7569E-32\\
Gaia DR2 5201074381775839872 &		3.6815E-297 &		3.1125E-298 &		3.1069E-298 &		3.1069E-298 &		3.1069E-298\\
Gaia DR2 5201074347416098560 &		3.6472E-78 &		2.2980E-78 &		1.4179E-79 &		4.7970E-80 &		5.5506E-83\\
Gaia DR2 5201073939393531648 &		2.3397E-81 &		1.5185E-81 &		2.8149E-83 &		1.4863E-83 &		6.8345E-88\\
Gaia DR2 5201073973752401664 &		1.7528E-212 &		6.1688E-215 &		1.3509E-217 &		6.8561E-220 &		7.4386E-228\\

\hline\hline
\end{tabular}
\end{table*}

\begin{table*}
\centering
\caption{Membership probability of all sources in the field derived independently using different $p_{in}$ values in the membership analysis conducted for Cha~II. (This table will be available in its entirety in machine-readable form.)
\label{tab_prob_Cha2}}
\begin{tabular}{cccccc}
\hline\hline
Source Identifier&probability&probability&probability&probability&probability\\
&($p_{in}=0.5$)&($p_{in}=0.6$)&($p_{in}=0.7$)&($p_{in}=0.8$)&($p_{in}=0.9$)\\
\hline\hline

Gaia DR2 5789212089299174016 & 	5.0757E-58 & 	2.3736E-59 & 	2.3849E-59 & 	9.2531E-58 & 	2.3516E-59\\
Gaia DR2 5789212089301993216 & 	3.7768E-293 & 	3.7767E-293 & 	3.7767E-293 & 	3.7767E-293 & 	3.7767E-293\\
Gaia DR2 5789211161587017600 & 		1.3585E-23 & 	9.7102E-24 & 	9.7266E-24 & 	2.9880E-24 & 	9.6661E-24\\
Gaia DR2 5789211157294203008 & 		1.7804E-10 & 	1.6458E-10 & 	1.6459E-10 & 	1.6232E-10 & 	1.6454E-10\\
Gaia DR2 5789210951131853824 & 	1.5332E-98 & 	8.5218E-100 & 	8.6591E-100 & 	7.8309E-89 & 	8.2448E-100\\
Gaia DR2 5789163882589260416 & 	6.2073E-21 & 	7.1975E-21 & 	7.2019E-21 & 	1.1882E-20 & 	7.1882E-21\\
Gaia DR2 5789211157294205056 & 		3.9985E-152 & 	3.8209E-153 & 	4.0248E-153 & 	2.7734E-139 & 	3.4379E-153\\
Gaia DR2 5789163500333081344 & 	1.7415E-61 & 	6.7563E-61 & 	6.8618E-61 & 	9.2221E-58 & 	6.5271E-61\\
Gaia DR2 5789211157290990080 & 		2.7402E-79 & 	1.3261E-79 & 	1.3710E-79 & 	4.7306E-84 & 	1.2402E-79\\
Gaia DR2 5789212085003280512 & 	5.2980E-204 & 	1.1604E-204 & 	1.3612E-204 & 	5.1170E-224 & 	8.4274E-205\\
Gaia DR2 5789210951131856128 & 	2.7245E-224 & 	2.1403E-224 & 	2.3742E-224 & 	1.1483E-201 & 	1.7413E-224\\
Gaia DR2 5789163672132325504 & 	1.9522E-32 & 	1.6995E-32 & 	1.7025E-32 & 	1.1393E-31 & 	1.6931E-32\\
Gaia DR2 5789210951131851008 & 	6.0084E-60 & 	6.2951E-61 & 	6.3458E-61 & 	3.4878E-62 & 	6.1709E-61\\
Gaia DR2 5789163672131796864 & 	9.6607E-121 & 	2.4770E-121 & 	2.6914E-121 & 	1.3474E-112 & 	2.0736E-121\\
Gaia DR2 5789163500333632256 & 	6.0395E-62 & 	1.5696E-61 & 	1.5891E-61 & 	1.8863E-57 & 	1.5279E-61\\
Gaia DR2 5789163878290242176 & 	8.1662E-297 & 	8.1659E-297 & 	8.1659E-297 & 	8.1659E-297 & 	8.1659E-297\\
Gaia DR2 5789162400824292224 & 	1.0405E-06 & 	2.5707E-07 & 	2.5707E-07 & 	2.6156E-07 & 	2.5709E-07\\
Gaia DR2 5789212020582705536 & 	2.7468E-34 & 	1.7392E-34 & 	1.7444E-34 & 	3.6916E-35 & 	1.7262E-34\\
Gaia DR2 5789211195946756736 & 		7.1236E-20 & 	5.6508E-20 & 	5.6523E-20 & 	5.8063E-20 & 	5.6475E-20\\
Gaia DR2 5789210951131859072 & 	1.7569E-108 & 	7.4057E-110 & 	7.7175E-110 & 	4.9623E-108 & 	6.8126E-110\\

\hline\hline
\end{tabular}
\end{table*}

\begin{table*}
\centering
\caption{Empirical isochrone of Cha~I inferred from our membership analysis. (This table will be available in its entirety in machine-readable form.)
\label{tab_isochrone_Cha1}}
\begin{tabular}{cc}
\hline\hline
$G_{RP}$&$G-G_{RP}$\\
(mag)&(mag)\\
\hline\hline

6.143&	-0.027\\
6.175&	-0.020\\
6.207&	-0.012\\
6.239&	-0.005\\
6.271&	0.003\\
6.303&	0.010\\
6.335&	0.018\\
6.366&	0.025\\
6.398&	0.033\\
6.430&	0.040\\
6.462&	0.048\\
6.494&	0.055\\
6.526&	0.063\\
6.558&	0.070\\
6.590&	0.078\\
6.622&	0.085\\
6.654&	0.093\\
6.686&	0.100\\
6.718&	0.107\\
6.750&	0.115\\

\hline\hline
\end{tabular}
\end{table*}

\begin{table*}
\centering
\caption{Empirical isochrone of Cha~II inferred from our membership analysis. (This table will be available in its entirety in machine-readable form.)
\label{tab_isochrone_Cha2}}
\begin{tabular}{cc}
\hline\hline
$G_{RP}$&$G-G_{RP}$\\
(mag)&(mag)\\
\hline\hline

10.528&	0.949\\
10.544&	0.949\\
10.561&	0.950\\
10.578&	0.950\\
10.595&	0.951\\
10.612&	0.951\\
10.629&	0.952\\
10.646&	0.952\\
10.663&	0.953\\
10.679&	0.953\\
10.696&	0.954\\
10.713&	0.954\\
10.730&	0.954\\
10.747&	0.955\\
10.764&	0.955\\
10.781&	0.956\\
10.798&	0.956\\
10.814&	0.957\\
10.831&	0.957\\
10.848&	0.958\\

\hline\hline
\end{tabular}
\end{table*}

\begin{table*}
\centering
\caption{Stellar parameters for the HRD analysis. (This table will be available in its entirety in machine-readable form.)
\label{tab_HRD}}
\scriptsize{
\begin{tabular}{ccccccccccc}
\hline\hline
Gaia-DR2 Identifier&$\alpha$&$\delta$&$J$&$A_{J}$&SpT&T$_{eff}$&$L_{\star}$&$t_{BHAC15}$&$t_{SDF00}$&Subgroup\\
&(h:m:s) &($^{\circ}$ $^\prime$ $^\prime$$^\prime$)&&(mag)&(mag)&(K)&($L_{\odot}$)&(Myr)&(Myr)\\
\hline\hline

Gaia DR2 5201384581492347904 & 10 53 39.65 & -77 12 33.9 & $ 13.282 \pm 0.024 $& 0.63 & M2.75 & $ 3392 \pm 82 $& $ 0.047 ^{+ 0.002 }_{ -0.003 } $& $ 15.8 ^{+ 8.6 }_{ -6.5 } $& $ 13.6 ^{+ 5.4 }_{ -4.0 } $& Cha I (south) \\
Gaia DR2 5201001852664912512 & 10 55 09.54 & -77 30 54.2 & $ 11.881 \pm 0.024 $& 0.34 & M4.5 & $ 3020 \pm 120 $& $ 0.117 ^{+ 0.005 }_{ -0.006 } $& & $ 2.7 ^{+ 0.7 }_{ -0.0 } $& Cha I (south) \\
Gaia DR2 5201378641555926784 & 10 55 59.69 & -77 24 40.1 & $ 10.778 $& 0.79 & M0   & $ 3770 \pm 62 $& $ 0.583 ^{+ 0.010 }_{ -0.009 } $& $ 1.1 ^{+ 0.3 }_{ -0.0 } $& $ 1.8 ^{+ 0.2 }_{ -0.1 } $& Cha I (south) \\
Gaia DR2 5201536661993294720 & 10 56 16.28 & -76 30 53.1 & $ 12.542 \pm 0.024 $& 0.00 & M5.6 & $ 2900 \pm 100 $& $ 0.049 ^{+ 0.002 }_{ -0.002 } $& $ 1.1 ^{+ 1.5 }_{ -0.0 } $& & Cha I (south) \\
Gaia DR2 5201388292344009984 & 10 56 30.28 & -77 11 39.4 & $ 9.969 \pm 0.024 $& 0.23 & M0.5 & $ 3700 \pm 62 $& $ 0.706 ^{+ 0.025 }_{ -0.023 } $& & $ 1.4 ^{+ 0.2 }_{ -0.2 } $& Cha I (south) \\
Gaia DR2 5201191037382571904 & 10 57 53.66 & -77 24 49.7 & $ 11.615 \pm 0.023 $& 0.34 & M4   & $ 3160 \pm 120 $& $ 0.146 ^{+ 0.007 }_{ -0.006 } $& $ 1.4 ^{+ 0.5 }_{ -0.0 } $& $ 2.9 ^{+ 0.5 }_{ -0.5 } $& Cha I (south) \\
Gaia DR2 5201387776948008320 & 10 58 05.86 & -77 11 50.1 & $ 13.405 \pm 0.028 $& 0.45 & M5.25 & $ 2988 \pm 100 $& $ 0.034 ^{+ 0.002 }_{ -0.002 } $& $ 3.4 ^{+ 1.4 }_{ -2.2 } $& $ 4.0 ^{+ 2.7 }_{ -0.0 } $& Cha I (south) \\
Gaia DR2 5201196599363180928 & 10 59 00.95 & -77 22 40.9 & $ 10.135 \pm 0.023 $& 0.34 & K8   & $ 3940 \pm 22 $& $ 0.741 ^{+ 0.028 }_{ -0.026 } $& $ 1.3 ^{+ 0.2 }_{ -0.2 } $& $ 2.0 ^{+ 0.3 }_{ -0.1 } $& Cha I (south) \\
Gaia DR2 5201226389256838528 & 10 59 06.87 & -77  01 40.3 & $ 8.462 \pm 0.032 $& 0.00 & K2   & $ 4760 \pm 92 $& $ 3.159 ^{+ 0.151 }_{ -0.141 } $& & $ 2.6 ^{+ 0.7 }_{ -0.7 } $& Cha I (south) \\
Gaia DR2 5201553704423697792 & 11  00 40.14 & -76 19 28.0 & $ 11.859 \pm 0.026 $& 0.20 & M3.75 & $ 3210 \pm 120 $& $ 0.111 ^{+ 0.005 }_{ -0.005 } $& $ 2.0 ^{+ 1.3 }_{ -0.6 } $& $ 3.7 ^{+ 1.0 }_{ -0.7 } $& Cha I (north) \\
Gaia DR2 5201194816953813504 & 11  01 13.58 & -77 22 38.7 & $ 13.064 \pm 0.024 $& 0.27 & M5.25 & $ 2988 \pm 100 $& $ 0.040 ^{+ 0.002 }_{ -0.002 } $& $ 2.9 ^{+ 1.3 }_{ -0.0 } $& $ 4.0 ^{+ 2.1 }_{ -0.0 } $& Cha I (south) \\
Gaia DR2 5201185574182015744 & 11  01 19.09 & -77 32 38.7 & $ 13.096 \pm 0.033 $& 0.45 & M7.25 & $ 2630 \pm 58 $& $ 0.040 ^{+ 0.003 }_{ -0.003 } $& & & Cha I (south) \\
Gaia DR2 5201218864474119808 & 11  01 31.93 & -77 18 25.0 & $ 14.631 \pm 0.039 $& 0.00 & M8   & $ 2570 \pm 50 $& $ 0.007 ^{+ 0.001 }_{ -0.000 } $& & & Cha I (south) \\
Gaia DR2 5225599096011296640 & 11  02 19.17 & -75 36 57.6 & $ 12.131 \pm 0.027 $& 0.27 & M4.5 & $ 3020 \pm 120 $& $ 0.088 ^{+ 0.004 }_{ -0.004 } $& $ 1.5 ^{+ 1.0 }_{ -0.0 } $& $ 3.2 ^{+ 0.8 }_{ -0.0 } $& Cha I (north) \\
Gaia DR2 5225743475632752128 & 11  02 26.01 & -75  02 40.7 & $ 11.761 \pm 0.026 $& 0.14 & M4.75 & $ 2950 \pm 100 $& $ 0.098 ^{+ 0.005 }_{ -0.004 } $& & $ 1.8 ^{+ 1.4 }_{ -0.0 } $& Cha I (north) \\
Gaia DR2 5201182413088266624 & 11  02 41.71 & -77 24 24.6 & $ 12.804 \pm 0.026 $& 0.72 & M5   & $ 2880 \pm 100 $& $ 0.067 ^{+ 0.003 }_{ -0.004 } $& & & Cha I (south) \\
Gaia DR2 5201206048291386880 & 11  02 54.91 & -77 21 50.8 & $ 11.565 \pm 0.026 $& 0.00 & M4.5 & $ 3020 \pm 120 $& $ 0.105 ^{+ 0.004 }_{ -0.004 } $& $ 1.1 ^{+ 0.8 }_{ -0.0 } $& $ 2.8 ^{+ 0.7 }_{ -0.0 } $& Cha I (south) \\
Gaia DR2 5201181313576638208 & 11  03 41.74 & -77 26 52.1 & $ 12.997 \pm 0.024 $& 0.61 & M5.5 & $ 2925 \pm 100 $& $ 0.057 ^{+ 0.003 }_{ -0.003 } $& $ 1.4 ^{+ 1.2 }_{ -0.0 } $& $ 0.5 ^{+ 4.0 }_{ -0.0 } $& Cha I (south) \\
Gaia DR2 5201206361825924992 & 11  03 47.54 & -77 19 56.5 & $ 11.313 \pm 0.023 $& 0.68 & M5   & $ 2880 \pm 100 $& $ 0.260 ^{+ 0.010 }_{ -0.011 } $& & & Cha I (north) \\
Gaia DR2 5201335481425778560 & 11  04 04.14 & -76 39 33.0 & $ 12.953 \pm 0.023 $& 0.74 & M4.25 & $ 3090 \pm 120 $& $ 0.063 ^{+ 0.003 }_{ -0.003 } $& $ 2.7 ^{+ 1.7 }_{ -1.1 } $& $ 4.6 ^{+ 1.4 }_{ -1.6 } $& Cha I (south) \\

\hline\hline
\end{tabular}
}
\tablefoot{We provide for each star the Gaia-DR2 identifier and position, $J$-band photometry and extinction in this band, spectral type, effective temperature, bolometric luminosity derived in this study, ages estimates inferred from the \citet[][BHAC15]{BHAC15} and \citet[][SDF00]{Siess2000} models, and the subgroup to which the star belongs. We used the spectral types and extinctions compiled by \citet{Esplin2017} and \citet{Spezzi2008} for Cha~I and Cha~II stars, respectively.   }
\end{table*}

\end{appendix}
\end{document}